\documentclass[pra,twocolumn,showpacs,nofootinbib,superscriptaddress,notitlepage]{revtex4-1}
\usepackage{amsmath,empheq}
\usepackage{amssymb,bm}
\usepackage{amsthm,color,xcolor,dsfont}
\usepackage{graphicx} 
\usepackage{xcolor}
\usepackage{epstopdf}
\usepackage[colorlinks=true, hyperindex, breaklinks, linkcolor=blue, urlcolor=blue, citecolor=blue]{hyperref} % Physical Review style
\usepackage{ulem}
\usepackage{caption}
\usepackage[labelformat=simple]{subcaption}

\usepackage{cleveref}
\usepackage{mathtools}
\usepackage{comment}

\input{Qcircuit}

\DeclarePairedDelimiter\ceil{\lceil}{\rceil}

\newcommand{\dyad}[2]{\ket{#1}\bra{#2}} % dyad
\begin{document}

\title{Overhead analysis of universal concatenated quantum codes}

\author{Christopher Chamberland}
\email{c6chambe@uwaterloo.ca}
\affiliation{
    Institute for Quantum Computing and Department of Physics and Astronomy,
    University of Waterloo,
    Waterloo, Ontario, N2L 3G1, Canada
    }
\author{Tomas Jochym-O'Connor}
\email{trjochym@uwaterloo.ca}
\affiliation{
    Institute for Quantum Computing and Department of Physics and Astronomy,
    University of Waterloo,
    Waterloo, Ontario, N2L 3G1, Canada    
    }
\affiliation{
Walter Burke Institute for Theoretical Physics and Institute for Quantum Information and Matter,
California Institute of Technology, 
Pasadena, California 91125, USA
}
\author{Raymond Laflamme}
\affiliation{
    Institute for Quantum Computing and Department of Physics and Astronomy,
    University of Waterloo,
    Waterloo, Ontario, N2L 3G1, Canada
    }
\affiliation{Perimeter Institute, Waterloo, Ontario, N2L~2Y5, Canada}
\affiliation{Canadian Institute For Advanced Research, Toronto, Ontario, M5G~1Z8, Canada}

\begin{abstract}
We analyze the resource overhead of recently proposed methods for universal fault-tolerant quantum computation using concatenated codes. Namely, we examine the concatenation of the 7-qubit Steane code with the 15-qubit Reed-Muller code, which allows for the construction of the 49 and 105-qubit codes that do not require the need for magic state distillation for universality. We compute a lower bound for the adversarial noise threshold of the 105-qubit code and find it to be $8.33\times 10^{-6}.$ We obtain a depolarizing noise threshold for the 49-qubit code of $9.69\times 10^{-4}$ which is competitive with the 105-qubit threshold result of $1.28\times 10^{-3}$. We then provide lower bounds on the resource requirements of the 49 and 105-qubit codes and compare them with the surface code implementation of a logical $T$~gate using magic state distillation. For the sampled input error rates and noise model, we find that the surface code achieves a smaller overhead compared to our concatenated schemes. 
\end{abstract}


\maketitle

\section{Introduction}

Fault-tolerant computations provide a means to control and suppress error rates to arbitrarily low levels, without a detrimental overhead in terms of the number of qubits and computation time. However, estimating the additional resources that would be required for such computations is an important area of study as physical architectures begin to approach the realms of scalability~\cite{WHL05, SJ09, DFS+09, vMLFY10, JvMF12, YJG+12, DS13, DSMN13, BKM+14, CGM+14, KBF+15, TCM+16}. As such, it has become increasingly important to evaluate whether different architectures have particular advantages over one another with respect to targeting the implementation of a given algorithm with required accuracy.

The early proposals for fault-tolerant architectures were given by concatenated quantum error correcting codes, where qubits forming an error correcting code are re-encoded for further protection. This provides a means to reduce the error rate in a double-exponential manner as the number of concatenation levels increase, assuming the physical error rate is below some threshold value, deemed the fault-tolerance threshold~\cite{AB97, Preskill98}. Subsequently, topological quantum codes were proposed, beginning with the surface code~\cite{Kitaev97}. There, logical information is stored in highly non-local degrees of freedom, while using local stabilizer checks, thus providing the ability to increase the protection of the code by increasing the size of the physical lattice encoding the information. One of the primary advantages of schemes such as the surface code is its high threshold value in comparison to concatenated code schemes. For depolarizing noise on each gate and memory location, the threshold value of the surface code is on the order of~$10^{-2}$ \cite{DKLP02,FMMC12} in comparison to~$10^{-3}$ \cite{AGP06, PR12} for most concatenated schemes. Moreover, stabilizer syndrome checks are simpler as the weight of the checks remains fixed as the distance increases, unlike in the case of concatenated architectures. 

The goal of this work is to estimate the overhead with a recently proposed scheme for fault-tolerant computation using concatenated codes that allows for the implementation of a universal set of gates without the need for magic state distillation~\cite{JL14, CJL16}. Magic state distillation forecasts to be a challenge for logical computation on 2D~topological codes, such as the surface codes. As such, many recent developments in the area of quantum error correction have focused on circumventing no go theorems regarding the implementation of universal quantum logic using transversal gates~(the simplest form of fault-tolerant operation)~\cite{EK09, ZCC11, PY15}. The 105-qubit scheme circumvents these no go theorems by concatenating complementary sets of transversal gates and lead to a fault-tolerant threshold of~$1.28 \times 10^{-3}$ for depolarizing noise. While this threshold rate compares favourably with other concatenated methods, it is still an order of magnitude below that of the surface code, thus will require higher distance iterations to reach a given target error rate when compared to the surface code. However, the primary advantage of the universal scheme is to avoid magic state distillation and as such this potential reduction in complexity would allow for the concatenated model to have a reduced overall overhead. The main result of this work is to provide a lower bound on the number of qubits and gates that would be required for the universal concatenated scheme to reach particular target error rates, given a physical depolarizing error rate. In order to do so, the overhead in state preparation needed for Steane error correction~\cite{Steane97} as well as the suppression of the logical error rate as a function of concatenation level are determined. 

In this work we make no assumptions on the locality of the code, treating each location and gate with equal weighting in terms of accessibility and error probability. We acknowledge that this may be unrealistic for many current realizations of physical quantum computing experiments, although there are some exceptions~\cite{Haffneretal05}, however in order to asses whether such a scheme would provide a benefit in terms of overhead with respect to local codes such as the surface code, it is necessary to treat them on equal footing. If a non-local scheme were to show significant improvements over schemes that are local, then this would motivate the experimental community to optimize for better performance of long range gates. However, if codes such as the surface code are shown to be more efficient in even this non-realistic non-local error model, then this would further solidify the status of such local codes as the most promising physical schemes.

In Section~\ref{sec:Fault-tolerant scheme} we review the concepts behind Steane error correction and outline important parameters for the counting of resources related to the logical ancilla state preparation. In Section~\ref{sec:Malignant set counting overview} we review the concepts of malignant set counting~\cite{AGP06} which allows for the establishment of a lower bound for the theoretical threshold. In Section~\ref{sec:UniversalCodes} we review the 105-qubit code that allows for universal fault-tolerant computation, and discuss a reduction of this code to 49~qubits~\cite{NSZ16}. We additionally present models for decoding in these codes given a particular error syndrome, highlighting important differences needed depending on the logical gate implemented. Additionally, we provide new logical ancilla state preparation circuits in order to reduce circuit depth allowing for increased success probability and higher threshold values and we present the adversarial noise threshold. In Section~\ref{sec:Concatenated 49-qubit thresholds} we review and present the depolarizing noise threshold results for the 105-qubit and 49-qubit codes, respectively. In Section~\ref{sec:Resource overhead 49} we calculate a lower bound on the qubit and gate resource overhead for the implementation of the Hadamard and CNOT gates under depolarizing noise as they dominate the resource requirements in the universal concatenated method. We compare these results with an estimate for the surface code qubit resource overhead using magic state distillation. We conclude the article with a review of the results and their impact for quantum architectures, and provide open questions and targets for future universal quantum codes.

\section{Fault-tolerant scheme for error correction} \label{sec:Fault-tolerant scheme}

\begin{figure}[h]%[htbp]
\centering
\includegraphics[width=0.5\textwidth]{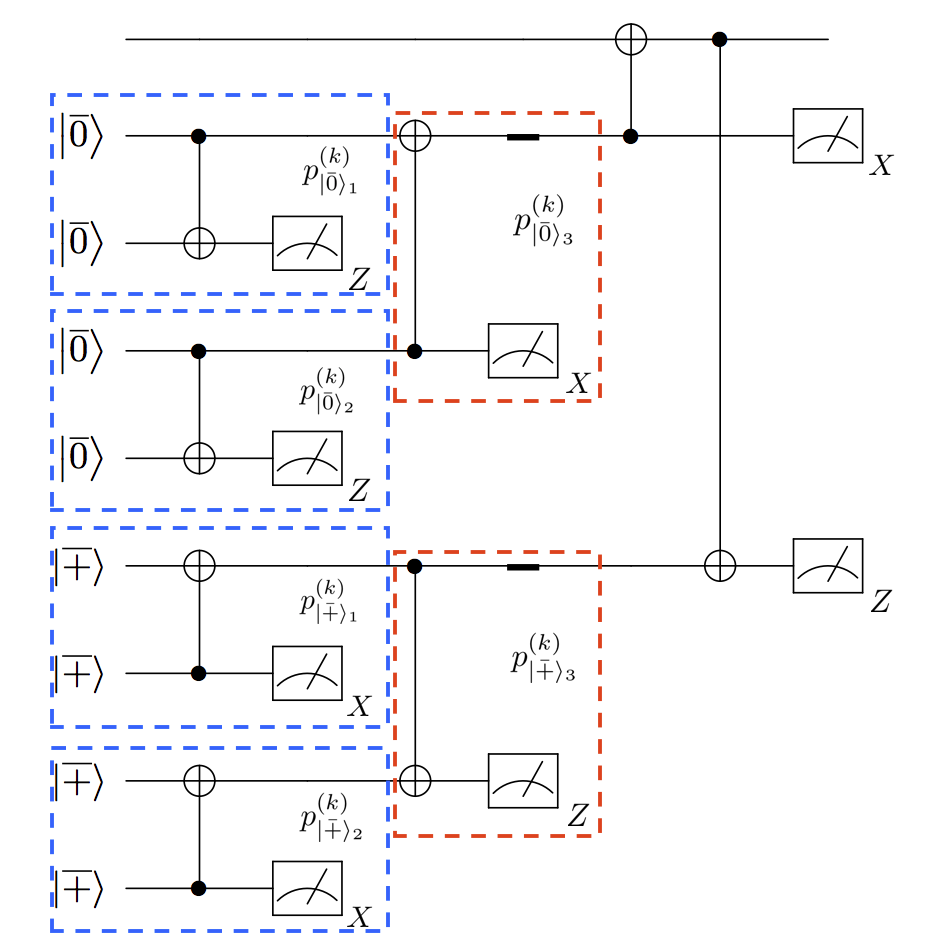}
\caption{Illustration of Steane's error correction circuit. The encoded $\ket{\overline{0}}$ and $\ket{\overline{+}}$ ancilla states are used to detect and correct $Z$ and $X$ errors. However, because the encoding circuits are not fault-tolerant, extra $\ket{\overline{0}}$ and $\ket{\overline{+}}$ verifier states are used to detect errors during the encoding step. If an error is detected (or the $-1$ eigenvalue of a logical Pauli operator is measured), the ancilla states are rejected and the error correction round starts over. The terms $p^{(k)}_{\ket{\overline{0}}_{i}}$ and $p^{(k)}_{\ket{\overline{+}}_{j}}$ (where $i,j \in \{1,2\} $) correspond to the probabilities that no $X$ ($Z$) errors are detected in the blocks $\ket{\overline{0}}_{i}$ $(\ket{\overline{+}}_{j})$.  Conditioned on acceptance of the previous blocks,   $p^{(k)}_{\ket{\overline{0}}_{3}}$ ($p^{(k)}_{\ket{\overline{+}}_{3}}$) denote the probabilities that no $Z$ ($X$) errors are detected in the last verifier blocks. These probabilities will be used in the depolarizing noise overhead calculations of the 49-qubit and 105-qubit codes.}
\label{fig:SteaneECcircuit}
\end{figure}

In this section we describe the fault-tolerant error correction scheme used in the implementation of the universal concatenated quantum codes considered in this paper. We use Steane's method for fault-tolerant error correction~\cite{Steane97,AGP06} which applies to CSS codes with stabilizer generators given by tensor products of all $X$ and $I$ operators ($X$ generators) or tensor products of all $Z$ and $I$ operators ($Z$ generators). 

In Steane's method, the ancilla qubits used for syndrome extraction are encoded using the same CSS code that protects the data qubits. Since the stabilizer generators are separated into $X$ and $Z$ generators, we can measure them separately. To measure the $Z$ generators, we prepare the ancilla in the encoded state $\ket{\overline{+}}=(\ket{\overline{0}}+\ket{\overline{1}})/\sqrt{2}$ and apply transversal CNOT gates with the data block as control and the ancilla block as target. This will propagate each $X$ error in the data block to the corresponding position in the ancilla block. The ancilla is encoded in the $\ket{\overline{+}}$ state since it is an eigenstate with eigenvalue one of the logical $X$ Pauli operator and thus the CNOT gate has no effect on the encoded state of the ancilla. If no errors occur in either the preparation of $\ket{\overline{+}}$ or the encoded CNOT gate, then measuring the ancilla block in the $Z$-basis detects the $X$ errors when considering the $Z$~stabilizers of the code which are products of the individual measurements. Similarly, to detect and correct $Z$ errors, we prepare an ancilla state in the $\ket{\overline0}$ basis and apply transversal CNOT gates with the data block as target and ancilla block as control. The sequence ends by performing a measurement of the ancilla in the $X$-basis. In general, the circuits used for encoding the $\ket{\overline{+}}$ and $\ket{\overline{0}}$ ancilla states are not fault-tolerant since a single fault can propagate badly leading to a high weight error at the output of the encoding circuit. In order to ensure fault-tolerance in Steane's error correction method, a verification step is needed to detect if errors occurred during the encoding of the ancilla states. For the $\ket{\overline{0}}$ state, $X$ errors can propagate from the ancilla to the data. Consequently, we encode a "verifier" state in the $\ket{\overline{0}}$ state and apply an encoded CNOT gate with the ancilla block as control and the verifier block as target. Any $X$ errors are then detected by performing a measurement in the $Z$ basis. Not only are $X$ errors detected but after a classical error correction step the eigenvalue of the encoded $\overline{Z}$ operator is also found. If a non-trivial syndrome is measured or the eigenvalue of  $\overline{Z}$ is found to be $-1$, all ancilla blocks are rejected and the error correction protocol starts over. The latter will play an important role when considering the resource overhead of the concatenated 49-qubit and 105-qubit code. 

\section{Malignant set counting overview} \label{sec:Malignant set counting overview}
 
\begin{figure}[h]
\centering
\includegraphics[width=0.4\textwidth]{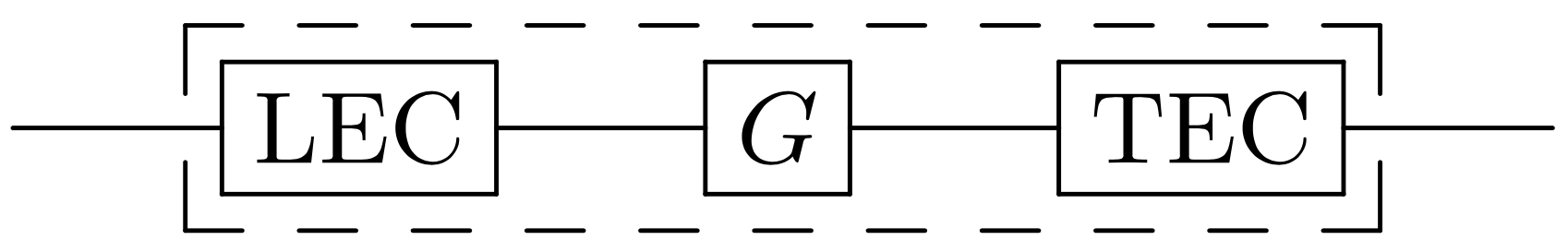}
\caption{Illustration of an extended rectangle consisting of a gate $G$ along with its leading and trailing error correction circuit. Extended rectangles are used in the fault-tolerant implementation of a logical gate. The leading and trailing error correction circuits are implemented using Steane's method (see Section \ref{sec:Fault-tolerant scheme}) which is a protocol for a fault-tolerant error correction.}
\label{fig:exRecCircuit}
\end{figure}

Following \cite{AGP06}, we define a $k$-Rec to be the encoded gate under consideration acting on codewords followed by a trailing error correction (TEC) circuit at the $k$-th level of concatenation. A $k$-exRec (where exRec stands for \textit{extended rectangle}) corresponds to a leading error correction (LEC) circuit followed by a $k$-Rec (see  Figure~\ref{fig:exRecCircuit}).

Consider an arbitrary 1-exRec. A location can be either a 0-preperation, 0-measurment or a gate (note that the identity gate corresponds to a resting qubit). We say that a set of $m$ locations is \textit{benign} if the 1-Rec contained in the 1-exRec is \textit{correct} for arbitrary faults occurring at those locations. Otherwise, the set of locations are defined to be \textit{malignant}. A 1-Rec is correct if it takes any input with no more than one error per block to an output with no more than one error per block. 

The idea behind malignant set counting is that for a fixed number of locations in the circuit, we would like to count all such sets of locations which are malignant. A 1-exRec will be \textit{bad} if it contains faults at a malignant set of locations, otherwise we will define it to be \textit{good}. We can generalize the definition of goodness and badness to a $k$-exRec (for $k>1$) by defining the $k$-exRec to be bad if it contains independent bad $(k - 1)$-exRecs at a malignant set of locations, otherwise we define it to be good.

In this section, we will consider an independent stochastic noise model where fault locations are independently and identically distributed. The operations at the chosen locations are arbitrary trace-preserving completely positive maps. The latter noise model is also known as adversarial noise. Since an arbitrary fault can be expanded in terms of Pauli operators, a set of locations will be benign if the 1-Rec contained in the 1-exRec is correct for all Pauli faults at those locations. Furthermore, if all of the locations are in the LEC of the 1-exRec, then by definition the 1-Rec contained in the 1-exRec is correct. Therefore, when counting malignant sets of locations, we will exclude the cases where all faults are in a LEC. As in Ref.~\cite{AGP06}, for a fixed set of locations, we insert all combinations of $X$ and $Z$ Pauli faults at those locations and propagate them through the 1-exRec. If for all combinations of Pauli faults the 1-Rec contained in the 1-exRec is correct, then the set of locations are benign. 

\begin{table}
\begin{centering}
\begin{tabular}{|c|c|}
\hline 
Level-0 location label & Types of level-0 locations\tabularnewline
\hline 
\hline 
1 & Rest during a gate cycle\tabularnewline
\hline 
2 & Rest during a measurement cycle\tabularnewline
\hline 
3 & Preparation of $\ket{0}$\tabularnewline
\hline
4 & Preparation of $\ket{+}$\tabularnewline
\hline 
5 & Measurement of $X$\tabularnewline
\hline 
6 & Measurement of $Z$\tabularnewline
\hline 
7 & CNOT gate\tabularnewline
\hline 
8 & $H$ gate\tabularnewline
\hline 
9 & $T$ gate\tabularnewline
\hline 
\end{tabular}
\par\end{centering}
\caption{\label{tab:Types-of-Locations}Types of level-0 locations present in the considered 1-exRec's.}
\end{table}

Consider a 1-exRec with a total of $L$ locations and suppose that the EC's have $L_{EC}$ locations. For a circuit simulating a $t$-qubit gate, we define 

\begin{align}
L_{n,t}\equiv{{L \choose n} - t{L_{EC} \choose n}},
\label{eq:LocationexRec1}
\end{align}
where $n$ is the number of faulty locations in the circuit. Similarly, define 

\begin{align}
A_{n,t}\equiv{f_{n}L_{n,t}},
\label{eq:Adefinition}
\end{align}
where $f_{n}=n_{mal}/N$ is the fraction of malignant locations containing $n$ faults. Due to limits in computation time, when calculating the noise threshold for a particular gate, we count malignant sets of locations for up to $m$ faults and assume that all $m$ + 1 sets of locations are malignant. For large enough $m$, the error in the truncation can be made very small \cite{AGP06}. Defining the probability of a $k$-exRec to be bad by $\varepsilon^{(k)}$, we can use the statistical independence of bad $(k-1)$-exRecs to calculate an upper bound on $\varepsilon^{(k)}$

\begin{align}
\varepsilon^{(k)}&\leq A_{2,t}(\varepsilon^{(k-1)})^{2} + A_{3,t}(\varepsilon^{(k-1)})^{3} + \cdots \nonumber \\  
&\qquad  + A_{m-1,t}(\varepsilon^{(k-1)})^{m-1} + L_{m,t }(\varepsilon^{(k-1)})^{m}.
\label{eq:Acalculation}
\end{align}

Note that in the above equation we have assumed that all level-0 locations in our circuit have the same fault rate $\varepsilon$. In our scheme, noisy circuits will be constructed from $\ket{0}$ and $\ket{+}$ initialization, Hadamard and CNOT gates as well as single qubit measurements in the $X$ and $Z$ eigenbasis. The level-0 locations used in our circuits are given in Table~\ref{tab:Types-of-Locations}. Since the wait time of a qubit during the application of a gate could differ from the wait time of a measurement, we can associate different failure rates for the two rest cycles. Generalizing Eq.~\ref{eq:Acalculation} to include different failure rates for distinct level-0 locations, we obtain 

\begin{align}
\varepsilon^{(k)}&\leq \sum_{j\leq i=1}^{l_{max}}\alpha_{ij,t}^{(2)}\varepsilon_{i}^{(k-1)}\varepsilon_{j}^{(k-1)} + \nonumber \\  
&\sum_{l\leq j\leq i=1}^{l_{max}}\alpha_{ijl,t}^{(3)}\varepsilon_{i}^{(k-1)}\varepsilon_{j}^{(k-1)}\varepsilon_{l}^{(k-1)} + \cdots + L_{m,t}(\varepsilon^{(k-1)})^{m},
\label{eq:GeneralAcalculation}
\end{align}
where $l_{max}$ depends on the types of locations in a particular exRec. Hence, exRec's excluding $H$ or $T$ gates would have $l_{max}=7$. The indices correspond to a particular level-0 location and are summed over all location types of the exRec under consideration. $\alpha_{ij,t}^{(2)}$ corresponds to the number of malignant pairs of types $i$ and $j$ for a gadget acting on $t$ qubits. For example, following the indexing from Table~\ref{tab:Types-of-Locations}, 5 is the label used for a measurement in the $X$ basis and 7 is the label for a CNOT gate. Therefore, $\alpha_{57,t}^{(2)}$ corresponds to the number of malignant pairs in the exRec where one location is a measurement in the $X$ basis and the other location is a CNOT gate. Generalizing to larger sets of locations, $\alpha_{i_{1}i_{2} \cdots i_{n},t}^{(n)}$ is the number of malignant sets of $n$ locations of type $i_{1},i_{2}, \cdots ,i_{n}$. For the remainder of this section, we will assume that all location types have the same failure probability, so that the upper bound in Eq.~\ref{eq:Acalculation} will be used for the threshold calculation (in this case  $A_{2,t}$ is simply the sum of all the elements of the $\alpha_{t}^{(2)}$ matrix, and $\varepsilon_{i}^{(k-1)}=\varepsilon_{j}^{(k-1)}=\varepsilon^{(k-1)}$).

From Eq.~\ref{eq:Acalculation}, it follows that   

\begin{align}
\varepsilon^{(k)}\leq A'_{t}(\varepsilon^{(k-1)})^{2},
\label{eq:EpsMal2}
\end{align}
where the threshold estimate is 

\begin{align}
\varepsilon \leq \varepsilon_{0}=(A'_{t})^{-1}.
\label{eq:EpsThresh}
\end{align}

$A'_{t}$ can be calculated from the polynomial equation 

\begin{align}
&(A'_{t})^{m}-A_{2,t}(A'_{t})^{m-1}-A_{3,t}(A'_{t})^{m-2}- \nonumber \\ 
&\qquad \cdots -A_{m-1,t}A'_{t}-L_{m,t}=0.
\label{eq:ApolyEqu}
\end{align}
Since the coefficients $A_{j,t}$ are strictly increasing for increasing $j$, there will only be one positive solution to Eq.~\ref{eq:ApolyEqu}. 

Recall that the ancilla blocks used to extract the error syndrome need to successfully pass a verification test, otherwise the ancilla's are rejected and the computation is started over. Instead, the failure probability for a $k$-exRec should be upper bounded conditioned on the acceptance of all ancilla blocks. In calculating the coefficients $A_{j,t}$, we count sets of locations such that faults at those locations cause the $k$-exRec to fail but lead to acceptance of all ancilla blocks. Hence, the upper bound in Eq.~\ref{eq:Acalculation} actually corresponds to the \textit{joint} probability of acceptance of all ancilla blocks and failure of the $k$-exRec. We can use Bayes' rule to obtained the conditional probability of failure given the acceptance of all ancillas. Using the notation from Ref.~\cite{AGP06}, we need to calculate the probability $P_{\ket{\overline{0}},\mathrm{accept}}^{(k)}$ and $P_{\ket{\overline{+}},\mathrm{accept}}^{(k)}$ that a level-$k$ encoded $\ket{\overline{0}}$ or $\ket{\overline{+}}$ passes the verification test. Distinction between $P_{\ket{\overline{0}},\mathrm{accept}}^{(k)}$ and $P_{\ket{\overline{+}},\mathrm{accept}}^{(k)}$ is important since the encoding circuits for $\ket{\overline{0}}$ and $\ket{\overline{+}}$ are not symmetric for the 15-qubit Reed-Muller code (since the code is not self-dual) and so they will contain a different number of locations. We define 

\begin{align}
P_{\mathrm{min},\mathrm{accept}}^{(k)}\equiv \mathrm{min}\{P_{\ket{\overline0},\mathrm{accept}}^{(k)},P_{\ket{\overline+},\mathrm{accept}}^{(k)}\},
\label{eq:Pmin}
\end{align}
which corresponds to the smallest of the two acceptance probabilities. In Steane's error correction protocol, there will always be an equal number of encoded $\ket{\overline0}$ and $\ket{\overline+}$ circuits in the $k$-EC's. We define $n_{anc}$ to be the number of encoded $\ket{\overline0}$ or $\ket{\overline+}$ circuits in the $k$-exRec under consideration. Using Bayes' rule, the probability of failure $\varepsilon^{(k)}$ for the $k$-exRec, conditioned on acceptance of all ancillas can be upper bounded as 

\begin{align}
\varepsilon^{(k)}\leq (P_{\mathrm{min},\mathrm{accept}}^{(k)})^{-n_{anc}}\varepsilon_{\mathrm{joint}}^{(k)},
\label{eq:Eq9}
\end{align}
where $\varepsilon_{\mathrm{joint}}^{(k)}$ is upper bounded by Eq.~\ref{eq:Acalculation}. Let $C_{\ket{\overline0}}$ and $C_{\ket{\overline+}}$ correspond to the number of locations in the encoding and verification circuits of $\ket{\overline0}$ and $\ket{\overline+}$ in the EC's. For an ancilla to be rejected, the previous encoding and verification circuits must contain at least one bad $(k-1)$-exRec. Therefore, a lower bound on $P_{\mathrm{min},\mathrm{accept}}$ is given by 

\begin{align}
P_{\mathrm{min},\mathrm{accept}}^{(k)}\geq 1-C_{\mathrm{max}}\varepsilon^{(k-1)},
\label{eq:Eq10}
\end{align}
where $C_{\mathrm{max}}\equiv \mathrm{max}\{C_{\ket{\overline0},\ket{\overline+}}\}$. Using \ref{eq:Pmin}--\ref{eq:Eq10} and assuming that $\varepsilon^{(k-1)}<(A'_{t})^{-1}$, we have that

\begin{align}
\varepsilon^{(k)}\leq \Big(1-\frac{C_{\mathrm{max}}}{A'_{t}}\Big)^{-n_{anc}}A'_{t}(\varepsilon^{(k-1)})^{2}.
\label{eq:Eq11}
\end{align}
Defining 

\begin{align}
A''_{t}\equiv \Big(1-\frac{C_{\mathrm{max}}}{A'_{t}}\Big)^{-n_{anc}}A'_{t},
\label{eq:AdoublePrime}
\end{align}
the threshold estimate is then given by 

\begin{align}
\varepsilon_{0}=(A''_{t})^{-1}.
\label{eq:EpsThreshold}
\end{align}

\section{The 105 and 49-qubit codes}
\label{sec:UniversalCodes}

In this section we will review the construction of the 105-qubit concatenated code that allows for universal fault-tolerant logical gate implementations, as well as a simplification of this construction to 49~qubits~\cite{NSZ16}. We shall then discuss the different ways we implement decoding and error correction given syndrome measurements. Our methods for constructing the ancilla state preparation circuits can be found in \cref{app:StatePrepCircuits}.

\subsection{Fault-tolerant universal concatenated quantum codes}
The idea behind the 105-qubit construction is to use two different error correcting codes, with different sets of transversal gates (logical gates that can be implemented by applying individual gates to each qubit composing the code), in concatenation in order to implement a universal set of fault-tolerant logical operations~\cite{JL14}. In the construction of the 105-qubit code, the outer code is designated as the 7-qubit code and therefore contains transversal Clifford operations. The inner code, that is each qubit composing the 7-qubit code, is the 15-qubit code which has transversal logical $CNOT$ and~$T$~gates, where~$T = \dyad{0}{0} +e^{i \pi /4} \dyad{1}{1}$ and completes the universal gate set when combined with Clifford operations. Since the $T$~gate cannot be implemented transversally for the 7-qubit code, any logical construction will necessarily have to couple different qubits in the code (in the case of the 105-qubit code, coupling qubits corresponds to coupling blocks of qubits composing the code). There exists a construction for the implementation of the $T$~gate using a sequence of $CNOT$~gates and a~$T$~gate on the qubits of the 7-qubit code. However, since each of these operations are implemented at the logical level from the perspective of the 15-qubit code, they will all be transversal with respect to this code. Therefore, any single qubit fault occurring during the action of this sequence of gates may lead to a propagation of the faults, but in a controlled manner to only a single location to each of the codeblocks. Therefore, the logical gate remains fault-tolerant as any single fault remains correctable. 

The logical Hadamard is implemented by applying the logical Hadamard gate on each of the seven encoded codeblocks (as the Hadamard is transversal for the 7-qubit code). However, the Hadamard is not transversal on each 15-qubit codeblock, and as such a single error may spread to form a logical fault on an individual codeblock. However, since only one codeblock is corrupted, overall the error will remain correctable as such an error can be detected and corrected by the 7-qubit outer code syndrome. This complication will have important consequences for decoding, as explained further in this section.

An important observation is that the re-encoding of the qubits into 15-qubit codeblocks is only important to protecting the blocks that have active gates in the implementation of the logical $T$~gate. Therefore, it is only necessary to encode three codeblocks, corresponding to the set of blocks that would correspond to a logical Pauli operator for the 7-qubit code~\cite{NSZ16}. Therefore, the total qubit count can be reduced to be $3 \times 15 + 4 = 49$~qubits. While the overall distance of the code will be lower, reduced from a distance 9 to distance 5 code, the logical operations will still be able to correct for arbitrary single qubit faults, just as in the case of the 105-qubit code. This idea can be generalized further to any construction using as an outer code a 2D~color code, which has transversal Clifford operations, and choosing a set of qubits that contain both logical Pauli~operators and re-encoding these qubits in a 3D~color code containing the transversal~$CNOT$ and $T$~gates required from the original scheme~\cite{AD13private}.

\subsection{Decoding the 105-qubit code}
As we have previously explained, the 105-qubit code is a distance~9 quantum error correcting code whose distance is sacrificed for the  implementation of the non globally-transversal $H$ and $T$~gates. However, the decoder should be designed such that for the $CNOT$~gate any weight-4 error can be corrected. The 105-qubit syndromes consist of the 15-qubit syndromes on each of the codeblocks composing the outer code, as well as the 6~syndromes of the 7-qubit code which correspond to syndromes across 4~blocks (since all stabilizers of the 7-qubit code have weight~4). 

The most basic decoding scheme is the greedy decoder, where each of the 15-qubit codeblocks are corrected according to their measured syndromes individually, and then the outer 7-qubit code is corrected independently according to the remaining syndrome~\footnote{Note that in Steane error correction, all of the syndromes are measured in parallel for both the 15-qubit codeblocks as well as the outer 7-qubit code. Any correction made at the 15-qubit level that would modify the 7-qubit syndrome results can be accounted for in software.}. The greedy decoder will fail to correct for all weight-4 errors. For example, consider the case when two 15-qubit codeblocks each have a weight-2~$Z$ error. Each codeblock is then corrected with the identified single qubit recovery (since the 15-qubit code is weight-3 for $Z$~errors, it can only correct weight-1 errors) resulting in a logical error on the two 15-qubit codeblocks. Then, the outer 7-qubit stabilizers will identify this weight-2 logical fault with a logical fault on a third codeblock, and thus ``correct" by implementing a logical~$Z$ on the third codeblock. As such, the final state will undergo a global logical~$Z$ error from the composition of the 3~logical codeblock errors. Therefore, having corrected each of the 15-qubit codeblocks, when correcting the 7-qubit codeblocks, information from the 15-qubit syndromes will have to be used in the correction of the outer 7-qubit code. 

The decoding of the 105-qubit code will be implemented using the following steps:
\begin{enumerate}
\item Correct all of the 15-qubit codeblocks individually, and store which codeblocks underwent any correction.
\item Update the 7-qubit syndromes according to the corrections from the 15-qubit codeblocks.
\item If the 7-qubit syndrome is trivial (no syndrome identifies an error) then correction complete.
\item If the 7-qubit syndrome identifies an error on a codeblock that did not undergo a 15-qubit correction, and in addition there is a set of complementary blocks~\footnote{The complementary blocks are the two blocks that along with the identified block would form a logical error for the 7-qubit code. For example, $Z_1 Z_2 Z_3$ forms a logical error for the 7-qubit code, therefore complementary blocks for block~1 are (2,3), yet also blocks~(4,5) and~(6,7), since they are logically equivalent.} that were corrected at the 15-qubit level, then perform further logical operations on the complementary blocks, then correction complete.
\item Otherwise, correct the identified codeblock by applying a logical correction.
\end{enumerate} 

We illustrate the advantage of this decoding scheme by highlighting the example that the greedy decoder failed to correct. Consider weight-2 $Z$~errors on the first and second codeblocks. Each of these codeblocks are corrected by applying a weight-1 correction, resulting in logical~$Z$ errors on each of the codeblocks. Then, the 7-qubit syndrome would identify an error on codeblock~3, since the syndrome associated with~$Z_1 Z_2$ is equivalent to~$Z_3$. Since the 15-qubit decoder did not make any corrections on codeblock 3, yet did make corrections to codeblocks 1~and~2, which are complementary to codeblock~3, logical~$Z$ corrections are applied to each of these codeblocks, therefore correcting all of the errors. One can verify that all weight-4 errors will be corrected by this scheme, thus achieving the promised distance of the code. Of course, for the implementation of the logical~$H$ and~$T$ gates, not all weight-4 errors will be corrected as the gates are not globally transversal. However, all weight-1 errors will still be corrected by our decoding scheme.

\subsection{Decoding the 49-qubit code}

The 49-qubit code sacrifices the full distance of the 105-qubit code by only encoding three codeblocks, therefore reducing the overall distance to be 5. As such, any decoder will be able to correct at most any weight-2 error. The correction scheme implemented for the logical~$CNOT$ and $T$ gate is as follows (as described below, a different decoder is used for the logical~$H$):
\begin{enumerate}
\item Correct the three 15-qubit codeblocks, tracking which blocks contained errors.
\item Update the 7-qubit syndromes according to the corrections from the 15-qubit codeblocks.
\item If the 7-qubit syndrome identifies a 15-qubit codeblock that had a trivial syndrome in Step~1, then perform a weight-2 correction on the complementary single qubit blocks, then correction complete.
\item For any other non-trivial 7-qubit syndrome, correct according to the identified single or 15-qubit codeblock.
\end{enumerate}

For the transversal $CNOT$, any weight-1 error will either be corrected originally by the 15-qubit syndrome measurement if it occurs on a codeblock or at the 7-qubit level if it occurs on an unencoded block. If two errors occur there are four possibilities. If both errors occur on the same encoded codeblock, then they are corrected at the 15-qubit level, potentially resulting in a logical fault on that codeblock. However, such a logical fault will be detected by the 7-qubit stabilizers and as such will be corrected according to the protocol. If the two errors occur on different encoded 15-qubit codeblocks, they will each be correctly identified at the 15-qubit level. If one error occurs on a codeblock and one on an unencoded block, then the error on the 15-qubit codeblock will be corrected, and only the error on the single qubit will remain when measuring the 7-qubit syndromes and will be correctly identified. The tricky case comes when errors occur on two different unencoded blocks, therefore not identified by the 15-qubit stabilizers. In this case, the codeblock that is complementary to these two errors will be identified when the 7-qubit stabilizers are measured, however since no error had been identified on that codeblock at the 15-qubit level, the error complementary to that block is corrected on single qubits resulting in an ``error" that will be logically equivalent to the identity.

In the case of the $T$~gate, some of the distance is sacrificed in order to implement the non-transversal $T$~gate, yet remains fault-tolerant and can correct for any weight-1 error. Any weight-1 error is corrected equivalently to a Greedy decoder in the case of our protocol, and as such is corrected in the same manner as in the case of~$CNOT$. The only difference in potential realizations is that a weight-1 error on any of the codeblocks may spread to a weight-1 error on the other codeblocks, yet they all remain correctable through the protocol by correcting at the 15-qubit level.

Unlike in the case of the 105-qubit code, it is important to point out that a different decoder must be used in the case of the logical Hadamard. Due to the non fault-tolerant construction of the logical Hadamard on a given 15-qubit codeblock, a single fault on such a codeblock could result in a logical fault on that codeblock at the conclusion of the gate. Such an error would be corrected in the wrong fashion according to the above decoder, as it would go unrecognized by the 15-qubit stabilizers and would falsely identify a correction on the complementary single qubits. As such, the solution is to correct in a greedy fashion at the 7-qubit level, thus resulting in a correction on the single 15-qubit block. Therefore, the resulting protocol would be to correct in a greedy manner at the 15-qubit level and the 7-qubit level, and weight-2 errors on two single-qubit codeblocks would therefore result in a logical fault unlike in the case of the transversal~$CNOT$.  As mentioned previously, not being able to correct all weight-2 errors is due to sacrificing the full code distance for the implementation of the logical~$H$.

\subsection{Adversarial noise threshold results for the 105-qubit code} \label{Threshold results for adversarial noise}

We present the results of the threshold analysis for the adversarial noise model described in section~\ref{sec:Malignant set counting overview}. We will begin by computing the noise threshold for the Hadamard-exRec which turns out to be the circuit that provides a lower bound for the threshold of the 105-qubit code. From the analysis leading to Eq.~\ref{eq:Acalculation}, we first need to compute the coefficients $A_{j,t}$ where $j\in \{2,3,\cdots,m\}$ for some cutoff value $m$. Given the large number of locations in the exRec's for the 105-qubit code, it is computationally impractical to count all malignant sets of locations. In order to overcome this difficulty, we use a Monte Carlo method following the ideas of Ref.~\cite{Aliferis07}. Suppose we want to count all malignant locations for a set of $j$ faults, i.e. the coefficient $A_{j,t}$. Instead of looking at every combination of $j$ locations within the full exRec, we can uniformly sample the set of all fault paths for a fixed set of $j$ faults. We thus obtain an estimate of the fraction $\hat{f}_{j,t}$ of malignant faults for $j$ sets of locations which in turn provides an estimate of the exact coefficient $A_{j,t}$. The standard error is obtained from the relation

\begin{align}
\sigma_{j,t}=\sqrt{\hat{f}_{j,t}(1-\hat{f}_{j,t})/N},
\label{eq:StandardError}
\end{align}
where $N$ is the sample size. 
Due to computational limits, we set the cutoff value to $m=6$ and choose a sample size of $N=10^{7}$. Since the Hadamard-exRec only contains one LEC, $t=1$ and from Table~\ref{tab:NumLocationsHadamard}

\begin{table}
\begin{centering}
\begin{tabular}{|c|c|}
\hline 
$A_{j,1}$ coefficients for Hadamard 1-exRec & Monte Carlo estimate \tabularnewline
\hline 
\hline 
$A_{2,1}$ & $(6.80\pm 0.24)\times10^{4}$\tabularnewline
\hline 
$A_{3,1}$ & $(1.73\pm 0.03)\times10^{9}$\tabularnewline
\hline 
$A_{4,1}$ & $(1.30\pm 0.05)\times10^{13}$\tabularnewline
\hline
$A_{5,1}$ & $(6.44\pm 0.20)\times10^{16}$\tabularnewline
\hline 
$A_{6,1}$ & $(2.42\pm 0.06)\times10^{20}$\tabularnewline
\hline 
$L_{7,1}$ & $3.47\times10^{25}$\tabularnewline
\hline 
\end{tabular}
\par\end{centering}
\caption{\label{tab:AcoefTable}Value of Hadamard-exRec $A_{j,1}$ coefficients obtained from a Monte Carlo simulation which are used in the threshold calculation for adversarial noise.}
\end{table}

\begin{align}
L_{n,1}={15067 \choose n}-{7110 \choose n}.
\label{eq:HadLn}
\end{align}
The coefficients $A_{j,1}$ can be computed from 

\begin{align}
A_{j,1}=\hat{f}_{j,1}L_{j,1}.
\label{eq:Aj1}
\end{align}
Our Monte Carlo implementation was written in Matlab.  Solving Eq.~\ref{eq:ApolyEqu} with the coefficients from Table~\ref{tab:AcoefTable}, we find that $A'_{1}=(8.92\pm 0.23)\times10^{4}$. Next, we need to compute $C_{\mathrm{max}}$ by counting the number of locations in the encoding and verification circuits of $\ket{\overline0}$ and $\ket{\overline+}$ in the EC's which can be found in \cref{app:StatePrepCircuits}. The encoding and verification circuits contain four encoded state preparations ($\ket{\overline0}$ or $\ket{\overline+}$), three CNOT gates and three measurement locations. Hence, we have that $C_{\ket{\overline+}}=3254$ and $C_{\ket{\overline0}}=3226$ so that $C_{\mathrm{max}}=3254$. Since there are eight $\ket{\overline+}$ preparation circuits, then $n_{anc}=8$. Using Eqs.~\ref{eq:AdoublePrime} and~\ref{eq:EpsThreshold}, we calculate the noise threshold for the Hadamard-exRec to be 

\begin{align}
\varepsilon_{0}=(8.33\pm 0.28)\times 10^{-6}.
\label{eq:AdversarialThresholdHad}
\end{align}

We can repeat the same calculations leading to Eq.~\ref{eq:AdversarialThresholdHad} for the CNOT 1-exRec (which contains 28545 locations compared to 15067 locations for the Hadamard 1-exRec). For $N=10^{7}$ iterations, we found that the fraction of malignant events $\hat{f}_{2,2}$ to $\hat{f}_{9,2}$ were all zero. Clearly, for a large enough number of iterations some of these coefficients would be nonzero. However, given that the Hadamard-exRec circuit is more sensitive to input noise which will lower bound the threshold for the 105-qubit code, and due to limits in computation time, it is impractical to compute the coefficients $A_{j,2}$ for $N>10^{7}$. 

In an adversarial noise model calculation, locations fail independently and at random with the constraint that the error at each location is chosen by an adversary. The error could be correlated with errors at other failing locations. Although the threshold computed in Eq.~\ref{eq:AdversarialThresholdHad} is quite low, it is important to keep in mind that malignant set counting is overly pessimistic for noise models such as depolarizing noise. As pointed out in \cite{PR12}, malignant set counting is independent of the underlying noise model and so ignores large quantities of information. For the remainder of this paper, we will focus on depolarizing noise and compute thresholds for the 49-qubit code. For the 105-qubit code, we will use the threshold results that were calculated in \cite{CJL16} using similar methods. The protocol for our simulation will be explained in section~\ref{sec:Concatenated 49-qubit thresholds}. Using the computed threshold results for both codes, we will obtain the resource overhead for performing encoded gates and compare the results with magic state distillation techniques applied to the surface code.

\section{Concatenated 49 and 105-qubit code thresholds} \label{sec:Concatenated 49-qubit thresholds}

The goal of designing fault-tolerant architectures is to allow arbitrary long computations to be performed on large-scale quantum computers where the probability of failure can be made as small as desired. For concatenated coding schemes, fault-tolerant architectures give rise to \textit{asymptotic thresholds} which corresponds to the error rate $p_{th}$ such that for physical error rates $p<p_{th}$, the logical error rate can be made as small as desired for sufficiently large number of concatenation levels. Furthermore, a quantum circuit containing $A$ gates can be simulated with probability of error at most $\epsilon$ with a space/time overhead which scales as $\mathcal{O}(\mathrm{poly}(\log A/\epsilon)A)$ \cite{NC00}. 

In this study, fault-tolerant syndrome measurement and error correction is implemented using Steane's method (see Section~\ref{sec:Fault-tolerant scheme} for a detailed description of the method) in order to take advantage of the CSS structure of the codes. Error correction steps are interleaved between the implementation of each fault-tolerant gate. Following Ref.~\cite{AGP06}, at the first level of concatenation, each logical gate is represented by a 1-exRec as illustrated in Fig~\ref{fig:exRecCircuit}. Hence, the components in a level-1 logical gate will consist of state preparation and measurement, physical gates and memory locations.  We use a recursive simulation for higher levels of concatenation where a fault-tolerant gate at level $k$ is constructed by replacing each level-0 location in the level-$(k-1)$ logical gate by the corresponding level-1 rectangle. 

The noise threshold for the 49 and 105-qubit code is computed by considering a depolarizing noise model for each physical (level-0) location. The depolarizing channel for a single qubit is define by the quantum operation 

\begin{align}
\varepsilon(\rho)=(1-\frac{3p}{4})\rho + \frac{p}{4}(X\rho X + Y\rho Y + Z\rho Z),
\label{eq:DepolarizingChannel}
\end{align}
where $p$ will be referred to as the physical error rate. Using the same parameters as Ref.~\ref{eq:DepolarizingChannel}, the depolarizing channel is generalized to all forms of quantum operations as follows:
\begin{enumerate}
   \item A noisy CNOT gate is modelled as applying a CNOT gate followed by, with probability $\frac{15p}{16}$, a two-qubit Pauli error drawn uniformly and independently from $\{I,X,Y,Z\}^{\otimes 2}\setminus \{I\otimes I\}$. 
   \item A noisy preparation of the $\ket{0}$ state is modelled as the ideal preparation of the $\ket{0}$ state with probability $1-\frac{p}{2}$ and $\ket{1}=X\ket{0}$ with probability $\frac{p}{2}$ (we use $\frac{p}{2}$ instead of $\frac{p}{4}$ since $Y$ errors have the same effect as $X$ errors). Similarly, the noisy preparation of the $\ket{+}$ state is modelled as the ideal preparation of the $\ket{+}$ state with probability $1-\frac{p}{2}$ and $\ket{-}=Z\ket{+}$ with probability $\frac{p}{2}$.
   \item A noisy measurement in the $Z$-basis is modelled by applying a Pauli $X$ error with probability $\frac{p}{2}$ followed by an ideal measurement in the $Z$-basis. Similarly, a noisy measurement in the $X$-basis is modelled by applying a Pauli $Z$ error with probability $\frac{p}{2}$ followed by an ideal measurement in the $X$-basis.
   \item A single-qubit gate error or storage error is modelled by applying the ideal gate (identity gate for a resting qubit) with probability $1-\frac{3p}{4}$. With probability $\frac{3p}{4}$, the ideal gate is implemented followed by a Pauli error chosen uniformly from the set  $\{ X,Y,Z \}$. 
\end{enumerate}
To determine the probability of having a logical fault at the output of an exRec, we first define the notion of a malignant error event. Let $\ket{\psi_{1}}$ be a single or two-qubit logical state obtained by applying ideal decoders immediately after the LEC circuit and $\ket{\psi_{2}}$ the logical state obtained by applying ideal decoders immediately after the TEC. The event $\mathrm{mal}_{E}$ is defined as $\ket{\psi_{2}}=EU\ket{\psi_{1}}$ where $E$ is a single or two-qubit error and $U$ is the desired gate. We now describe our simulation protocol to obtain estimates of the event $\mathrm{mal}_{E}$ for various logical gates.
\begin{enumerate}
   \item Given a 1-exRec encoding a particular gate, we fix a particular value of $p$ and $N$, where $N$ corresponds to the total number of iterations that the depolarizing channel is applied to the 1-exRec.
   \item We would like to calculate the probability of the event $\mathrm{mal}_{E}$ \textit{conditioned} on acceptance of all ancilla states in the LEC and TEC circuits. For every location in the ancilla state preparation and verification circuits, we insert Pauli errors according to the depolarizing error model described above. We propagate the errors through the encoding and verification circuits of $\ket{\overline0}$ and $\ket{\overline+}$. If no errors are detected at the ancilla measurement locations of the EC's, we record the errors that lead to acceptance in the matrices $M_{\ket{\overline0}}$ and $M_{\ket{\overline+}}$. Each row of $M_{\ket{\overline0},\ket{\overline+}}$ will correspond to an error of the form $e=[k,i,l,t]$ where $k$ corresponds to the error type, $i$ and $l$ encode the logical and physical qubit number and $t$ is the particular time step where the error occurred. Note that since an ensemble of errors can combine leading to acceptance of the ancillas, the rows of $M_{\ket{\overline0},\ket{\overline+}}$ will be grouped into several blocks. We repeat this process until the number of blocks of $M_{\ket{\overline0},\ket{\overline+}}$ reaches a predefined size (in most cases, we chose the size to be $10^{6}$ as we believe it to be an accurate representation of the noise for the state preparation circuits). 
   \item For all the remaining locations of the 1-exRec (excluding the state preparation encoding and verification circuits), we insert errors according to the depolarizing noise model. Furthermore, we randomly pick a block from the matrices $M_{\ket{\overline0}}$ and $M_{\ket{\overline+}}$. The combined errors propagate through the 1-exRec and we project the final output errors back onto the codespace. Steps 1-3 are repeated $N$ times.
   \item For single qubit gates, the logical errors are recorded into the vector $v_{1}=[a_{X},a_{Y},a_{Z}]$ where, for example, $a_{X},$ corresponds to the number of logical $X$ errors that occurred after $N$ iterations. For a two-qubit gate, the errors are recorded into a vector with 15 columns, one for each error type. For a gate $G$ at a physical error rate $p$, the estimate of the probability of the event $\mathrm{mal}_{E}$ is given by $\mathrm{Pr}[\mathrm{mal}_{E}|G,p]=a_{E}/N$. Hence, larger values of $N$ will lead to better estimates of $\mathrm{Pr}[\mathrm{mal}_{E}|G,p]$ by reducing the standard deviation. 
\end{enumerate}
For a 1-exRec encoding a logical gate $G$, the pseudo-threshold is defined as the crossing point $p=p_{G}^{1}(p)$, where $p_{G}^{1}(p)=\sum_{E_{i}}\mathrm{Pr}[\mathrm{mal}_{E_{i}}|G,p]$ for all possible logical errors $E_{i}$ for a given logical gate $G$. The pseudo-threshold thus corresponds to the physical error rate below which the logical error rate is smaller than the physical error rate. To obtain the asymptotic threshold, we first upper bound $\mathrm{Pr}[\mathrm{mal}_{E}^{(1)}|G,p]$ (the probability of a malignant event $\mathrm{mal}_{E}$ at the first level of concatenation) by 

\begin{align}
\mathrm{Pr}[\mathrm{mal}_{E}^{(1)}|G,p]\le \sum_{k=\lceil \frac{d^{*}}{2}\rceil}^{L_{G}}c(k)p^{k}\equiv \Gamma_{G,E}^{(1)} ,
\label{eq:DefGamma}
\end{align}
where the coefficients $c(k)$ are positive integers that parametrize the number of possible weight-$k$ errors that can lead to a logical fault, $L_{G}$ is the total number of circuit locations in the logical gate $G$ and $d^{*}$ characterizes the minimal distance of  a given logical gate. As was shown in Ref.~\cite{PR12}, $\Gamma_{G,E}^{(1)}(p)$ is a polynomial that is monotonically increasing as a function of the physical error rate and serves as an upper bound for the failure probability at the first level of concatenation. 
Following Refs.~\cite{PR12,CJL16}, to obtain the probability of having the event $\mathrm{mal}_{E}$ at the second level of concatenation, we can treat each level-1 exRec in the level-2 simulation as a physical location with a modified noise model (no longer depolarizing) given by the $\Gamma_{G,E}^{(1)}(p)$ terms. This procedure can be generalized to the $k$-th level of concatenation, enabling the upper bound on $\mathrm{Pr}[\mathrm{mal}_{E}^{(k)}|G,p]$ to be given by 
\begin{align}
\mathrm{Pr}[\mathrm{mal}_{E}^{(k)}|G,p]\le \sum_{l=\lceil \frac{d^{*}}{2}\rceil}^{L_{G}}c(l)(\Gamma_{G,E}^{(k-1)})^{l}\equiv \Gamma_{G,E}^{(k)} ,
\label{eq:DefGammaLevelk}
\end{align}
where the $c(l)$ coefficients are the same as those in Eq.~\ref{eq:DefGamma}. It was then showed in Ref.~\cite{CJL16} that for physical error rates $p$ smaller than the crossing point between $\Gamma_{G,E}^{(1)}$ and $\Gamma_{G}^{(2)}$ (which we define to be $p_{th,G}$), the logical error rates for the $m$-th level of concatenation ($m\ge 2$) could be upper bounded by

\begin{align}
\mathrm{Pr}[\mathrm{mal}_{E}^{(m)}|G,p]\le \Gamma_{G,E}^{(m)}\le \epsilon^{{\lceil \frac{d^{*}}{2} \rceil}^{m-2}+1}\Gamma_{G,E}^{(1)}.
\label{eq:AssymThreshCross}
\end{align}
Due to the exponential suppression seen in Eq.~\ref{eq:AssymThreshCross} of the logical error rate for $p\le p_{th,G}$, $p_{th,G}$, this serves as a lower bound for the asymptotic threshold of the logical gate $G$. 

\begin{figure}[h]%[htbp]
\centering
\begin{subfigure}{0.4\textwidth}
\includegraphics[width=\textwidth]{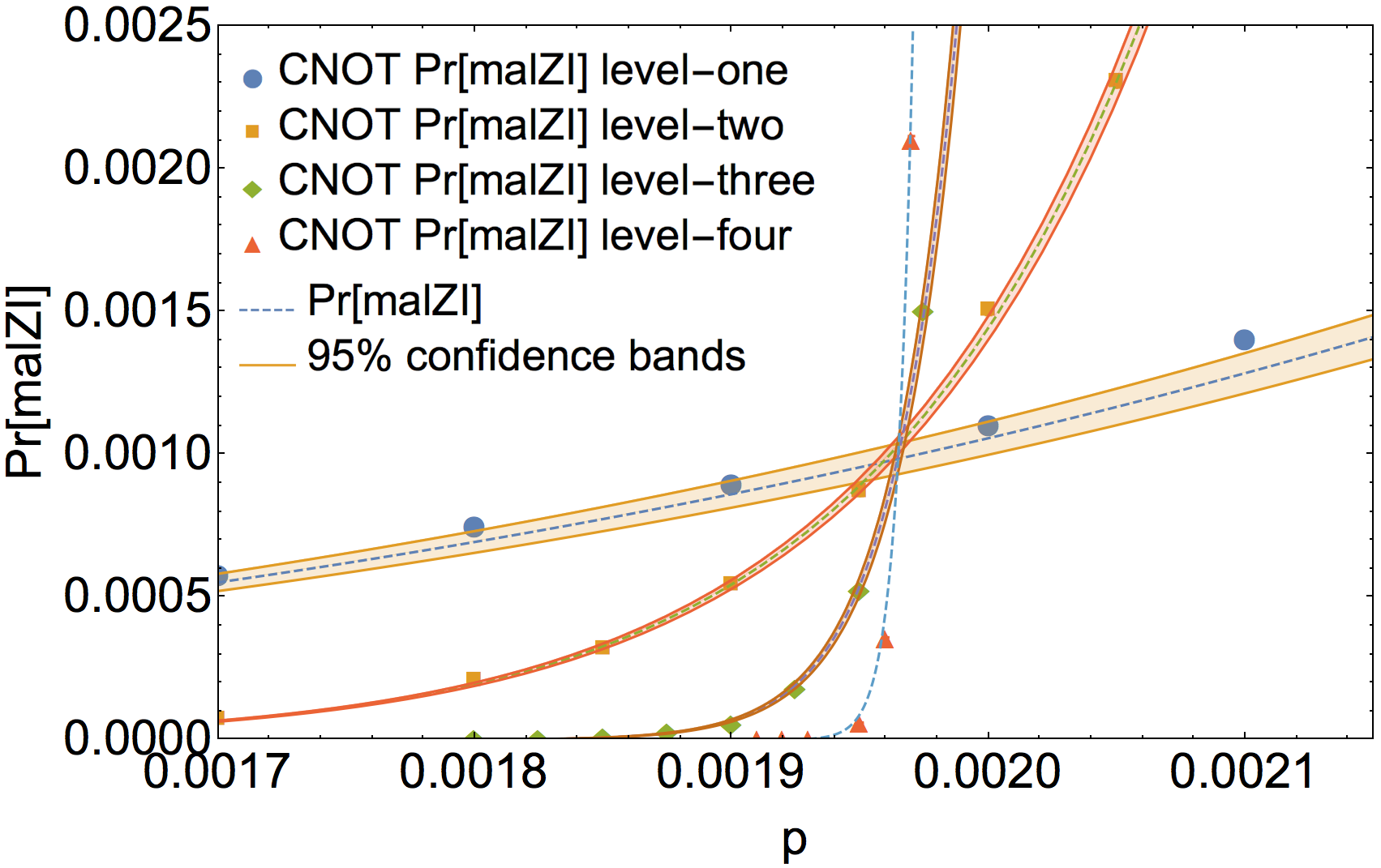}
\caption{}
\label{fig:AsymptoticCNOT}
\end{subfigure}
\begin{subfigure}{0.4\textwidth}
\includegraphics[width=\textwidth]{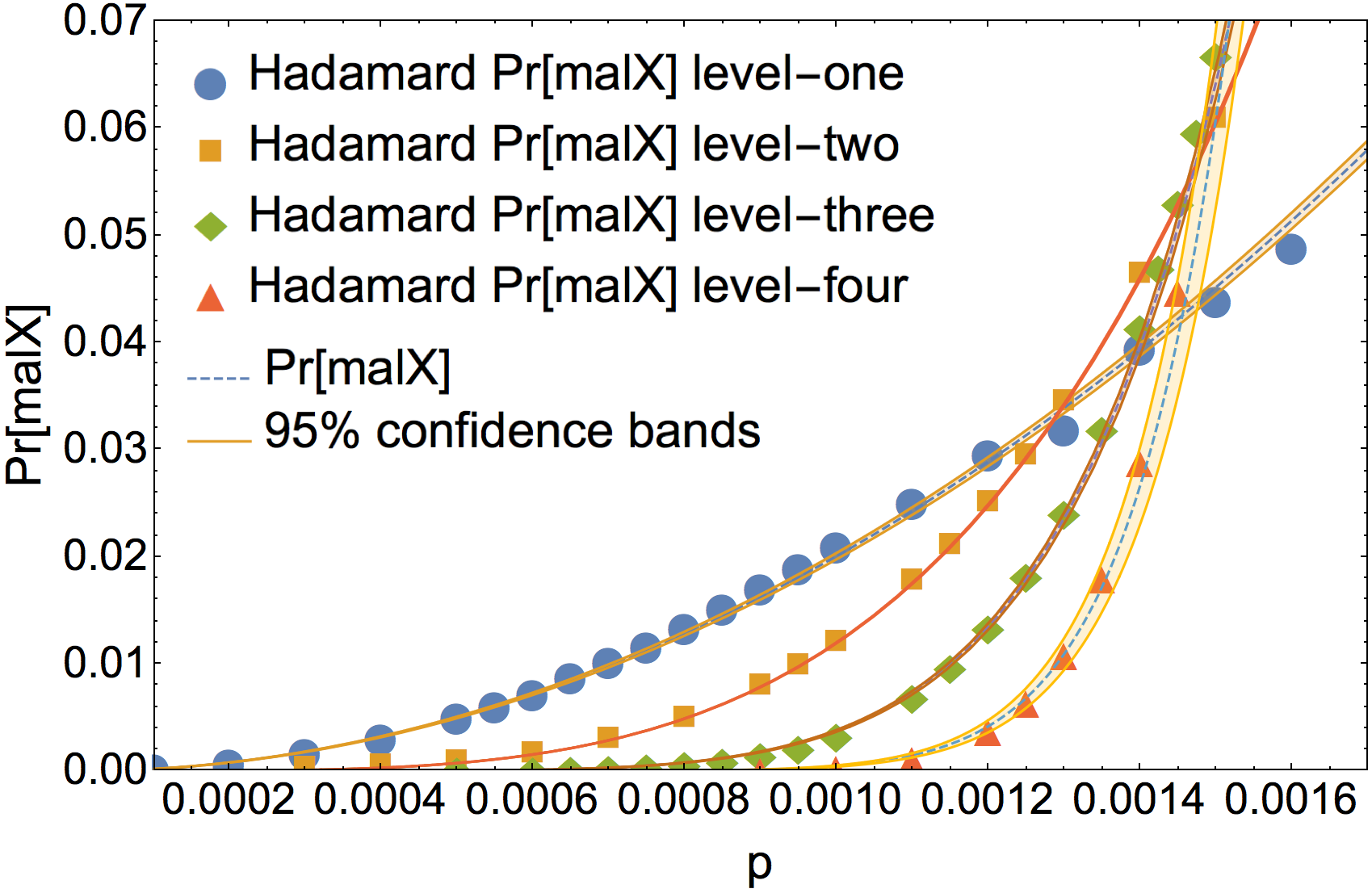}
\caption{}
\label{fig:AsymptoticHadamard}
\end{subfigure}
\caption{Probability $\mathrm{Pr}[\mathrm{mal}_{E}]$ of logical error as function of physical error rate $p$ for the level-1, level-2 and level-3 logical \subref{fig:AsymptoticCNOT}~CNOT and \subref{fig:AsymptoticHadamard}~Hadamard gates of the 105-qubit code. The crossing point between the level-1 and level-2 curves allows for the determination of a lower bound for the asymptotic threshold for each of the logical gates. The CNOT gate exhibits a much lower logical error rate than the Hadamard at the first level.}
\label{fig:AsymptoticThresh105}
\end{figure}

In Ref.~\cite{CJL16}, the pseudo and asymptotic thresholds for the 105-qubit code were calculated for the $H$, $T$ and CNOT gates. Note that $\Gamma_{G,E}^{(m)}$ was also calculated for all other location types (storage, measurement and state-preparation) and were shown to have much higher threshold than the logical gates. The results are summarized in Table~\ref{tab:Pseudo-and-asymptotic}, and Fig.~\ref{fig:AsymptoticThresh105} illustrates the noise behaviour of the Hadamard and CNOT gates for several concatenation levels.
\begin{table}
\begin{centering}
\begin{tabular}{|c|c|c|}
\hline 
 & Pseudo-Threshold & Asymptotic threshold\tabularnewline
\hline 
\hline 
CNOT gate & $\left(2.11\pm0.02\right)\times10^{-3}$ & $\left(1.95\pm0.01\right)\times10^{-3}$\tabularnewline
\hline 
$T$ gate & $\left(4.89\pm0.11\right)\times10^{-4}$ & $\left(1.58\pm0.02\right)\times10^{-3}$\tabularnewline
\hline 
Hadamard gate & $\left(4.47\pm0.29\right)\times10^{-5}$ & $\left(1.28\pm0.02\right)\times10^{-3}$\tabularnewline
\hline 
\hline
{\bf 105-qubit} & $\mathbf{\left(4.47\pm0.29\right)\times10^{-5}}$ & $\mathbf{\left(1.28\pm0.02\right)\times10^{-3}}$\tabularnewline
\hline 
\end{tabular}
\par\end{centering}
\caption{\label{tab:Pseudo-and-asymptotic}Lower bounds for the pseudo and asymptotic threshold results for the Hadamard, $T$ gate and CNOT gates of the 105-qubit code. The Hadamard asymptotic-threshold is larger than its pseudo-threshold resulting from the double protection of the CNOT gates as seen by the high CNOT pseudo-threshold.}
\end{table}
An important feature of these results is that, for the $H$ and $T$ gates, noise can be further suppressed by several orders of magnitude even for physical error rates above the pseudo-threshold. To understand this type of noise behaviour, it is important to point out that since the CNOT gate is transversal in both the 7 and 15-qubit codes, it receives a double protection from both codes. The double protection results in a much larger pseudo-threshold for the CNOT gate compared to the pseudo-thresholds for the $H$ and $T$ gates. Furthermore, the asymptotic threshold of the CNOT is comparable to its pseudo-threshold. Another important feature is that CNOT gates are the gates that are most present in the Steane's~EC circuits as well as in the logical $H$ and $T$ gate circuits. Consequently, when going to higher levels of concatenation, and for error rates below the CNOT asymptotic threshold, all the CNOT gates will be less likely to fail compared to the previous level of concatenation. Even if the physical $H$ and $T$ gate locations are more likely to fail, the lower logical failure rates of the CNOT gates will compensate, resulting in overall further noise suppression. 

\begin{figure*}[htbp]
\begin{center}
\begin{subfigure}{0.4\textwidth}
\includegraphics[width = \textwidth]{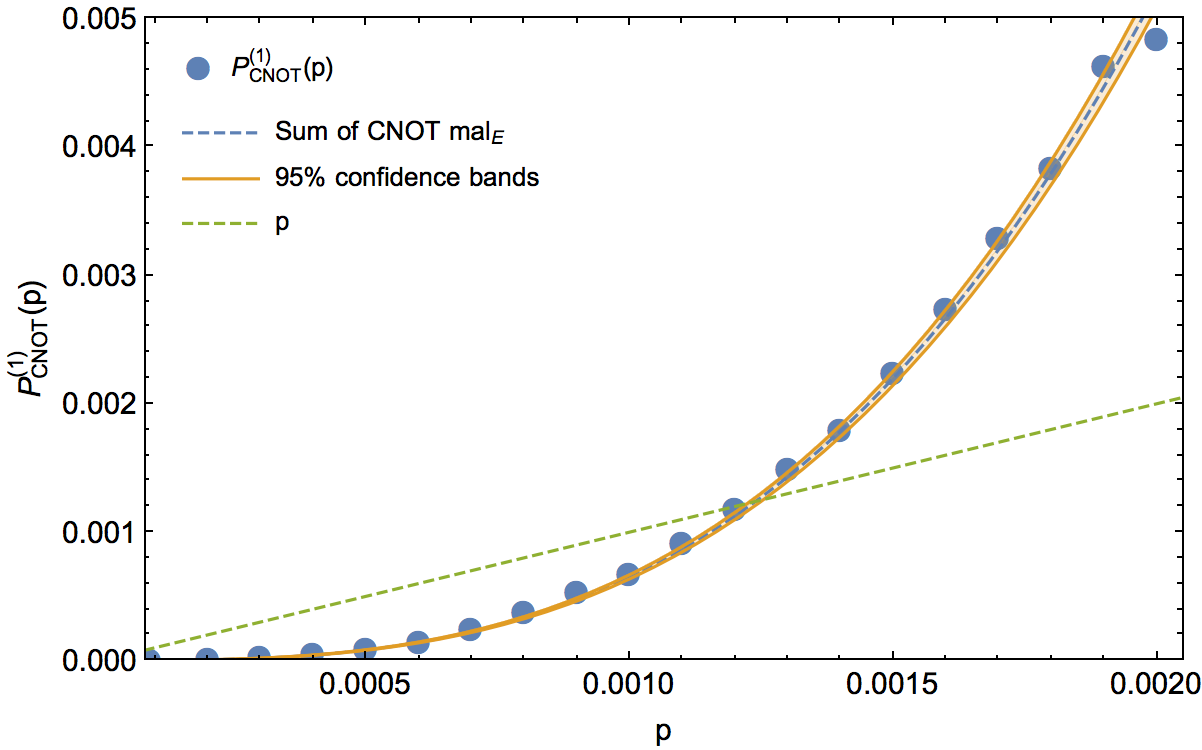}
\caption{}
\label{fig:a}
\end{subfigure}
\begin{subfigure}{0.4\textwidth}
\includegraphics[width =\textwidth]{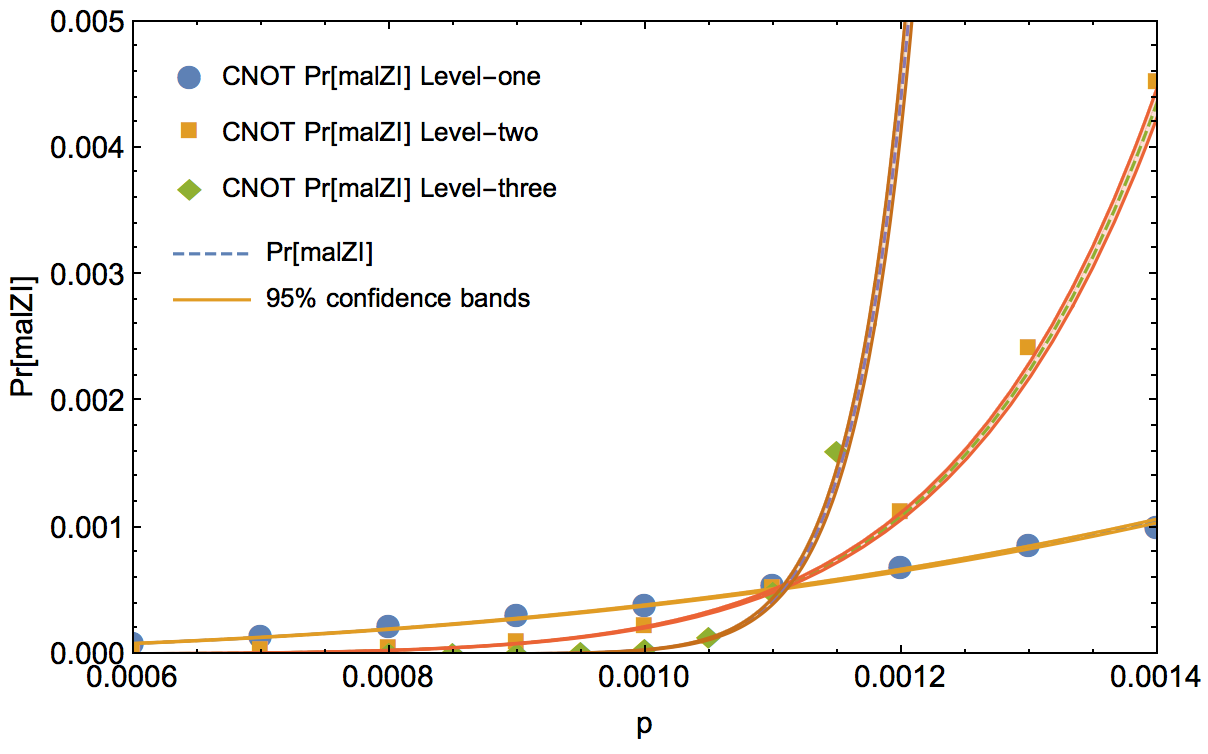}
\caption{}
\label{fig:b}
\end{subfigure}
\begin{subfigure}{0.42\textwidth}
\includegraphics[width = \textwidth]{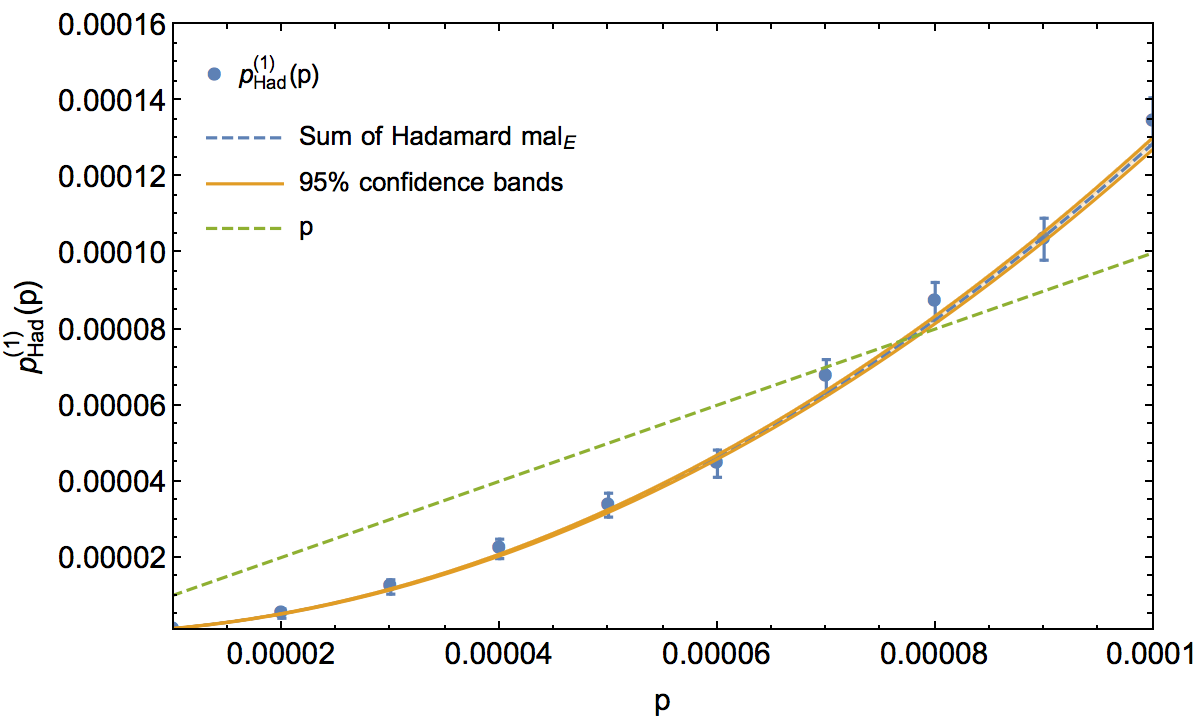}
\caption{}
\label{fig:c}
\end{subfigure}
\begin{subfigure}{0.4\textwidth}
\includegraphics[width = \textwidth]{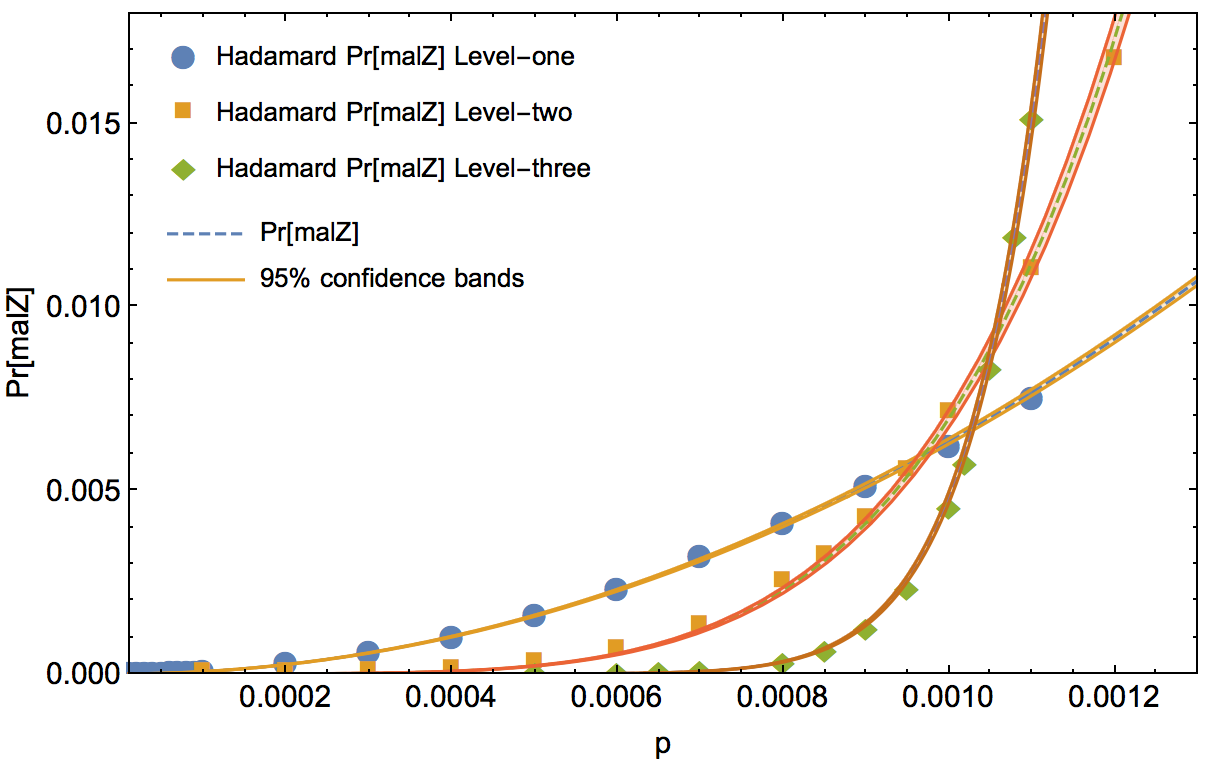}
\caption{}
\label{fig:d}
\end{subfigure}
\begin{subfigure}{0.42\textwidth}
\includegraphics[width = \textwidth]{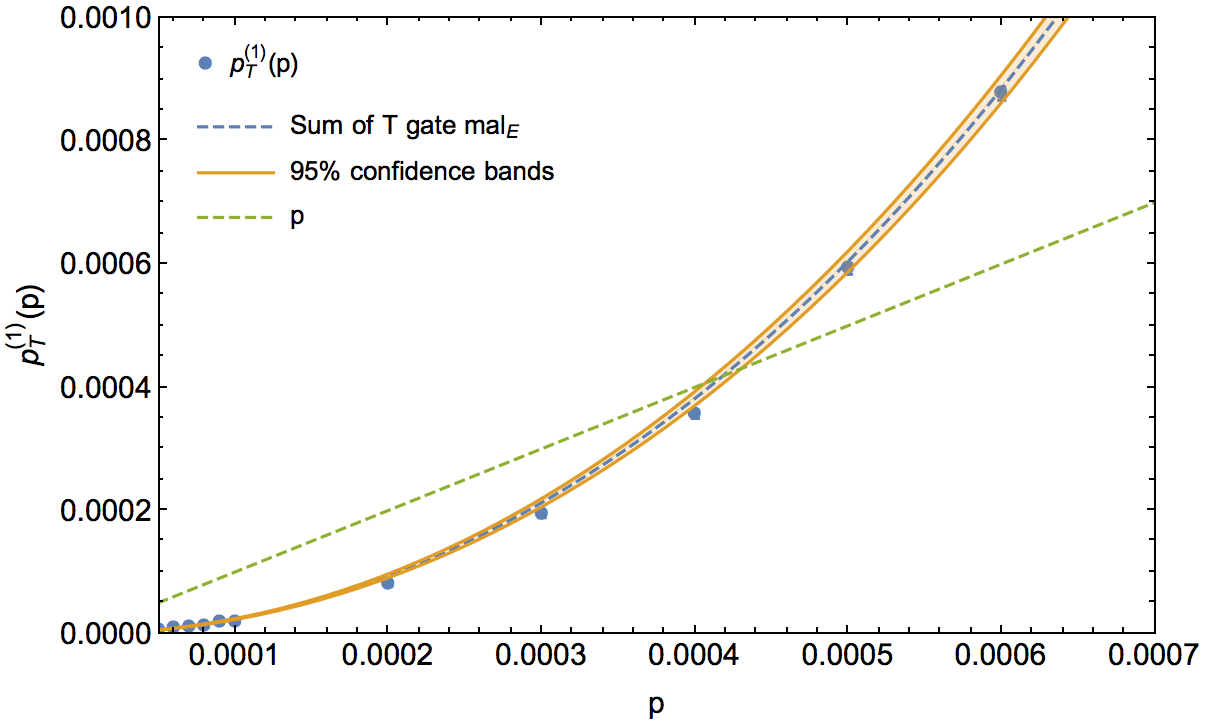}
\caption{}
\label{fig:e}
\end{subfigure}
\begin{subfigure}{0.4\textwidth}
\includegraphics[width = \textwidth]{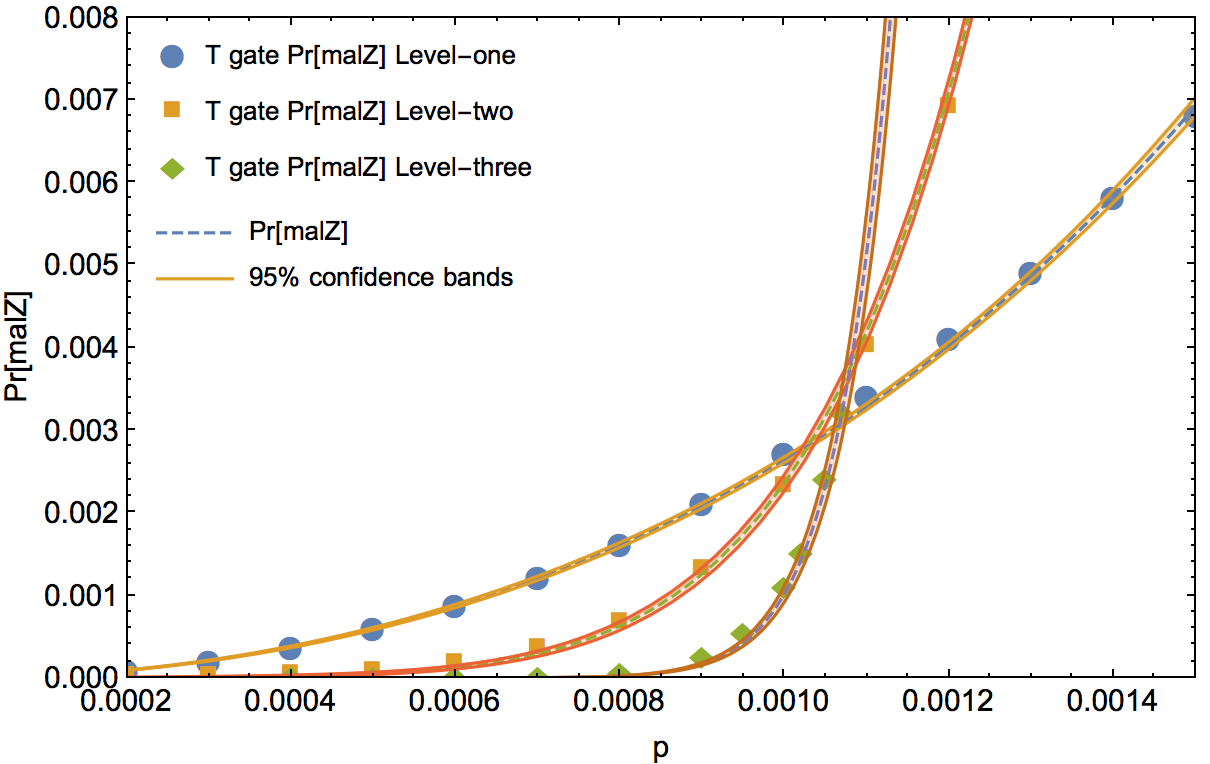}
\caption{}
\label{fig:f}
\end{subfigure}
\caption{All plots in Fig.~\ref{fig:PseudoAndLevelThreeThresholds} pertain to the 49-qubit code. The plots on the left column illustrate the probability of logical error as function of physical error rate for logical \subref{fig:a}~CNOT, \subref{fig:c}~Hadamard and \subref{fig:e}~$T$~gate. The crossing point of the fitted curve allows for the determination of the level-1 pseudo-threshold for each of the logical gates. The CNOT pseudo-threshold is the largest among all three gates due to the double protection of the 7-qubit and 15-qubit code. The plots on the right column illustrate the polynomials upper bounding the probability of obtaining a logical error $E$ for the first, second and third level of concatenation. The crossing point between the level-one and level-two polynomials determine the asymptotic threshold for the gate under consideration. For the logical CNOT gate~\subref{fig:b}, it is the event $\text{mal}_{ZI}$ which limits the threshold value. The same behaviour holds for the 105-qubit code. However, given that the effective distance of the 49-qubit code is 5 compared to 9 for the 105-qubit code (when the gate is transversal in both the 7 and 15-qubit code), the 105-qubit code offers greater suppression for logical error rates below threshold. For the logical gate $H$~\subref{fig:d} and $T$~gate~\subref{fig:f}, $\text{mal}_{Z}$ limits the threshold value. Note that for the 105-qubit code, $\mathrm{mal}_{X}$ limited the Hadamard threshold value. See \ref{sec:Concatenated 49-qubit thresholds} for more details.}
\label{fig:PseudoAndLevelThreeThresholds}
\end{center}
\end{figure*}

We now provide the threshold results for the 49-qubit code. The plots in Fig.~\ref{fig:PseudoAndLevelThreeThresholds} illustrate $ \Gamma_{G,E}^{(m)}$ for the Hadamard, $T$ and CNOT gates. As was the case for the 105-qubit code, the pseudo-threshold of the 49-qubit code is limited by the Hadamard gate and is given by $p_{Had}^{(1)}=(7.76\pm 0.17)\times 10^{-5}$. A lower bound on the asymptotic threshold was found to be $p_{th}=(9.69\pm 0.28)\times 10^{-5}$. However, Fig.~\ref{fig:c} shows a few noticeable differences with the 49-qubit code. First, logical $Z$ errors are what limit the Hadamard asymptotic threshold for the 49-qubit code, not logical $X$ errors as in the 105-qubit code. Second, the Hadamard circuit is less sensitive to input $Z$ errors than in the 105-qubit code. Furthermore, the 49-qubit Hadamard pseudo-threshold is larger than for the 105-qubit code. To understand these results, we first point out that the 49-qubit code Hadamard circuit contains four blocks consisting only of physical $H$ and storage gates (no CNOT gates). As was explained in section \ref{Ancilla prep section}, the cascading sequence of CNOT gates in the 15-qubit Hadamard circuit makes it very sensitive to input $Z$ errors since any $Z$ error that lands on the target of the CNOT gates (appearing before the physical $H$ gate) will lead to a logical $X$ error. Furthermore, $Z$ errors occurring at CNOT or storage locations within the 15-qubit Hadamard circuit also play a dominant role in producing a logical $X$ fault at the output of the circuit. The four blocks consisting only of physical $H$ and storage gates will treat $X$ and $Z$ errors on the same footing and these blocks contain much fewer locations were $Z$  errors can occur. A numerical simulation also showed that many of the errors leading to a logical $Z$ fault were $Z$ errors that occurred on CNOT gates \textit{after} the physical $H$ gate combined with Y errors on one of the single-qubit codeblocks. We also note that for higher concatenation levels, given the smaller number of locations in a storage gate exRec, storage gates are less likely to fail than CNOT gates and so the single qubit codeblocks are less likely to acquire a logical fault.  Consequently, the 49-qubit Hadamard circuit will produce less logical $X$ errors compared to the circuit for the 105-qubit code. Since the 15-qubit code offers less protection against $Z$ errors, logical $Z$ errors become the dominant source of error for the Hadamard circuit.

\begin{table}
\begin{centering}
\begin{tabular}{|c|c|c|}
\hline 
 & Pseudo-Threshold & Asymptotic threshold\tabularnewline
\hline 
\hline 
CNOT gate & $\left(1.21\pm0.04\right)\times10^{-3}$ & $\left(1.10\pm0.01\right)\times10^{-3}$\tabularnewline
\hline 
$T$ gate & $\left(4.18\pm0.24\right)\times10^{-4}$ & $\left(1.03\pm0.03\right)\times10^{-3}$\tabularnewline
\hline 
Hadamard gate & $\left(7.76\pm0.17\right)\times10^{-5}$ & $\left(9.69\pm0.28\right)\times10^{-4}$\tabularnewline
\hline 
\hline
{\bf 49-qubit} & $\mathbf{\left(7.76\pm0.17\right)\times10^{-5}}$ & $\mathbf{\left(9.69\pm0.28\right)\times10^{-4}}$\tabularnewline
\hline 
\end{tabular}
\par\end{centering}
\caption{\label{tab:Pseudo-and-asymptotic49}Lower bounds for the pseudo and asymptotic threshold results for the Hadamard, $T$ gate and CNOT gates of the 49-qubit code. The Hadamard asymptotic-threshold is larger than its pseudo-threshold resulting from the double protection of the CNOT gates as seen by the high CNOT pseudo-threshold.}
\end{table}
Note that as in the 105-qubit code, the double protection of the CNOT gates from both the 7 and 15-qubit code results in a larger pseudo and asymptotic threshold relative to the $H$ and $T$ gates. The threshold results for the 49-qubit code are summarized in Table~\ref{tab:Pseudo-and-asymptotic49}. Although the 49-qubit code Hadamard pseudo-threshold is larger than the 105-qubit code pseudo-threshold, the asymptotic threshold is slightly smaller due to the lower CNOT pseudo and asymptotic threshold. 

\section{Resource overhead for the 49 and 105-qubit code} \label{sec:Resource overhead 49}

\subsection{Raw qubit overhead} \label{Raw qubit overhead} 

In the simulation of a particular gate (say $H$ or CNOT), we would like to obtain the resource overhead required to achieve a particular target logical error rate $p_{target}$. The overhead can be measured in several ways. The raw qubit overhead measures how many physical qubits are required in the simulation of a logical gate to achieve a particular target logical error rate. The gate overhead measures the total number of gates used in the simulation of a logical gate in order to achieve a particular target logical error rate. 

We begin with the analysis of the raw qubit overhead. Recall that Steane error correction~(EC) is implemented by the circuit in Fig.~\ref{fig:SteaneECcircuit} where the ancilla states $\ket{\overline{0}}$ and $\ket{\overline{+}}$ (for the 49 and 105-qubit code) are prepared using the circuits of section \ref{Ancilla prep section}.

\begin{figure}%[htbp]
\centering
\begin{subfigure}{0.35\textwidth}
\includegraphics[width=\textwidth]{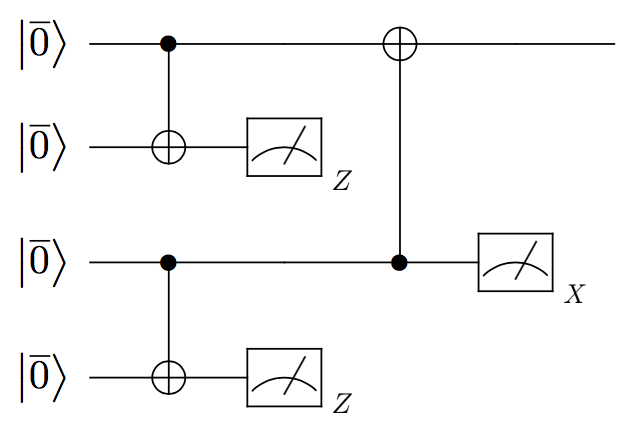}
\caption{}
\label{fig:nOprep}
\end{subfigure}
\begin{subfigure}{0.35\textwidth}
\includegraphics[width=\textwidth]{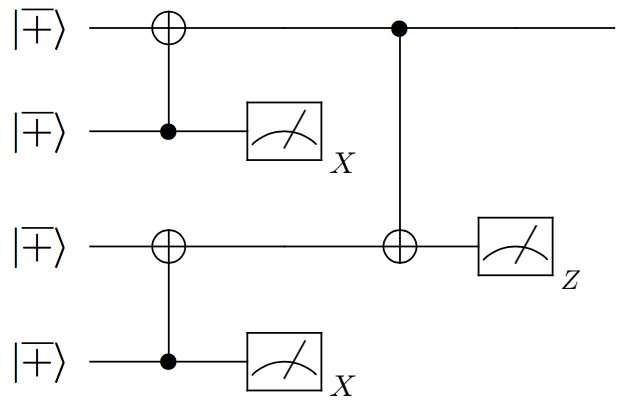}
\caption{}
\label{fig:nPlusPrep}
\end{subfigure}
\caption{\subref{fig:nOprep}~State preparation, including verification, for the ancilla state~$\ket{\overline 0}$. \subref{fig:nPlusPrep}~State preparation circuit for the state~$\ket{\overline +}$.}
\label{fig:nPrep}
\end{figure}
We assume that the qubits used in the preparation of the ancilla states can be reused at each time step in the computation prior to a measurement in the $X$ or $Z$ basis. Before proceeding with the overhead calculation, we provide a few definitions. Let $\ket{\overline{0}}^{(k)}$ and $\ket{\overline{+}}^{(k)}$ correspond to the state-preparation circuits of $\ket{0}$ and $\ket{+}$ at the $k$-th level of concatenation. In the first part of the Steane~EC circuit, the ancilla states are verified for errors by entangling a pair of ancilla states of the same type and performing a measurement in the appropriate basis (see section~\ref{Ancilla prep section} for more details). We define $n_{\ket{\overline{0}}}^{(k)}$ and $n_{\ket{\overline{+}}}^{(k)}$ to be the number of qubits required for the ancillas $\ket{\overline{0}}^{(k)}$ and $\ket{\overline{+}}^{(k)}$ to pass the verification test in the circuits of Fig.~\ref{fig:nOprep} and Fig.~\ref{fig:nPlusPrep} at the $k$-th level of concatenation. If an error is detected from the syndrome measurement, the ancilla is rejected and the process is repeated using a fresh batch of qubits which we assume are readily available. The probabilities of acceptance $p_{\ket{\overline0}_{1}}^{(k)}$ and $p_{\ket{\overline+}_{1}}^{(k)}$ were computed from a Monte Carlo algorithm. For an error correcting code with $n$ physical qubits, we have at the first level of concatenation

\begin{align}
n_{\ket{\overline{0}}}^{(1)}=\frac{2n(\frac{1}{p_{\ket{\overline0}_{1}}^{(1)}}+\frac{1}{p_{\ket{\overline0}_{2}}^{(1)}})}{p_{\ket{\overline0}_{3}}^{(1)}}, \ \label{eq:nOPrepLevel1}\\
n_{\ket{\overline{+}}}^{(1)}=\frac{2n(\frac{1}{p_{\ket{\overline+}_{1}}^{(1)}}+\frac{1}{p_{\ket{\overline+}_{2}}^{(1)}})}{p_{\ket{\overline+}_{3}}^{(1)}}.
\end{align}
Note that the term ${2n}/{p_{\ket{\overline0}_{j}}^{(1)}}$ (where $j\in \{1,2\}$) corresponds the the expected number of qubits for preparing a $\ket{\overline0}$ state free of $X$ errors (the first step in the ancilla verification test). Hence Eq.~\ref{eq:nOPrepLevel1} corresponds to the expected number of qubits for the full $\ket{\overline0}$ ancilla verification test. The analogous equation holds for $\ket{\overline +}$ state preparation. Defining $n_{EC}^{(k)}$ to be the raw qubit overhead for a Steane EC circuit at level-$k$ (excluding the overhead from the data qubits), we have that 

\begin{align}
n_{EC}^{(1)}= n_{\ket{\overline{0}}}^{(1)} + n_{\ket{\overline{+}}}^{(1)}.
\label{eq:nECk}
\end{align}

Since a logical CNOT gate consists of four EC circuits and two logical qubits, the level-1 overhead for the CNOT gate is

\begin{align}
q_{\mathrm{CNOT}}^{(1)}= 2n_{EC}^{(1)}+2n.
\label{eq:CNOTlevel1}
\end{align}
Note that the factor of $2$ and not $4$ in front of the $n_{EC}$ term. This is due to the fact that we assume that the qubits used in the LEC circuit can be reused in the TEC circuits. Similarly, the level-1 Hadamard qubit overhead is given by   

\begin{align}
q^{(1)}_{\mathrm{Had}} = n^{(1)}_{\mathrm{EC}} + n.
\label{eq:HadLevel1}
\end{align}

\begin{figure}[h]%[htbp]
\centering
\includegraphics[width=0.4\textwidth]{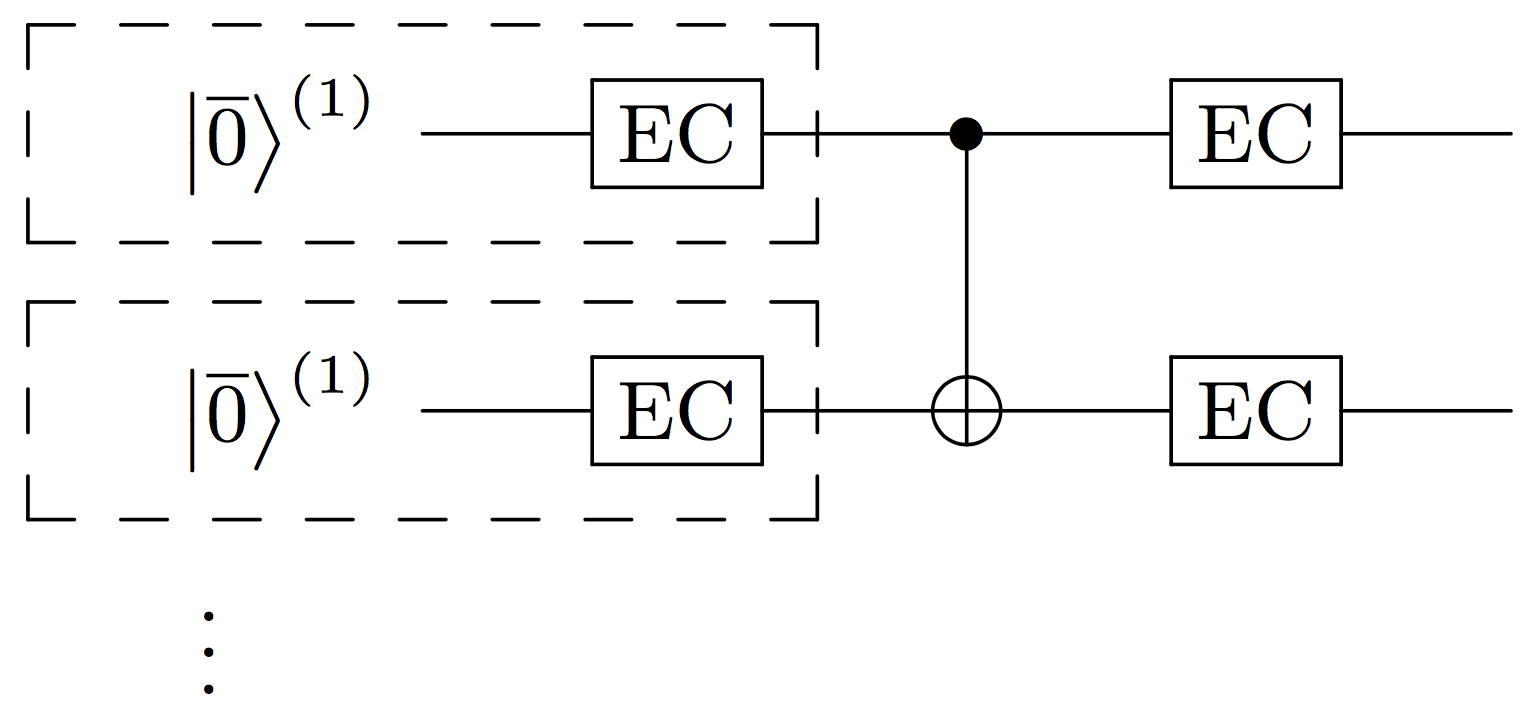}
\caption{Typical segment in a level-2  $\ket{\overline{0}}^{(2)}$ or  $\ket{\overline{+}}^{(2)}$ illustrating the overlapping EC circuits between the $\ket{\overline{0}}^{(1)}$ (could also be $\ket{\overline{+}}^{(1)}$) and the following logical gate (a CNOT gate in this example). The overhead of the first EC circuit needs to be taken into account in the calculation of $n_{\ket{\overline{0}}}^{(2)}$ and $n_{\ket{\overline{+}}}^{(2)}$. Each dashed box has an overhead of $n_{EC}^{(1)} + n$.}
\label{fig:TypicalO2Circuit}
\end{figure}

At the second level of concatenation, $\ket{\overline{0}}^{(2)}$ will contain $\ket{\overline{0}}^{(1)}$ and $\ket{\overline{+}}^{(1)}$ state preparation circuits which will be followed by an EC circuit. However, all other gates (CNOT's and storage) will also contain their respective LEC and TEC circuits (since recall that for a level-2 simulation each physical gate is replaced by a level-1 exRec). The EC circuits of $\ket{\overline{0}}^{(1)}$ and $\ket{\overline{+}}^{(1)}$ will overlap with the LEC circuit of the gate that follows (see Fig.~\ref{fig:TypicalO2Circuit}) and so it is important to take into account the overhead of the overlapping EC circuit. The full EC circuits (including the data qubits) each have an overhead of $n_{EC}^{(1)} + n$. We assume that the qubits that were used in the EC circuit following $\ket{\overline{0}}^{(1)}$ and $\ket{\overline{+}}^{(1)}$ can be reused for all other EC circuits that follows. Hence, we only take into account the overhead of the \textit{first} EC circuit. The entire $\ket{\overline{0}}^{(2)}$ circuit will have an overhead of $n(n_{EC}^{(1)} + n)$.  Generalizing to the $k$-th concatenation level (for $k\ge 2$), we have the recursive relation
\begin{align}
n_{\ket{\overline{0}}}^{(k)}=\frac{2n(n_{EC}^{(k-1)}+n^{(k-1)})(\frac{1}{p_{\ket{\overline0}_{1}}^{(k)}}+\frac{1}{p_{\ket{\overline0}_{2}}^{(k)}})}{p_{\ket{\overline0}_{3}}^{(k)}}, \ \\
n_{\ket{\overline{+}}}^{(k)}=\frac{2n(n_{EC}^{(k-1)}+n^{(k-1)})(\frac{1}{p_{\ket{\overline+}_{1}}^{(k)}}+\frac{1}{p_{\ket{\overline+}_{2}}^{(k)}})}{p_{\ket{\overline+}_{3}}^{(k)}}.
\label{eq:nPrepLevelk}
\end{align}
The EC circuit at level-$k$ has an overhead given by 
\begin{align}
n_{EC}^{(k)}=n_{\ket{\overline{0}}}^{(k)}+n_{\ket{\overline{+}}}^{(k)}.
\label{eq:nECoverheadLevelk}
\end{align}
The CNOT and Hadamard overhead at level-$k$ are then given by 
\begin{align}
&q_{CNOT}^{(k)}=2n_{EC}^{(k)} + 2n^k, \ \label{eq:nECoverheadLevelk1}\\
&q_{Had}^{(k)}=n_{EC}^{(k)} + n^k.
\label{eq:nECoverheadLevelk2}
\end{align}

\begin{figure}%[htbp]
\centering
%\begin{align*}
\includegraphics[width=0.5\textwidth]{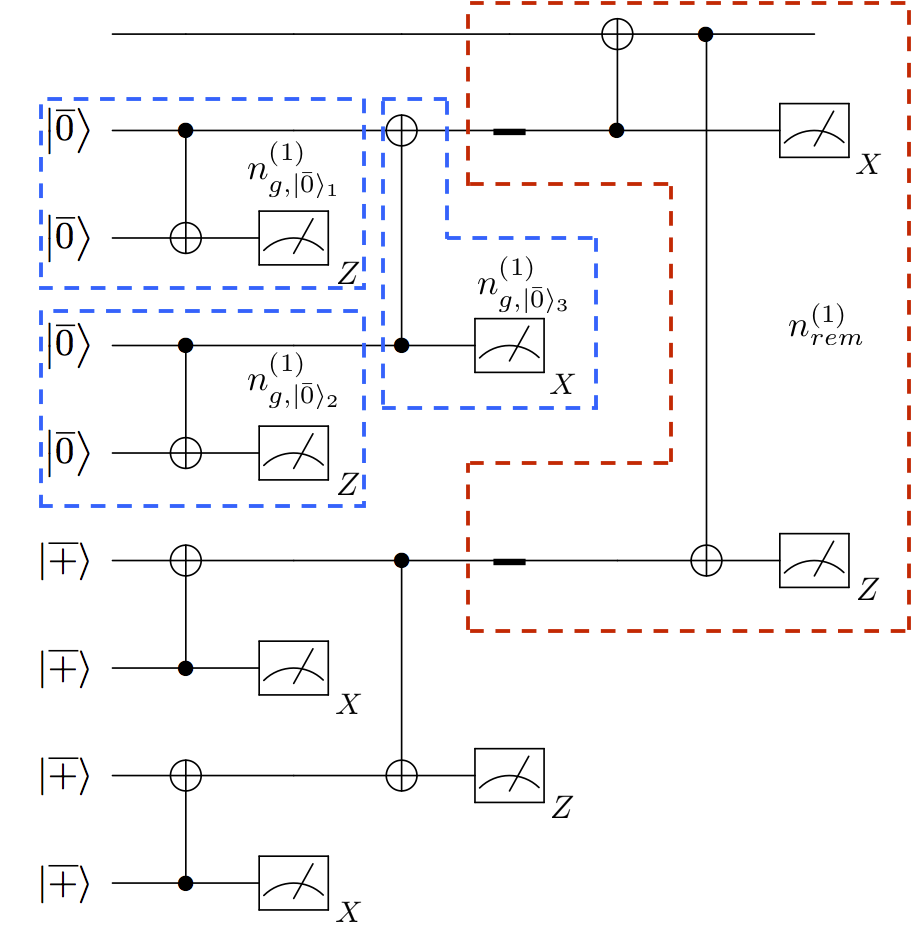}
%\end{align*}
\caption{Circuit illustrating the various definitions for the gate overhead calculation. $n_{g,\ket{\overline0}_{1}}^{(1)}$ corresponds to the gate overhead for the first part in the $\ket{\overline0}$ ancilla verification test at the first level of concatenation. Note that $n_{g,\ket{\overline0}_{1}}^{(1)}=n_{g,\ket{\overline0}_{2}}^{(1)}$. The term $n_{g,\ket{\overline0}_{3}}^{(1)}$ corresponds to the gate overhead for the final step in the $\ket{\overline0}$ ancilla verification test. The $n_{g,\ket{\overline+}_{j}}^{(1)}$ terms are defined in an analogous way. Lastly, $n_{\mathrm{rem}}^{(1)}$ is the gate overhead for the remaining part of the Steane~EC circuit.}
\label{fig:gateOverheadCircuit}
\end{figure}

\subsection{Gate overhead} \label{Gate overhead} 
In this subsection we focus on the gate overhead for the simulation of a logical Hadamard and logical CNOT gate. We define $g_{\ket{\overline0}}^{(k)}$ and $g_{\ket{\overline+}}^{(k)}$ to be the number of gates in a $\ket{\overline0}$ and $\ket{\overline+}$ circuit at level-$k$. Since $\ket{\overline0}$ and $\ket{\overline+}$ circuits consist of physical $\ket{0}$ and $\ket{+}$ states as well as CNOT and storage gates, we can define $\alpha_{j}$ to be the number of locations of type $j$ in a $\ket{\overline0}^{(1)}$ circuit and $\beta_j$ to be the number of locations of type $j$ in a $\ket{\overline+}^{(1)}$ circuit. In this case we can write 

\begin{align}
g_{\ket{\overline0}}^{(1)}&=\alpha_{\mathrm{CNOT}} + \alpha_{\mathrm{mem}} + \alpha_{\ket{0}} + \alpha_{\ket{+}}, \ \\
g_{\ket{\overline+}}^{(1)}&=\beta_{\mathrm{CNOT}} + \beta_{\mathrm{mem}} + \beta_{\ket{0}} + \beta_{\ket{+}}.
\label{eq:gDefsLevel1}
\end{align}
To obtain the gate overhead of a Steane EC circuit at the first level of concatenation, we can divide the EC circuit into several components and compute the overhead for each component. We define $n_{g,\ket{\overline0}_{j}}^{(1)}$ to be the gate overhead for the $\ket{\overline0}$ ancilla verification components, $n_{g,\ket{\overline+}_{j}}^{(1)}$ to be the gate overhead for the $\ket{\overline+}$ ancilla verification components and $n_{\mathrm{rem}}^{(1)}$ to be the gate overhead for the remaining EC circuit, see Fig.~{\ref{fig:gateOverheadCircuit}} for a circuit description. 

Since CNOT and measurement locations can be implemented transversally, they each contribute a factor of $n$ to the gate overhead at the first level so that 
\begin{align}
n_{g,\ket{\overline0}_{1}}^{(1)}&=n_{g,\ket{\overline0}_{2}}^{(1)}=2(g_{\ket{\overline0}}^{(1)}+n), \ \\
n_{g,\ket{\overline+}_{1}}^{(1)}&=n_{g,\ket{\overline+}_{2}}^{(1)}=2(g_{\ket{\overline+}}^{(1)}+n), \ \\
n_{g,\ket{\overline0}_{3}}^{(1)}&=n_{g,\ket{\overline+}_{3}}^{(1)}=2n, \ \\
n_{\mathrm{rem}}^{(1)}&=6n.
\label{eq:nECoverheadLevelk}
\end{align}
Taking into account that the ancilla states are rejected if a non-trivial error is detected at the measurement locations, we can compute the expected number of gates for a $\ket{\overline0}$ and $\ket{\overline+}$ verification test (defined as $n_{g,\mathrm{EC}\ket{\overline0}}^{(1)}$ and $n_{g,\mathrm{EC}\ket{\overline+}}^{(1)}$) according to their success probabilities as
\begin{align}
n_{g,\mathrm{EC}\ket{\overline0}}^{(1)}&=\frac{n_{g,\ket{\overline0}_{1}}^{(1)}\Big(\frac{1}{p_{\ket{\overline0}_{1}}^{(1)}}+\frac{1}{p_{\ket{\overline0}_{2}}^{(1)}}\Big)+n_{g,\ket{\overline0}_{3}}^{(1)}}{p_{\ket{\overline0}_{3}}^{(1)}}, \ \\
n_{g,\mathrm{EC}\ket{\overline+}}^{(1)}&=\frac{n_{g,\ket{\overline+}_{1}}^{(1)}\Big(\frac{1}{p_{\ket{\overline+}_{1}}^{(1)}}+\frac{1}{p_{\ket{\overline+}_{2}}^{(1)}}\Big)+n_{g,\ket{\overline+}_{3}}^{(1)}}{p_{\ket{\overline+}_{3}}^{(1)}}.
\label{eq:expectedNumberGatesVerificationTest}
\end{align}
From Eqs.~\ref{eq:nECoverheadLevelk} and~\ref{eq:expectedNumberGatesVerificationTest}, the total gate overhead~$n_{g,\mathrm{EC}}^{(1)}$ for an EC circuit at the first level of concatenation is given by 
\begin{align}
n_{g,\mathrm{EC}}^{(1)}=n_{g,\mathrm{EC}\ket{\overline0}}^{(1)}+n_{g,\mathrm{EC}\ket{\overline+}}^{(1)}+n_{\mathrm{rem}}^{(1)}.
\label{eq:TotalGateOverheadLevel1}
\end{align}
Using Eq.~\ref{eq:TotalGateOverheadLevel1}, the overhead of a level-1 CNOT and storage exRec is given by
\begin{align}
g_{\mathrm{CNOT}}^{(1)}&=4n_{g,\mathrm{EC}}^{(1)}+n, \ \\
g_{\mathrm{mem}}^{(1)}&=2n_{g,\mathrm{EC}}^{(1)}+n.
\label{eq:StorHadCNOToverheadLevel1}
\end{align}
The logical Hadamard circuit consists of storage, CNOT and physical Hadamard gates. Defining $\gamma_{j}$ to be the number of gates of type $j$ in the level-1 logical Hadamard circuit and using Eq.~\ref{eq:StorHadCNOToverheadLevel1}, the gate overhead for the level-1 Hadamard exRec is 
\begin{align}
g_{\mathrm{Had}}^{(1)}=2n_{g,\mathrm{EC}}^{(1)}+\gamma_{\mathrm{CNOT}}+\gamma_{\mathrm{mem}}+\gamma_{\mathrm{Had}}.
\label{eq:Level1HadamardexRec}
\end{align}

\begin{figure}%[htbp]
\centering
%\begin{align*}
\includegraphics[width=0.5\textwidth]{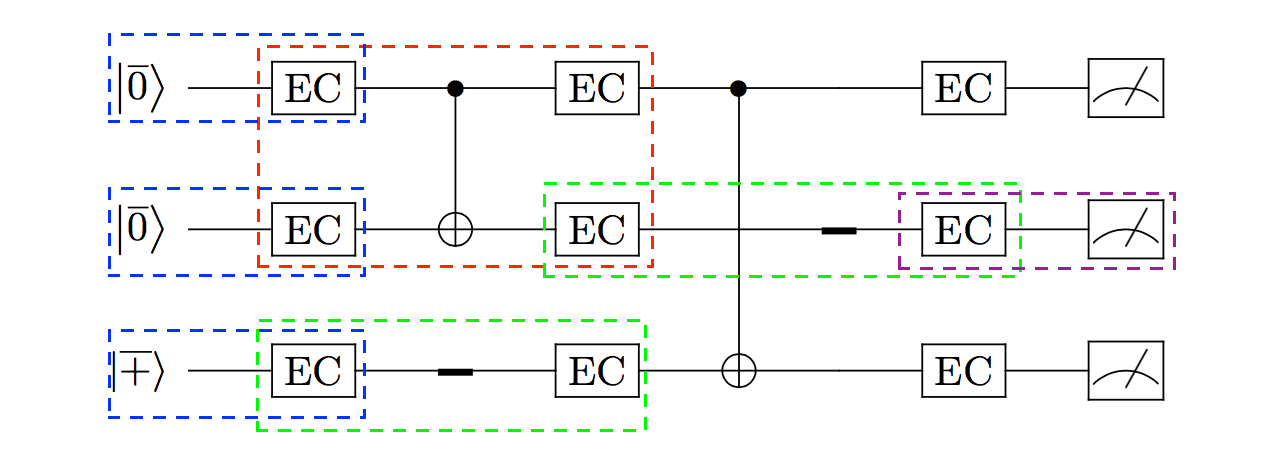}
%\end{align*}
\begin{align*}
\end{align*}
\caption{Typical sequence of gates in a level-$k$ exRec. The EC preceding the state-preparation circuit overlaps with the LEC of the following gate. The TEC of the last gate overlaps with the EC prior to a performing a measurement. In order to avoid over counting gates appearing in overlapping EC's, we use the overhead of logical gates with truncated LEC's in the counting procedure. The EC circuit arising from a measurement location will be included in the TEC of the previous gate.}
\label{fig:ECoverlaps}
\end{figure}
Before obtaining the overhead at higher levels of concatenation, it is important to point out that consecutive exRec's will have overlapping EC's. In a recursive calculation, if we were to use the overhead terms $g_{j}^{(k-1)}$ at the $k$-th concatenation level in the counting procedure, we would be over counting the gates appearing in overlapping EC's (see Fig.~\ref{fig:ECoverlaps}). The over counting can be avoided by ignoring the gate overhead in LEC's from the $(k-1)$-exRec's appearing in a $k$-exRec. For a $t$-qubit gate, we define 

\begin{align}
\tilde{g}_{j}^{(k)}=g_{j}^{(k)} - tn_{g,\mathrm{EC}}^{(k)}.
\label{eq:SingleQubitLECtruncation}
\end{align}
Therefore, in the overhead counting of a $k$-exRec, we will include the contributions from $\tilde{g}_{j}^{(k-1)}$ terms instead of $g_{j}^{(k-1)}$.

We now consider the gate overhead at the $k$-th concatenation for $k\ge 2$. The gate overhead in a $\ket{\overline0}$ and $\ket{\overline+}$ circuit at level-$k$ is given by 

\begin{align}
g_{\ket{\overline0}}^{(k)}&=\alpha_{\mathrm{CNOT}}\tilde{g}_{\mathrm{CNOT}}^{(k-1)} + \alpha_{\mathrm{mem}}\tilde{g}_{\mathrm{mem}}^{(k-1)} + \alpha_{\ket{0}}g_{\ket{\overline0}}^{(k-1)} \nonumber \\ 
&\qquad + \alpha_{\ket{+}}g_{\ket{\overline+}}^{(k-1)} + n_{g,\mathrm{EC}}^{(k-1)}, \ \\
g_{\ket{\overline+}}^{(k)}&=\beta_{\mathrm{CNOT}}\tilde{g}_{\mathrm{CNOT}}^{(k-1)} + \beta_{\mathrm{mem}}\tilde{g}_{\mathrm{mem}}^{(k-1)} + \beta_{\ket{0}}g_{\ket{\overline0}}^{(k-1)} \nonumber \\ 
&\qquad + \beta_{\ket{+}}g_{\ket{\overline+}}^{(k-1)} + n_{g,\mathrm{EC}}^{(k-1)}.
\label{eq:StatePrepGateOverheadLevelk}
\end{align}
Note that the $n_{g,\mathrm{EC}}^{(k-1)}$ term in Eq.~\ref{eq:StatePrepGateOverheadLevelk} was included to take into account the gate overhead from the level-$(k-1)$ EC~circuit that precedes $\ket{\overline0}^{(k)}$ and $\ket{\overline+}^{(k)}$.

In the first part of the $\ket{\overline0}$ ancilla verification test, we must include the contributions from the two $\ket{\overline0}^{(k)}$ circuits, the CNOT exRec and the measurement locations. Since the EC circuit prior to performing a measurement is included in the contribution from the TEC of the CNOT gate, level-$k$ measurement locations will always have a gate overhead of $n^k$ so that 

\begin{align}
n_{g,\ket{\overline0}_{1}}^{(k)}&=n_{g,\ket{\overline0}_{2}}^{(k)}=2g_{\ket{\overline0}}^{(k)}+n\tilde{g}_{\mathrm{CNOT}}^{(k-1)}+n^k, \ \\
n_{g,\ket{\overline+}_{1}}^{(k)}&=n_{g,\ket{\overline+}_{2}}^{(k)}=2g_{\ket{\overline+}}^{(k)}+n\tilde{g}_{\mathrm{CNOT}}^{(k-1)}+n^k, \ \\
n_{g,\ket{\overline0}_{3}}^{(k)}&=n_{g,\ket{\overline+}_{3}}^{(k)}=n\tilde{g}_{\mathrm{CNOT}}^{(k-1)} + n^k, \ \\
n_{\mathrm{rem}}^{(k)}&=2n(\tilde{g}_{\mathrm{CNOT}}^{(k-1)}+\tilde{g}_{\mathrm{mem}}^{(k-1)})+2n^k.
\label{eq:AncillaVerificationTestLevelk}
\end{align}
The overhead for the full ancilla verification procedure is obtained in the same way as in Eq.~\ref{eq:expectedNumberGatesVerificationTest}:
\begin{align}
n_{g,\mathrm{EC}\ket{\overline0}}^{(k)}&=\frac{n_{g,\ket{\overline0}_{1}}^{(k)}\Big(\frac{1}{p_{\ket{\overline0}_{1}}^{(k)}}+\frac{1}{p_{\ket{\overline0}_{2}}^{(k)}}\Big)+n_{g,\ket{\overline0}_{3}}^{(k)}}{p_{\ket{\overline0}_{3}}^{(k)}}, \ \\
n_{g,\mathrm{EC}\ket{\overline+}}^{(k)}&=\frac{n_{g,\ket{\overline+}_{1}}^{(k)}\Big(\frac{1}{p_{\ket{\overline+}_{1}}^{(k)}}+\frac{1}{p_{\ket{\overline+}_{2}}^{(k)}}\Big)+n_{g,\ket{\overline+}_{3}}^{(k)}}{p_{\ket{\overline+}_{3}}^{(k)}}.
\label{eq:AncillaVerificationTestLevelk}
\end{align}

The overhead for the complete EC circuit at level-$k$ is then the sum of the contributions from $n_{g,\mathrm{EC}\ket{\overline0}}^{(k)}$, $n_{g,\mathrm{EC}\ket{\overline+}}^{(k)}$ and $n_{\mathrm{rem}}^{(k)}$,
\begin{align}
n_{g,\mathrm{EC}}^{(k)} = n_{g,\mathrm{EC}\ket{\overline0}}^{(k)}+n_{g,\mathrm{EC}\ket{\overline+}}^{(k)}+n_{\mathrm{rem}}^{(k)}.
\label{eq:FullECgateOverheadLevelk}
\end{align}
It is then straightforward to obtain the overhead for the CNOT, storage and Hadamard exRecs, which we state below:
\begin{align}
g_{\mathrm{CNOT}}^{(k)}&=4n_{g,\mathrm{EC}}^{(k)}+n\tilde{g}_{\mathrm{CNOT}}^{(k-1)}, \ \label{eq:FinalGateOverheadLevelk1}\\
g_{\mathrm{mem}}^{(k)}&=2n_{g,\mathrm{EC}}^{(k)}+n\tilde{g}_{\mathrm{mem}}^{(k-1)}, \ \label{eq:FinalGateOverheadLevelk2}\\
g_{\mathrm{Had}}^{(k)}=&2n_{g,\mathrm{EC}}^{(k)}+\gamma_{\mathrm{CNOT}}\tilde{g}_{\mathrm{CNOT}}^{(k-1)} \nonumber \\
&\qquad + \gamma_{\mathrm{mem}}\tilde{g}_{\mathrm{mem}}^{(k-1)}+\gamma_{\mathrm{Had}}\tilde{g}_{\mathrm{Had}}^{(k-1)}
\label{eq:FinalGateOverheadLevelk3}
\end{align}

\subsection{49 and 105-qubit code overhead results} \label{49 and 105-qubit code overhead results} 

\begin{table*}[t!]
\centering
\begin{tabular}{|c|c|c|c|c|}
\hline 
Code/Gate & Concatenation level & Physical error rate & Qubit overhead & Gate overhead\tabularnewline
\hline 
\hline 
49-qubit Hadamard & 1 & N/A&N/A&N/A\tabularnewline
\hline 
 & 2 & $2.36\times 10^{-6}$ & $(1.75\pm0.38)\times 10^{5}$ & $(4.20\pm0.92)\times 10^{7}$\tabularnewline
\hline 
 & 3 & $8.47\times 10^{-5}$ & $(7.33\pm1.57)\times 10^{7}$ & $(2.14\pm0.66)\times 10^{11}$\tabularnewline
\hline
49-qubit CNOT & 1 & N/A&N/A&N/A\tabularnewline
\hline
& 2 & $5.24\times 10^{-5}$ & $(3.60\pm0.79)\times 10^{5}$ & $(8.38\pm0.18)\times 10^{7}$\tabularnewline
\hline 
& 3 & $3.92\times 10^{-4}$ & $(1.94\pm0.37)\times 10^{8}$ & $(5.45\pm1.50)\times 10^{11}$\tabularnewline
\hline
\hline 
105-qubit Hadamard & 1 & N/A&N/A&N/A\tabularnewline
\hline 
& 2 & $5.35\times 10^{-7}$ & $(1.14\pm0.22)\times 10^{6}$ & $(1.86\pm0.34)\times 10^{8}$\tabularnewline
\hline
& 3 & $1.60\times 10^{-5}$ & $(3.00\pm0.47)\times 10^{9}$ & $(4.82\pm0.73)\times 10^{12}$\tabularnewline
\hline
105-qubit CNOT & 1 & N/A&N/A&N/A\tabularnewline
\hline 
& 2 & $4.56\times 10^{-4}$ & $(2.27\pm0.44)\times 10^{6}$ & $(3.55\pm0.66)\times 10^{8}$\tabularnewline
\hline
& 3 & $1.39\times 10^{-3}$ & $(6.01\pm0.94)\times 10^{9}$ & $(9.28\pm1.40)\times 10^{12}$\tabularnewline
\hline
\end{tabular}
\caption{Overhead results for the 49 and 105-qubit codes. The first column indicates the code and corresponding logical gate for which the overhead is computed. The third column indicates the largest physical error rate $p$ that can be achieved for the particular concatenation level so that the logical error rate is below $p_{target}=10^{-15}$. The fourth and fifth columns give the qubit and gate overhead for the given physical error rate.}
\label{tab:OverheadTable}
\end{table*}

In this section we use the formalism of section \ref{Raw qubit overhead} and \ref{Gate overhead} to obtain the raw qubit and gate overhead results of the 49 and 105-qubit codes. Since the Hadamard gate limits the threshold value of both codes, we will focus on the overhead for performing a logical Hadamard and CNOT gate. Given a target logical error rate $p_{target}$, we can use the threshold results of Section~\ref{sec:Concatenated 49-qubit thresholds} to determine the appropriate level of concatenation to reach $p_{target}$. We can then calculate the raw qubit overhead and gate overhead from Eqs.~\ref{eq:nECoverheadLevelk1}--\ref{eq:nECoverheadLevelk2} and~\ref{eq:FinalGateOverheadLevelk1}--\ref{eq:FinalGateOverheadLevelk3}, respectively. The coefficients $\alpha_{j}$, $\beta_{j}$ and $\gamma_{j}$ are given in Section~\ref{Ancilla prep section}.

Figs.~\ref{fig:49QubitOverhead} and~\ref{fig:105QubitOverhead} illustrate the physical qubit and gate overhead for the 49 and 105-qubit codes. The key results are summarized in Table~\ref{tab:OverheadTable}. In the following discussion we focus on a target logical error rate of $p_{target}=10^{-15}$. For the 49-qubit logical Hadamard gate, if we limit the implementation to two levels of concatenation it would require a level of precision of~$2.36\times 10^{-6}$, far below the asymptotic threshold, in order to reduce the logical error rate to~$10^{-15}$. The raw qubit and gate overheads are given by $(1.75\pm0.38)\times 10^{5}$ and  $(4.20\pm0.92)\times 10^{7}$ in such a scenario. Going to a third level of concatenation allows for higher physical error rates, up to~$8.47\times 10^{-5}$, as the increase concatenation level will further reduce logical error rate. However there is a tradeoff to increasing the concatenation level as it requires further resources. The raw qubit and gate overheads for an error rate of~$8.47\times 10^{-5}$ are given by $(7.33\pm1.57)\times 10^{7}$ and $(2.14\pm0.66)\times 10^{11}$ (an increase of roughly three orders of magnitude). One aspect that is quite apparent in the overhead results is how well the 49-qubit logical Hadamard performs compared to the 105-qubit version. For the 105-qubit code, the logical Hadamard gate can achieve the desired target error rate at the second concatenation level for physical error rates below $5.35\times 10^{-7}$, as opposed to $2.36\times 10^{-6}$, due to the complexity of the 105-qubit Hadamard construction compared to that of the 49-qubit code. The third level of concatenation pushes this number to $1.60\times 10^{-5}$. Furthermore, for each concatenation level the overheads are roughly an order of magnitude larger than for the 49-qubit logical Hadamard. 

\subsection{Comparison to the surface code overhead using state distillation}

In this Section, we aim to estimate the qubit~(space) overhead for the implementation of fault-tolerant logic in the surface code. The primary obstacle to surface code implementations is the need to distill a special ancillary state for the purposes of implementing the $T$~gate through gate teleportation. High-fidelity logical state preparation of this special state is obtained through a process called state distillation. State distillation is implemented using only logical Clifford gates, resulting in a non-stabilizer state that can be used to implement the $T$~gate, called a magic state~\cite{BK05}.

The idea behind magic state distillation begins by assuming the initial magic state has an error rate~$p$. Then by using multiple noisy logical magic states and near-perfect Clifford operations (since the Clifford operations are performed using fault-tolerant logical operations in the surface code) a higher-fidelity magic state can be distilled from multiple noisy states. Depending on the distillation scheme used, the output state of the scheme will have a fidelity of~$c p^b$, for some constant value of $b$~and~$c$. The process can then be repeated with multiple copies of the newly distilled logical states to obtain a state with error rate~$c(cp^b)^b = c^{b+1}p^{b^2}$. Iterating this process $k$~times, the final logical output state will have an error rate:
\begin{align}
p_k = \dfrac{1}{c^{\frac{1}{b-1}}} \left( c^{\frac{1}{b-1}} p\right)^{b^k}.
\label{eq:MSDoutput}
\end{align}
The magic state distillation process can be probabilistic, yielding a distilled state based on the result of a set of measurements, and in general will succeed with probability~$1/r$ (where $r$ depends on the particular distillation scheme being used). Therefore, if~$n$~logical qubits are required for the distillation, and the probability of success for a given round is~$1/r$, the total number of logical qubits required for $k$~distillation rounds is given by~$(rn)^k$. The number of qubits required is thus exponential in the number of rounds, however the scheme remains theoretically efficient as the logical error rate is suppressed double-exponentially. Table~\ref{tab:MSDschemes} summarizes the different parameters for the magic state distillation schemes.

\begin{table}
\begin{centering}
\begin{tabular}{|c|c|c|c|}
\hline 
Distillation Type & Qubits & Error rate & Success prob. \tabularnewline
\hline 
\hline 
$\ket{T}$-type & 5 & $\frac{1}{5} (5p)^{2^k}$ & 1/6 \tabularnewline
\hline 
$\ket{H}$-type & 15 & $\frac{1}{\sqrt{35}} (\sqrt{35}p)^{3^k}$ & $\approx 1 - p/15$ \tabularnewline
\hline 
10-to-2 & 5 & $\frac{1}{9} (9p)^{2^k}$ & $\approx 1$ \tabularnewline
\hline
\end{tabular}
\par\end{centering}
\caption{\label{tab:MSDschemes}Parameters for different magic state distillation schemes. The $\ket{T}$~and~$\ket{H}$ schemes are given in Ref.~\cite{BK05}, while the 10-to-2 qubit scheme is presented in Ref.~\cite{MEK12}.}
\end{table}

\begin{figure}[htbp]
\centering
\begin{subfigure}{0.45\textwidth}
\includegraphics[width = \textwidth]{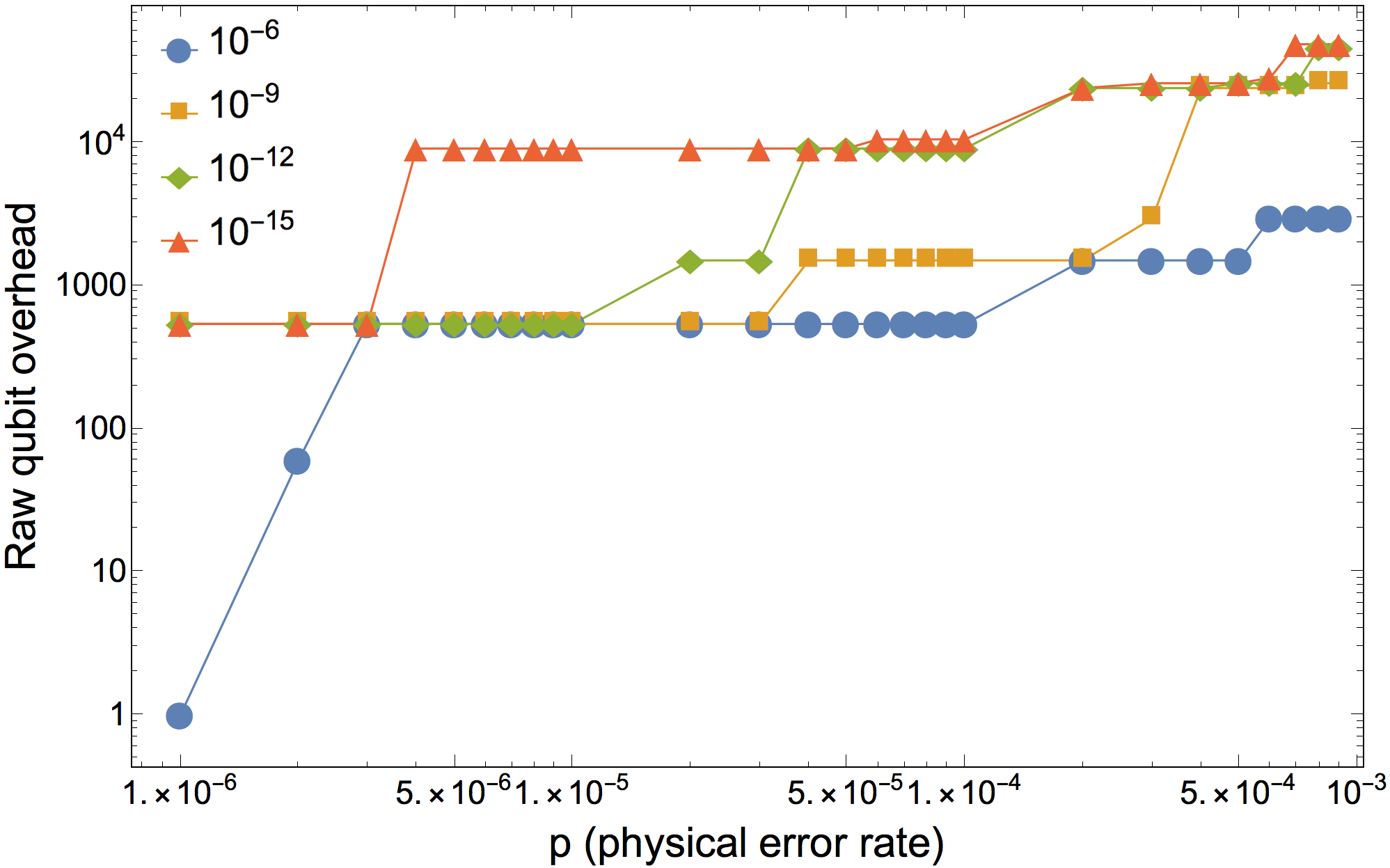}
\caption{}
\label{fig:HstateOverhead}
\end{subfigure}
\begin{subfigure}{0.453\textwidth}
\includegraphics[width =\textwidth]{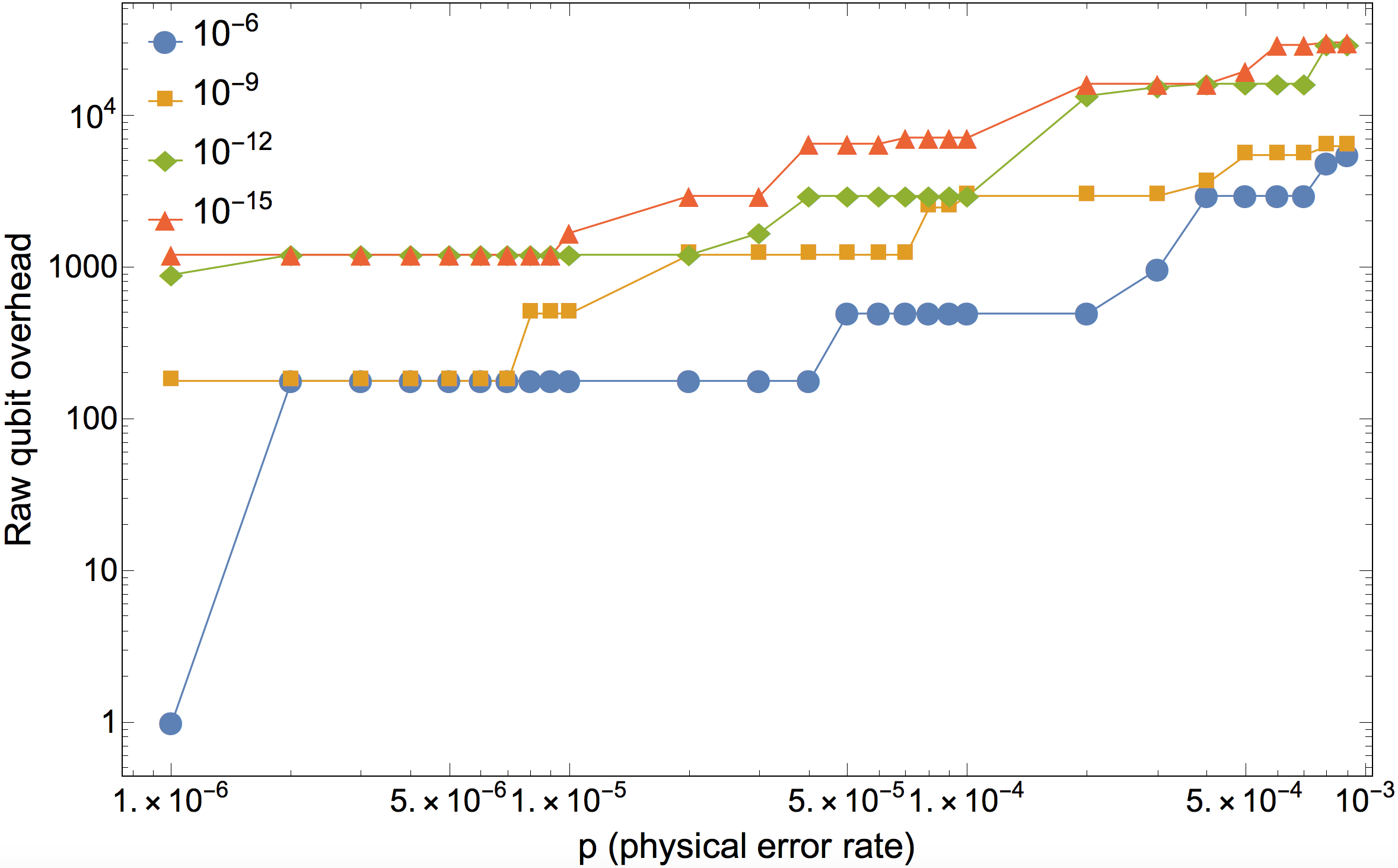}
\caption{}
\label{fig:MEstateOverhead}
\end{subfigure}
\caption{\subref{fig:HstateOverhead}~Physical qubit overhead for the distillation of the~$H$-type magic state to be used in the implementation of the $T$~gate for the surface code with different target gate error rates. \subref{fig:MEstateOverhead}~Distillation overhead for the Meier-Eastin 10-to-2 distillation scheme for the surface code. As can be seen in the above plots, for physical error rates of $p=10^{-5}$, the Meier-Eastin 10-to-2 distillation scheme requires roughly 1000 physical qubits to implement the $T$-gate. This is about an order of magnitude less than the physical qubits required for the distillation of the~$H$-type magic state.}
\label{fig:SurfaceCodeOverhead}
\end{figure}

Suppose we would like to obtain a lower bound on the number of physical qubits that are required to implement the logical~$T$ gate in the surface code. Given a target error rate~$p_{target}$, and a depolarizing physical error rate of~$p$, we must first determine the number of distillation levels required to obtain the desired target rate, that is choose a value of~$k$ from Eq.~\ref{eq:MSDoutput} such that $p_k < p_{target}$. Having determined the level of distillation, recall that we argued that the Clifford gates must have low levels of noise with respect to the target error rate. Therefore, for a given distillation level, we will need to choose logical gates that have a small enough logical error rate. For the surface code, by considering the probability of a logical string being created given that the syndrome extraction scheme is 8~time steps long, the logical gate error rate can be approximated as~\cite{FMMC12}:
\begin{align}
p_L(d) = d {d \choose \ceil{d/2}} (8p)^{\ceil{d/2}},
\label{eq:LogicalErrorSC}
\end{align}
where $d$ is the distance of the code. Therefore, if the output of a given distillation level is given by~$p_k$, as argued in Ref.~\cite{FMMC12}, the logical error rate must be small enough such that the resulting accumulation of errors from the distillation circuit does not negatively affect the distilled qubit, that is:
\begin{align}
8 \cdot 1.25 \cdot n_q \cdot d \cdot p_L(d) \le p_k,
\end{align}
where $n_q$ is the number of qubits in the distillation scheme. Therefore, the distance must be chosen large enough to reduce the logical error rate to sufficiently small levels (assuming we are always below threshold). Having distilled at a given level, the logical qubits can be enlarged through a fault-tolerant growing operation, in order to be of the required distance for the next distillation level. In our lower bound of the number of qubits required, we will assume that this operation is done error free, in reality this may slightly increase the number of physical qubits required. Therefore given the required distance at each level, and the fact that the number of physical qubits for a distance~$d$ surface code is~$(2d)^2$ (including the syndrome qubits), a lower bound of the distillation overhead can be obtained. Figure~\ref{fig:SurfaceCodeOverhead} illustrates the physical qubit overhead for two different distillation schemes used in the implementation of the $T$ gate for the surface code. It can be seen that for a target logical error rate of $p_{target}=10^{-15}$, both schemes yield an overhead on the order of $10^4$ physical qubits for input error rates $10^{-4}<p<10^{-3}$.

\section{Conclusion}

In this paper we reviewed the fault-tolerant construction of the 49 and 105-qubit codes obtained from concatenating Steane's 7-qubit code with the 15-qubit Reed-Muller code. One advantage of the concatenation scheme is that universal fault-tolerance can be achieved without using state distillation protocols. Taking advantage of the CSS structure of the 7 and 15-qubit codes, the error correction blocks were constructed in the framework of Steane error correction. In the study of the performance of the concatenated codes, we obtained the threshold for the 105-qubit code for an adversarial noise model using malignant set counting and found it to be~$(8.33\pm 0.28)\times 10^{-6}$, which is slightly below that of the concatenated 7-qubit Steane code assisted by magic state distillation~\cite{AGP06}. In order to obtain threshold estimates that would more accurately represent the noise seen in an actual experiments, we computed the threshold value of the 49-qubit code for depolarizing noise using the techniques developed in Ref. \cite{PR12, CJL16}. The 49-qubit code depolarizing threshold value was found to be $(9.69\pm0.28)\times 10^{-4}$ which is competitive with the 105-qubit code threshold of $(1.28\pm0.02)\times 10^{-3}$~\cite{CJL16}. As was the case for the 105-qubit code, the threshold for the 49-qubit code was limited by the logical Hadamard gate.

\begin{figure*}[htbp]
\begin{center}
\begin{subfigure}{0.48\textwidth}
\includegraphics[width = \textwidth]{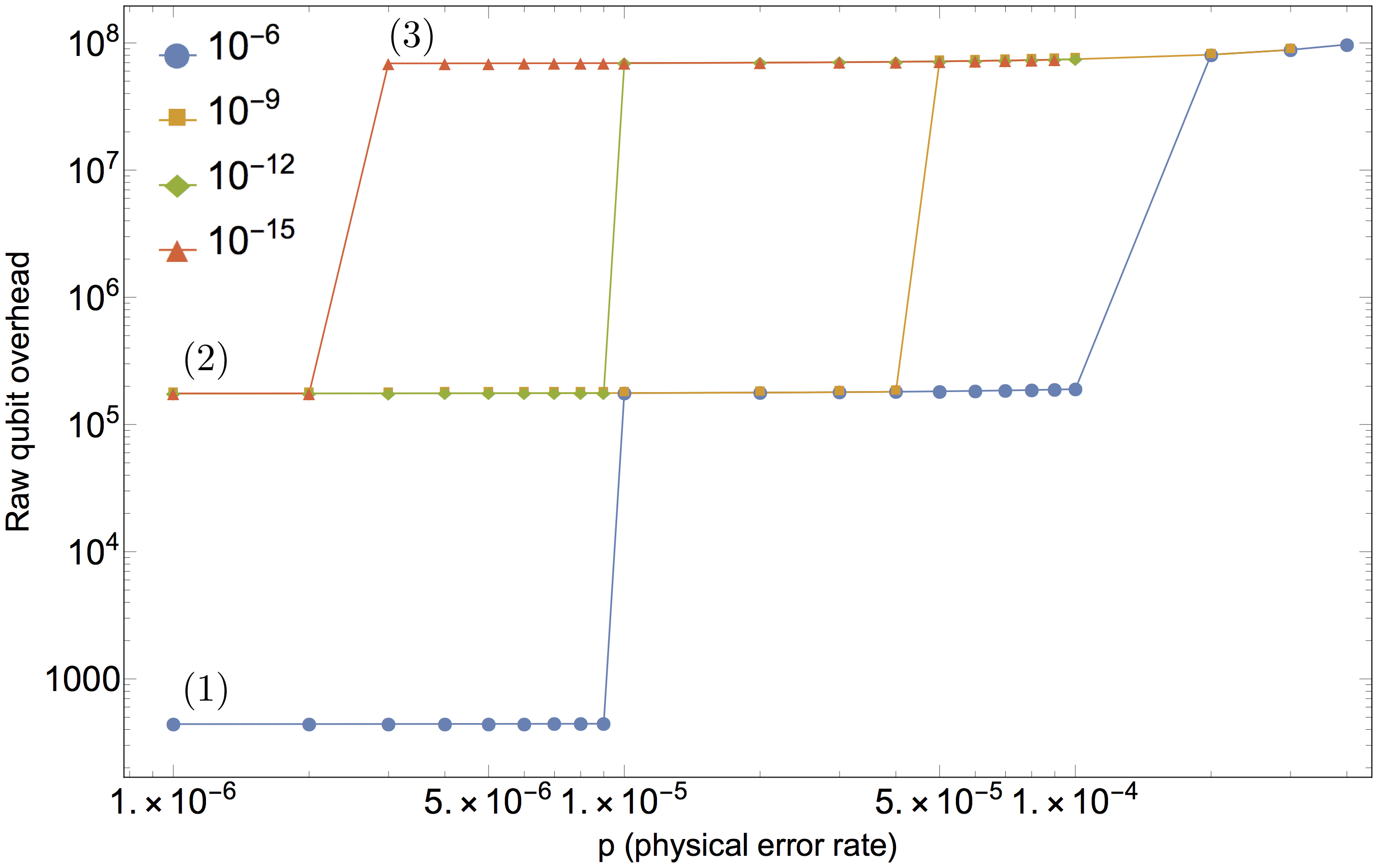}
\caption{}
\label{fig:aOverhead49}
\end{subfigure}
\begin{subfigure}{0.48\textwidth}
\includegraphics[width =\textwidth]{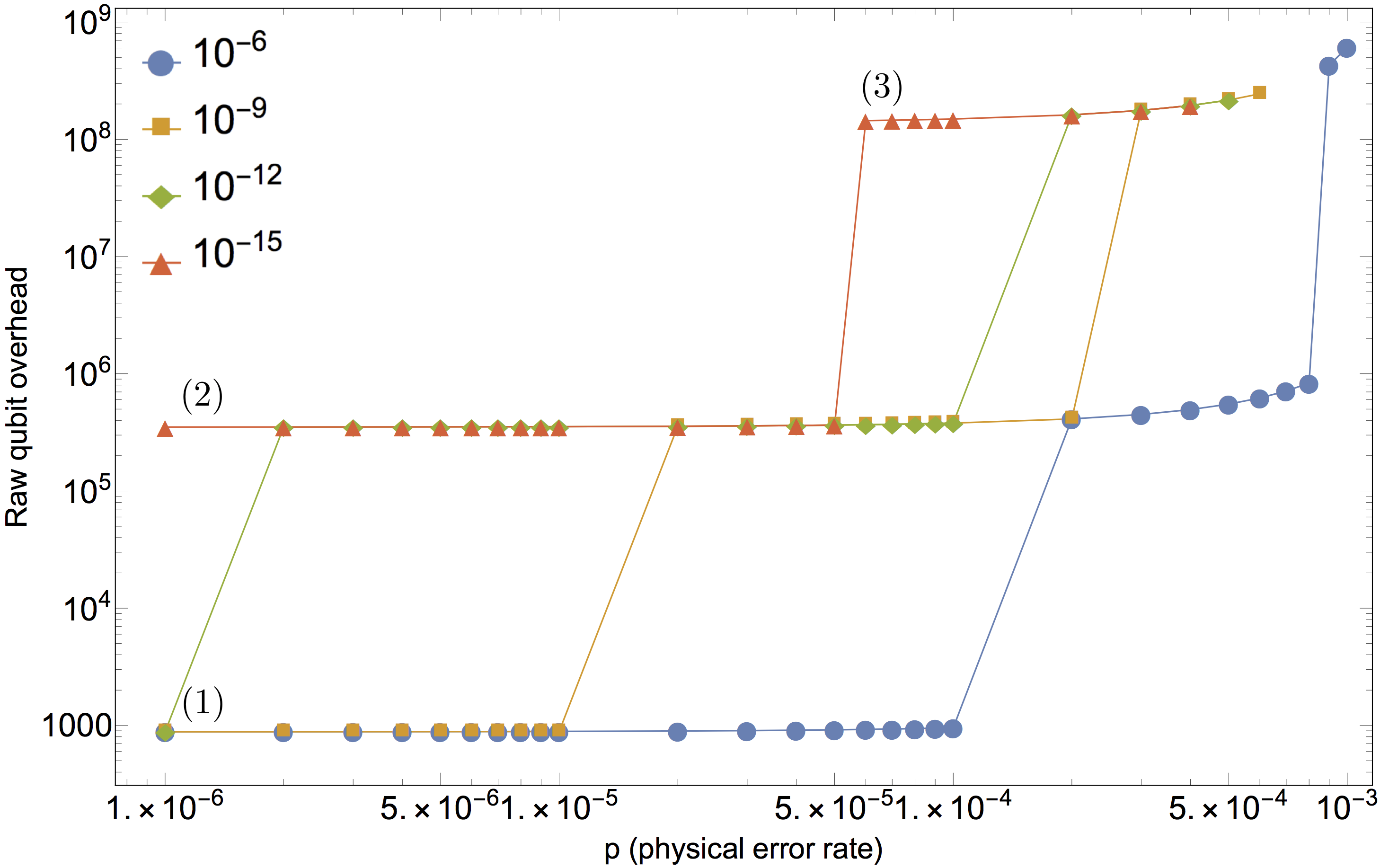}
\caption{}
\label{fig:bOverhead49}
\end{subfigure}
\begin{subfigure}{0.48\textwidth}
\includegraphics[width = \textwidth]{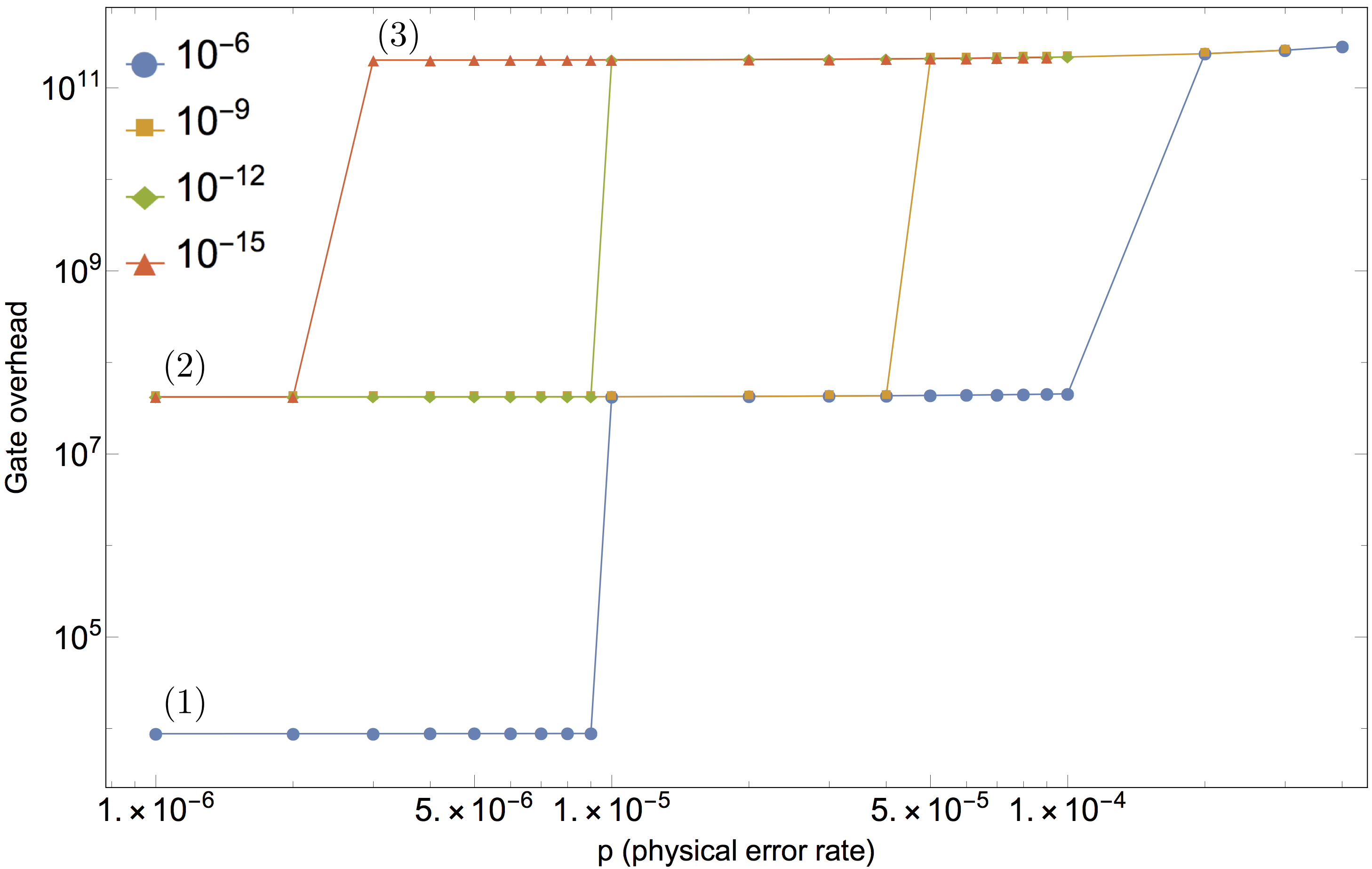}
\caption{}
\label{fig:cOverhead49}
\end{subfigure}
\begin{subfigure}{0.48\textwidth}
\includegraphics[width = \textwidth]{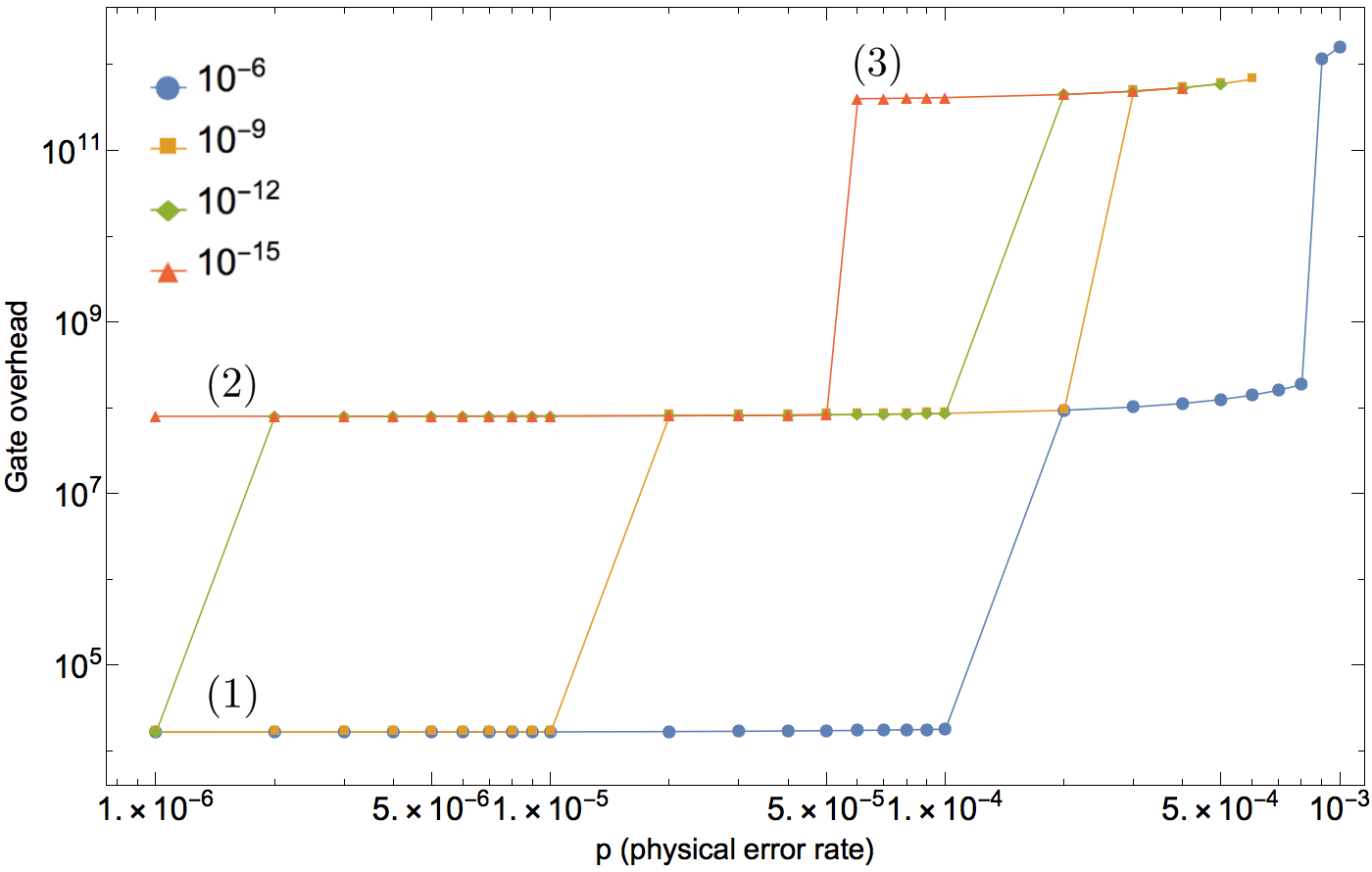}
\caption{}
\label{fig:dOverhead49}
\end{subfigure}
\caption{Physical qubit and gate overhead $\log$-$\log$ plots for the 49-qubit code. The plots in~(a) and~(c) illustrate the overhead for the Hadamard gate while the plots in~(b) and~(d) illustrate the overhead for the CNOT gate. The numbers in parenthesis indicate the level of concatenation required to achieve the target error rates shown in the legend. The plots have a step-like function behaviour since above certain physical error rates a higher level of concatenation is required to achieve the particular target error rate. For $p_{target}=10^{-15}$ and a physical error rate $p=10^{-5}$, roughly $10^{8}$ physical qubits are required to encode one logical Hadamard gate.}
\label{fig:49QubitOverhead}
\end{center}
\end{figure*}

\begin{figure*}[htbp]
\begin{center}
\begin{subfigure}{0.48\textwidth}
\includegraphics[width = \textwidth]{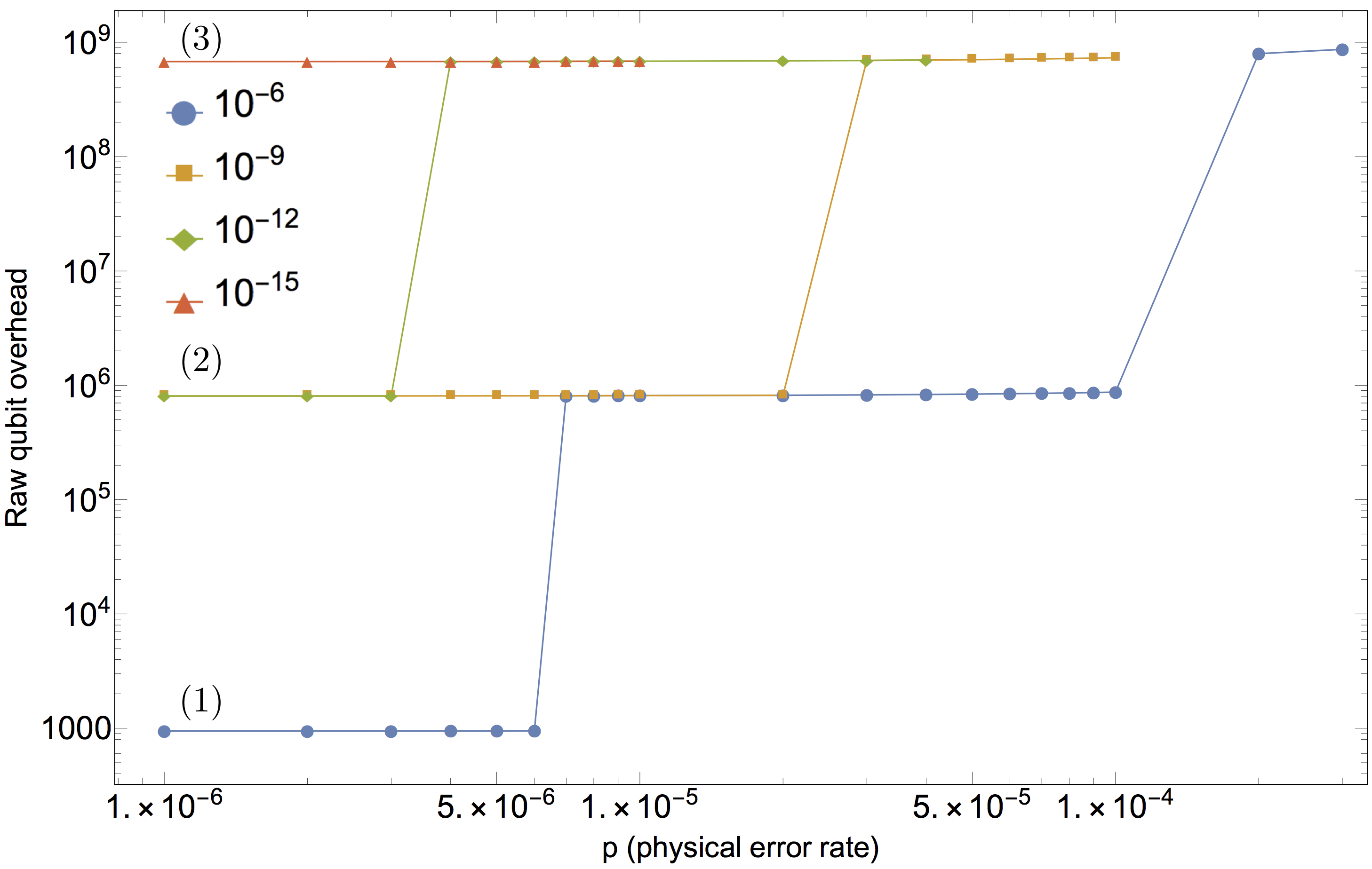}
\caption{}
\label{fig:aOverhead105}
\end{subfigure}
\begin{subfigure}{0.48\textwidth}
\includegraphics[width =\textwidth]{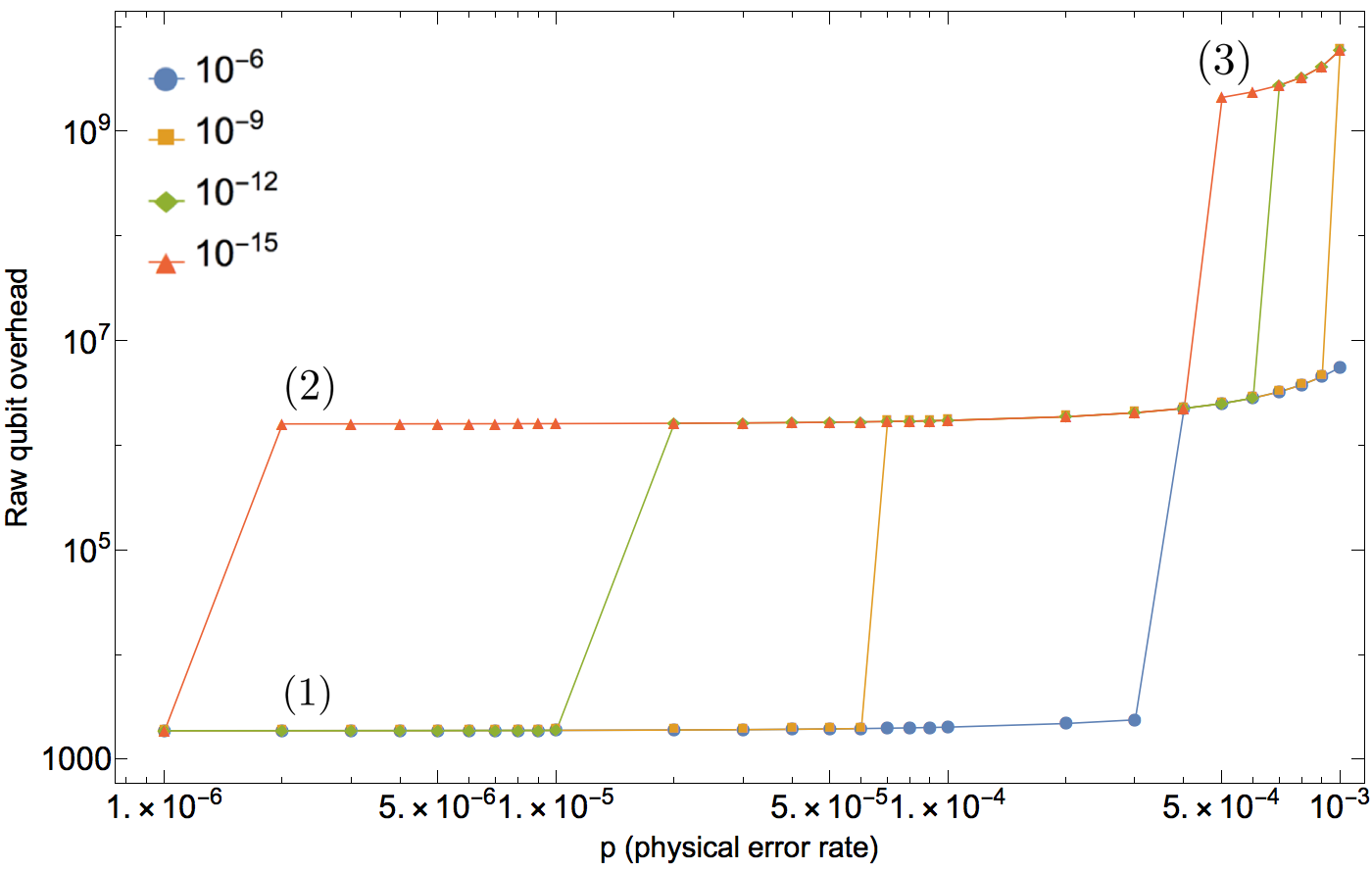}
\caption{}
\label{fig:bOverhead105}
\end{subfigure}
\begin{subfigure}{0.48\textwidth}
\includegraphics[width = \textwidth]{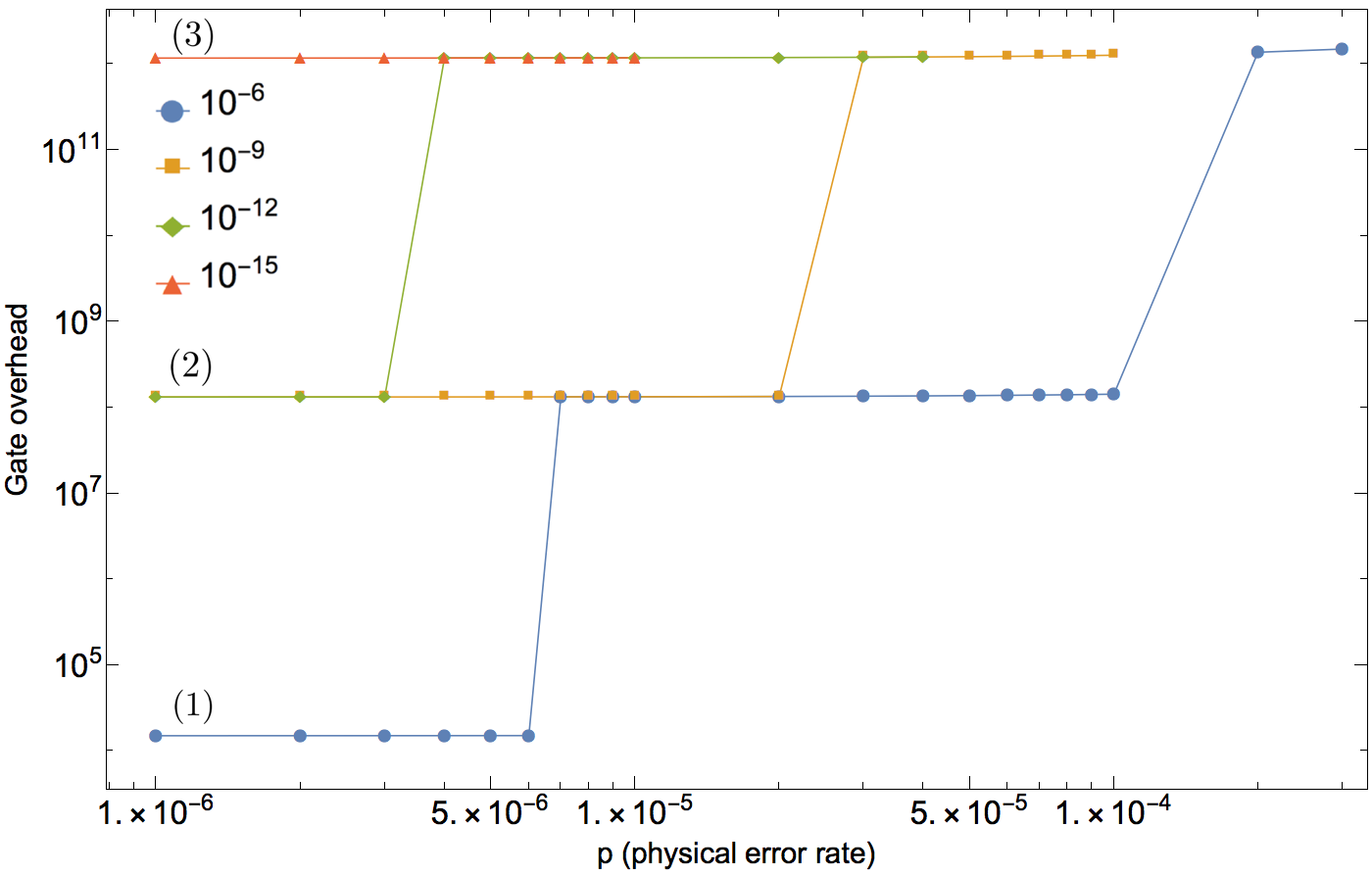}
\caption{}
\label{fig:cOverhead105}
\end{subfigure}
\begin{subfigure}{0.48\textwidth}
\includegraphics[width = \textwidth]{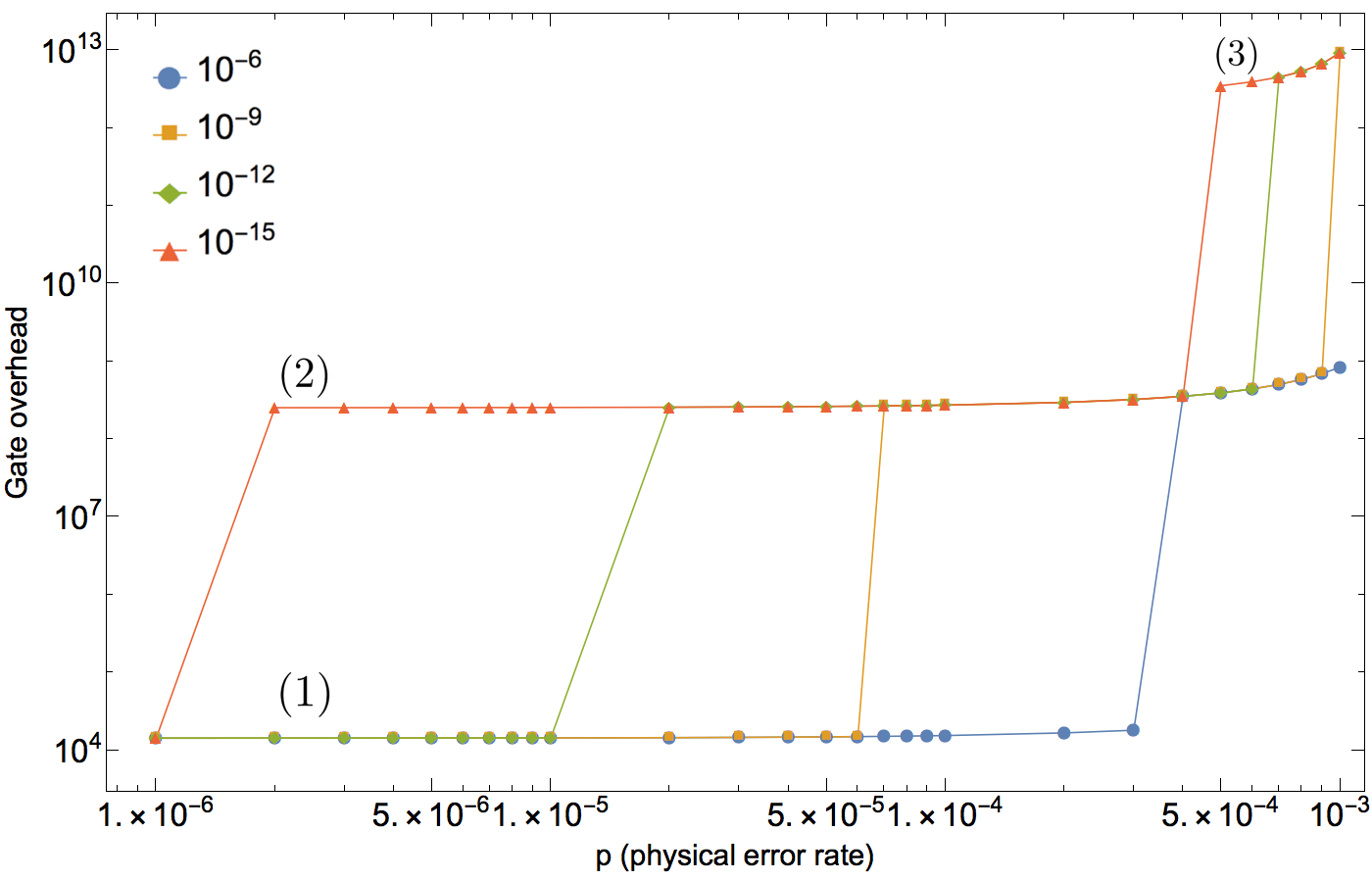}
\caption{}
\label{fig:dOverhead105}
\end{subfigure}
\caption{Physical qubit and gate overhead $\log$-$\log$ plots for the 105-qubit code. The plots in~(a) and~(c) illustrate the overhead for the Hadamard gate while the plots in~(b) and~(d) illustrate the overhead for the CNOT gate. The numbers in parenthesis indicate the level of concatenation required to achieve the target error rates shown in the legend. The plots have a step-like function behaviour since above certain physical error rates a higher level of concatenation is required to achieve the particular target error rate. For $p_{target}=10^{-15}$ and a physical error rate $p=10^{-5}$, roughly $10^{9}$ physical qubits are required to encode one logical Hadamard gate.}
\label{fig:105QubitOverhead}
\end{center}
\end{figure*}

We proceeded by developing general methods to compute the overhead of concatenated codes using Steane error correction and applied our methods to compute the physical qubit and gate overhead of the 49 and 105-qubit code. We also computed the physical qubit overhead for surface codes implementing the $T$~gate using the $H$-type magic state and the 10-to-2 distillation schemes~\cite{BK05, MEK12}. Comparing the plots of Fig.~\ref{fig:SurfaceCodeOverhead} with those of Fig.~\ref{fig:49QubitOverhead} and Fig.~\ref{fig:105QubitOverhead}, it is clear that surface code requires a smaller overhead than the 49 and 105-qubit codes given the sampled input error rates. For example, for an input error rate of $p=5\times10^{-5}$, the 49-qubit code requires more than $10^7$ physical qubits to implement a logical Hadamard gate compared to roughly $10^4$ physical qubits to implement the $T$~gate in the surface code. 

To explain the differences in overhead, it is first important to point out that the surface code thresholds are about an order of magnitude larger than the studied concatenated code thresholds. Hence, for comparable input error rates below threshold, the logical noise rate would be further suppressed for the surface codes requiring the use of fewer qubits to achieve a particular target logical error rate, even when using more rounds of distillation. Furthermore, the size of the EC blocks for Steane error correction represent a big drawback for concatenated codes. To illustrate this, it can be seen in Fig.~\ref{fig:aOverhead49} that the second level of concatenation requires the use of more than $10^5$ physical qubits for the 49-qubit code when implementing a logical Hadamard gate. Since the data qubits require the use $49^2=2401$ physical qubits, more than $97\%$ of the overhead comes from the size of the EC blocks. Consequently, even if the threshold for the 49 and 105 qubit codes were significantly improved, two levels of concatenation would require more than $10^5$ physical qubits which is more than the largest number of physical qubits used in the surface code implementation of the $T$~ gate for all sample error rates. The latter shows that in order to improve the overhead results of the studied concatenated codes, smaller EC blocks would need to be used that maintain the fault-tolerant properties of the concatenated scheme, even at the cost of lowering the asymptotic threshold. Additionally, it would be interesting to determine the resource overhead for the generalized construction of the concatenated model, using higher distance 2D and 3D~color codes as the base codes for the implementation of the fault-tolerant universal gate set.

Finally, in this work we explore an alternative to magic state distillation in the form of code concatenation, however other alternative schemes also exist~\cite{PR13, ADP14, BC15, JB16, JBH16, YTC16}. It remains unclear whether such schemes, typically using codes with transversal Clifford gates as their base codes along with different tricks to simulate the transversal action of the non-Clifford gate, could provide smaller overheads. Yet, while avoiding the primary obstacle of concatenation, such schemes may suffer in alternative ways, such as reduced threshold, increased ancilla space, longer measurement gadgets, etc.. A study of these alternative methods would be of great interest to the quantum error correction community and would likely involve a new set of tools to analyze the overall overhead.

\section{Acknowledgements}
T.~J.~would like to acknowledge the support of NSERC and the Vanier-Banting Secretariat through the Vanier~CGS. C. C. would like to acknowledge the support of QEII-GSST and to thank Steve Weiss for providing the necessary computational resources. This work was supported by CIFAR, NSERC, and Industry~Canada. 

\bibliographystyle{ieeetr}
\bibliography{bibtex_jochym}

\clearpage
\appendix

\section{Ancilla preparation and exRec circuits for the 105-qubit and 49-qubit code} \label{Ancilla prep section} 
\label{app:StatePrepCircuits}

In order to describe the ancilla states $\ket{\overline0}$ and $\ket{\overline+}$ for the 105 and 49-qubit codes, we first obtain their encoding circuits for the 7 and 15-qubit codes. Note that the 15-qubit Reed-Muller code is not self-dual which means that the encoding circuit $\ket{\overline+}_{15}$ cannot simply be obtained by reversing the direction of each CNOT gate and swapping the physical $\ket{0}$ and $\ket{+}$ states in $\ket{\overline0}_{15}$. As will be shown, this asymmetry will result in a larger number of physical locations in $\ket{\overline+}_{15}$ compared to $\ket{\overline0}_{15}$.

\begin{figure}[htbp]
\centering
\begin{subfigure}{0.4\textwidth}
\includegraphics[width=\textwidth]{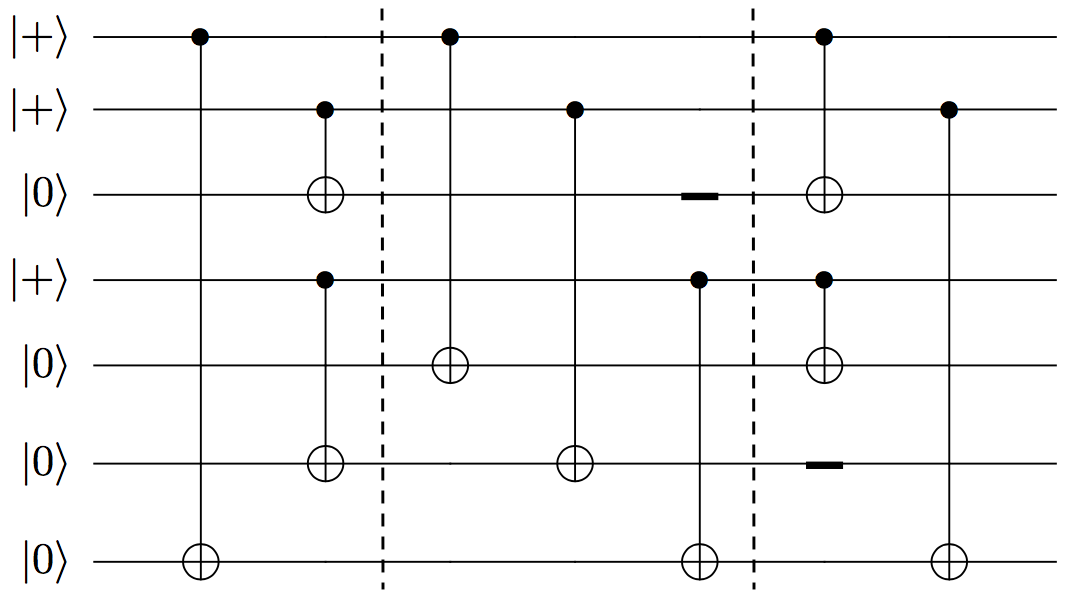}
\caption{}
\label{fig:NonOptimizedOPrep7App}
\end{subfigure}
\begin{subfigure}{0.4\textwidth}
\includegraphics[width=\textwidth]{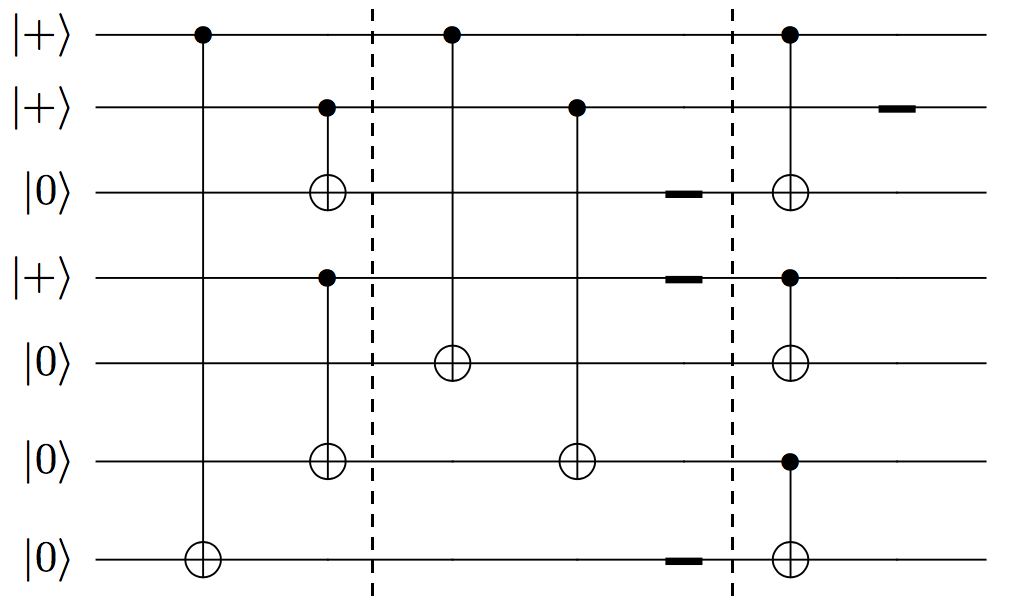}
\caption{}
\label{fig:optimizedOPrep7App}
\end{subfigure}
\caption{Encoding of $\ket{\overline0}_{7}$ for the 7-qubit Steane code. The circuit in \subref{fig:NonOptimizedOPrep7App} is obtained using Steane's Latin rectangle method and contains nine CNOT gates. The circuit in \subref{fig:optimizedOPrep7App} was obtained using the stabilizer overlap method and contains eight CNOT gates instead of nine. The dotted vertical lines are used to separate the time steps for which gates are applied in parallel. The bold dark lines represent resting qubits subject to storage errors. Note that we do not include a storage error on the fifth qubit in the first time step since it can be initialized in the second time step.}
\label{fig:First7qubitCircuitsApp}
\end{figure}

For CSS stabilizer codes, the encoded states can be obtained by solving a partial Latin rectangle using the codes stabilizer generators~\cite{Steane02}. However, as was shown in Ref.~\cite{PR12}, the circuits obtained from the previous method can be further optimized by considering overlaps in the codes stabilizer generators. As an example, consider the encoding circuit for the state~$\ket{\overline0}_{7}$ of the 7-qubit code. The circuit in Fig.~\ref{fig:NonOptimizedOPrep7App} (obtained via Steane's Latin rectangle method) contains nine CNOT gates. The logical state~$\ket{\overline0}_{7}$ can also be obtained by the circuit in Fig.~\ref{fig:optimizedOPrep7App} which uses eight CNOT gates, one fewer than the previous circuit. To see this, recall that the $X$ stabilizer generators for the 7-qubit code take the form $g_{1}=IIIXXXX$, $g_{2}=IXXIIXX$ and $g_{3}=XIXIXIX$. In Fig.~\ref{fig:NonOptimizedOPrep7App}, it can be seen that qubit seven is the target of qubits two and four and the corresponding stabilizer generators $g_{1}$ and $g_{2}$ overlap on qubits six and seven. Consequently, it is possible to replace the two CNOT's with control qubits two and four and target qubit seven with a CNOT having control on the sixth qubit and target on the seventh qubit. We use the circuit in Fig.~\ref{fig:optimizedOPrep7App} as the outer level of the state $\ket{\overline0}_{105}$. Since the 7-qubit code is self-dual, the $\ket{\overline{+}}_{7}$ can be obtained by reversing the direction of each CNOT gate and swapping the physical $\ket{0}$ and $\ket{+}$ states. 

\begin{figure}%[htbp]
\centering
\begin{subfigure}{0.4\textwidth}
\includegraphics[width=\textwidth]{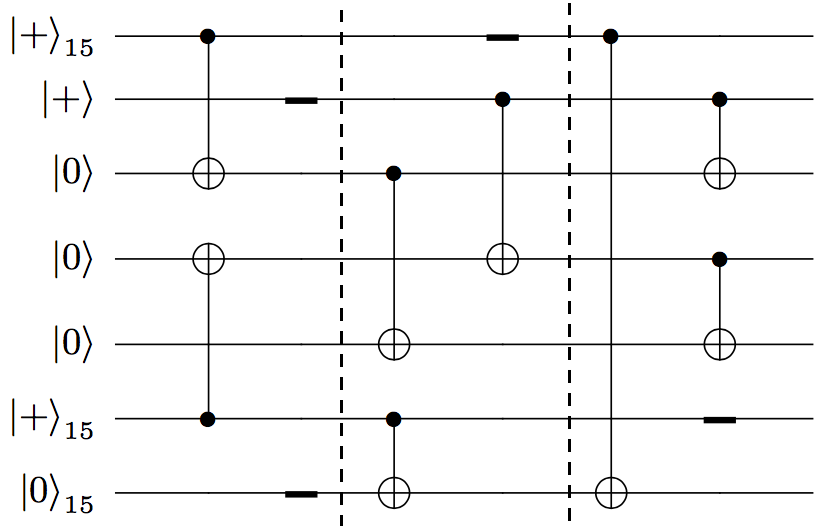}
\caption{}
\label{fig:O7Qubit49}
\end{subfigure}
\begin{subfigure}{0.4\textwidth}
\includegraphics[width=\textwidth]{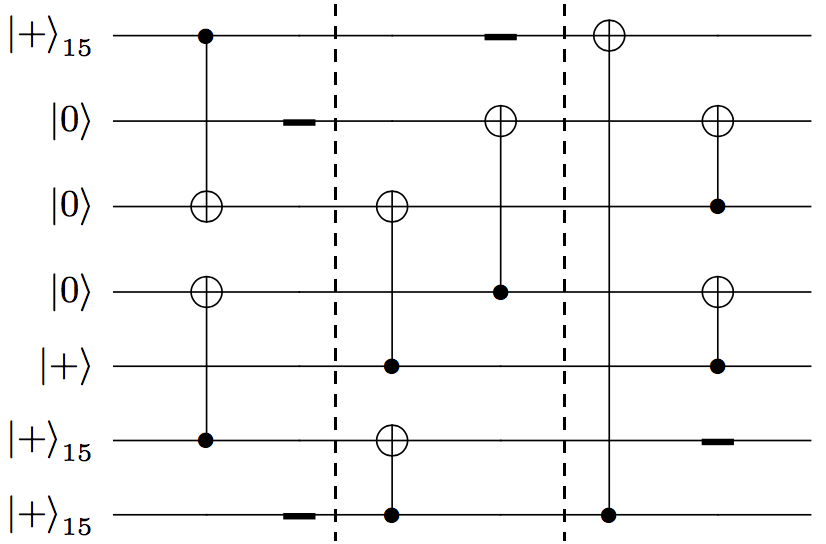}
\caption{}
\label{fig:Plus7Qubit49}
\end{subfigure}
\caption{$\ket{\overline0}_{49}$ and $\ket{\overline+}_{49}$ circuits for the 49-qubit code. These circuits have a lower space-time overhead compared to the one used in Fig.~\ref{fig:optimizedOPrep7App}. For example, the CNOT with block 1 as control and block 3 as target in \subref{fig:O7Qubit49} and \subref{fig:Plus7Qubit49} can be implemented in three time steps instead of seven since $\overline{Z}=Z_{1}Z_{2}Z_{3}$ is a logical $Z$ operator for the 15-qubit Reed-Muller code. Furthermore, any CNOT gate between blocks 1,6 and 7 can be implemented transversally.}
\label{fig:7QubitCircuit49}
\end{figure}

For the 49-qubit code, we use the circuits in Fig.~\ref{fig:7QubitCircuit49} at the outer level in order to minimize the total number of wait times experienced by each qubit. To see this, recall that the logical $Z$ operator for the 15-qubit code can take the form $\overline{Z}=Z_{1}Z_{2}Z_{3}$ (a weight 3 operator, since the first three bits of the codewords have even parity). Consequently, it is possible to apply a logical CNOT gate with the control lying on the first three qubits of a 15-qubit codeblock and target on a single-qubit codeblock. This would require the use of only three physical CNOT gates, one per time step. In this way, a $Z$ error on a single-qubit block would propagate to a logical $Z$ error on a 15-qubit codeblock. Since CNOT gates between 15-qubit codeblocks can be implemented transversally, a closer look at Fig.~\ref{fig:7QubitCircuit49} shows that the first two CNOT gates can be implemented in three time steps and all other CNOT gates can be implemented in a single time step. Hence the overall number of time steps required to implement the CNOT gates in Fig.~\ref{fig:7QubitCircuit49} is $3+1+1=5$. If instead we were to use the circuits in Fig.~\ref{fig:First7qubitCircuitsApp}, there are three CNOT gates coupling 15-qubit codeblocks to single qubit blocks. Two of these CNOT gates have their control on a 15-qubit codeblock encoding a $\ket{\overline +}$ state and so would be implemented in three time steps each. The CNOT gate coupling the $\ket{+}$ state to the 15-qubit $\ket{\overline 0}$ codeblock would require 7 time steps, since the minimum weight logical $X$ for the 15-qubit code is 7. The total number of time steps would have been $7+3+3=13$ instead of 5. 

\begin{figure}%[htbp]
%\centering
\begin{subfigure}{0.4\textwidth}
\includegraphics[width=\textwidth]{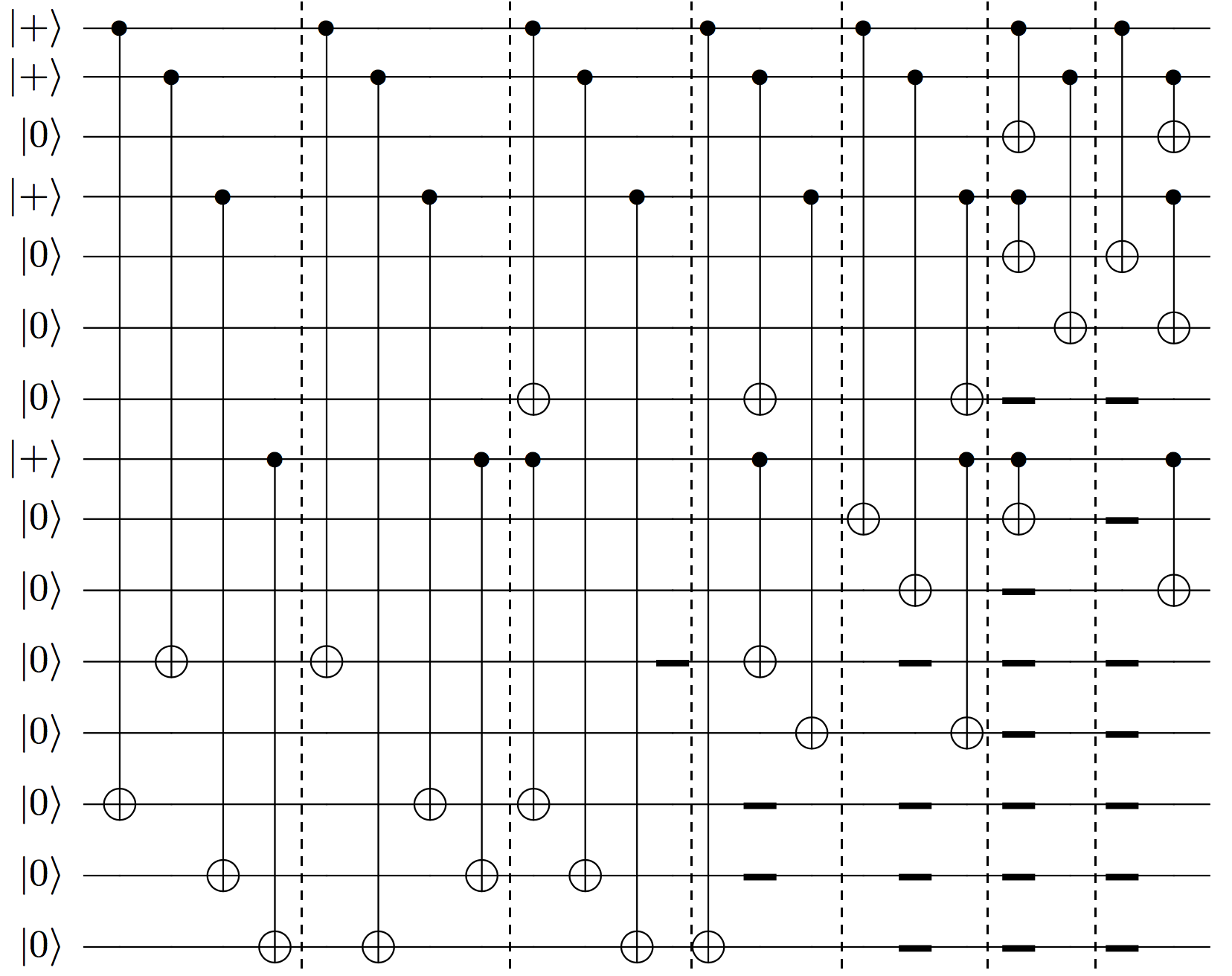}
\caption{}
\label{fig:Latin15QubitOprep}
\end{subfigure}
\begin{subfigure}{0.4\textwidth}
\includegraphics[width=\textwidth]{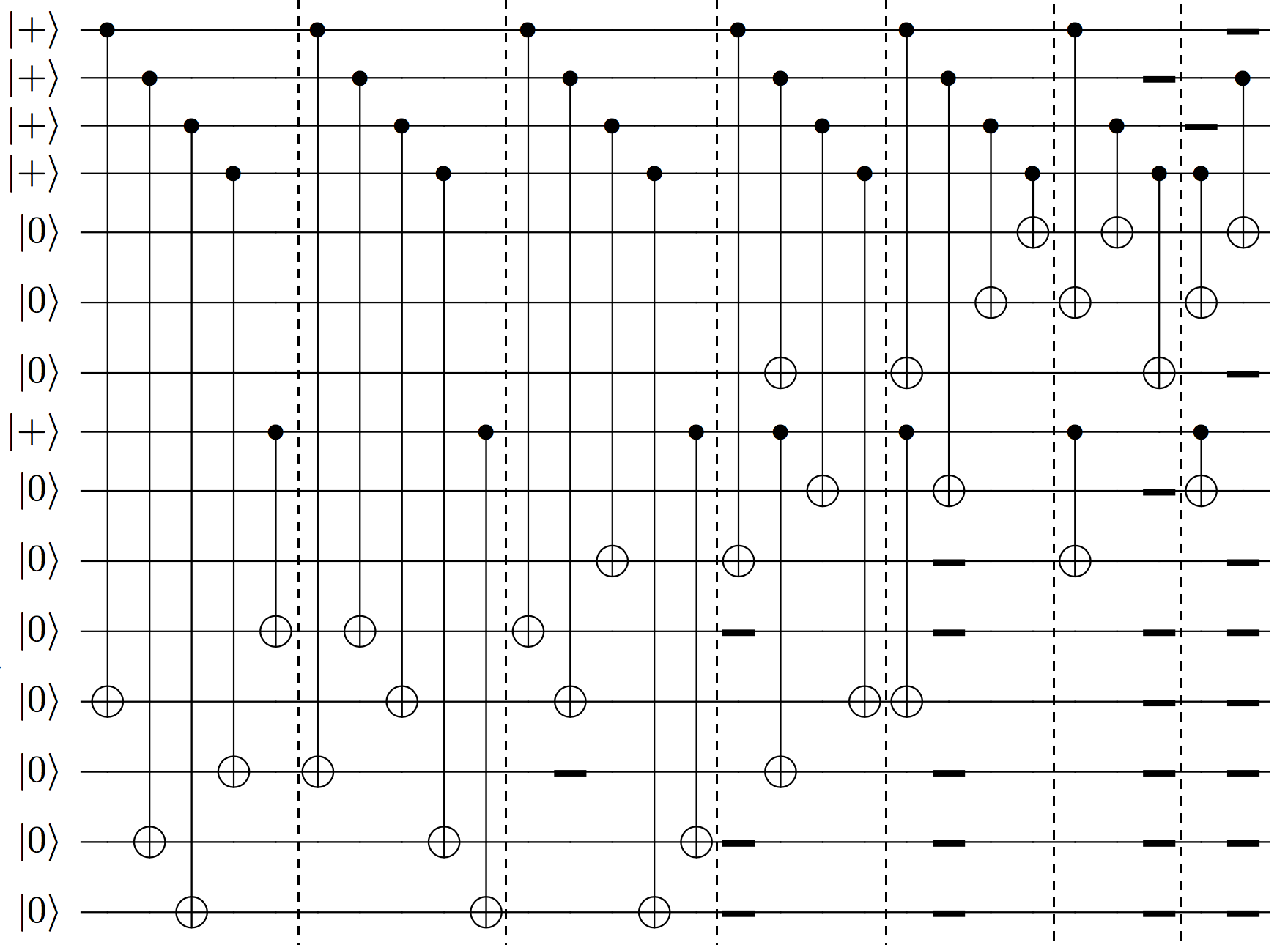}
\caption{}
\label{fig:Latin15QubitPlusprep}
\end{subfigure}
\caption{Encoding circuits $\ket{\overline0}_{15}$ and $\ket{\overline+}_{15}$ for the 15-qubit Reed-Muller code using Stean's Latin rectangle method. $\ket{\overline0}_{15}$ contains 28 CNOT gates and 21 resting qubit locations. $\ket{\overline+}_{15}$ contains 32 CNOT gates and 25 resting qubit locations.} 
\label{fig:15QubitCircuits}
\end{figure}

\begin{figure}%[htbp]
%\centering
\begin{subfigure}{0.4\textwidth}
\includegraphics[width=\textwidth]{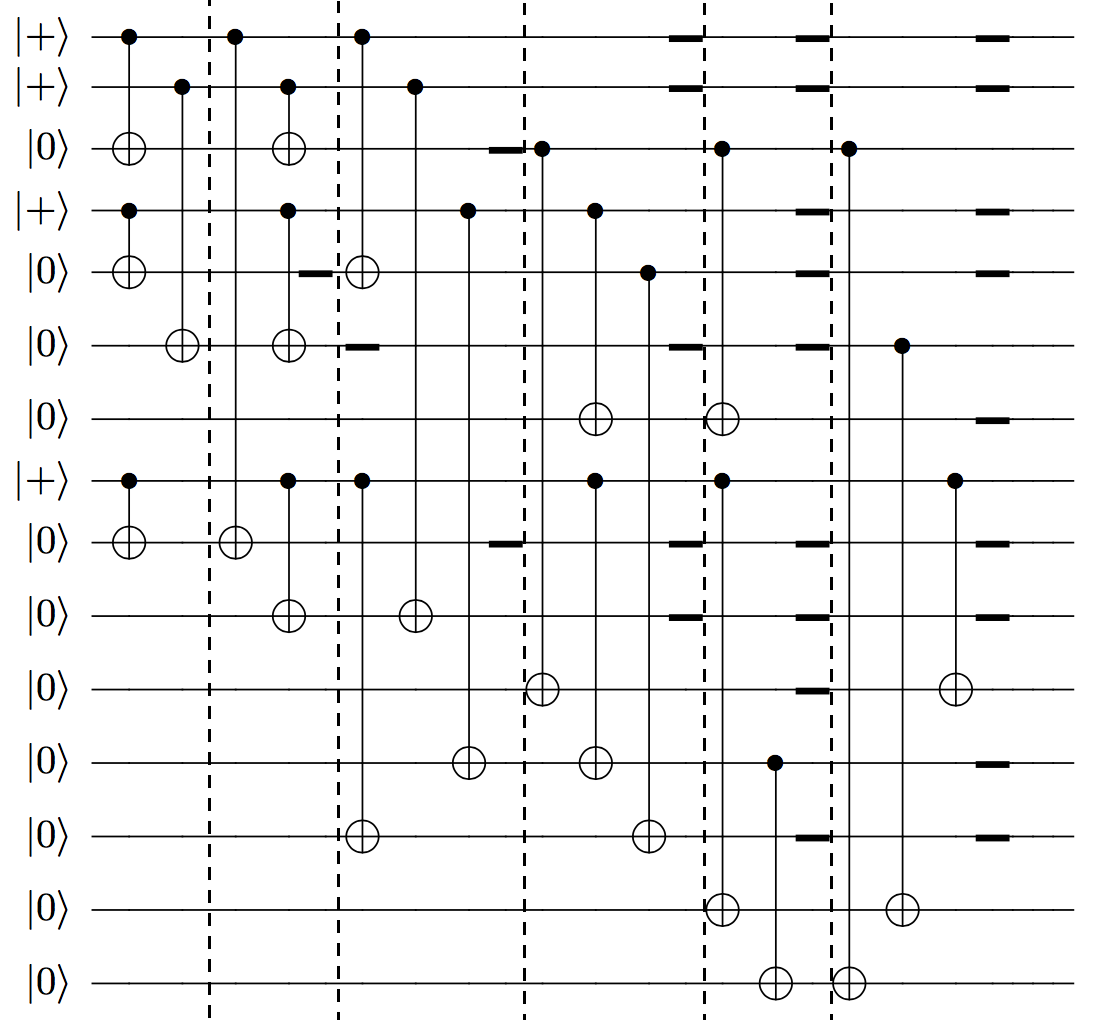}
\caption{}
\label{fig:Optimized15QubitOprep}
\end{subfigure}
\begin{subfigure}{0.4\textwidth}
\includegraphics[width=\textwidth]{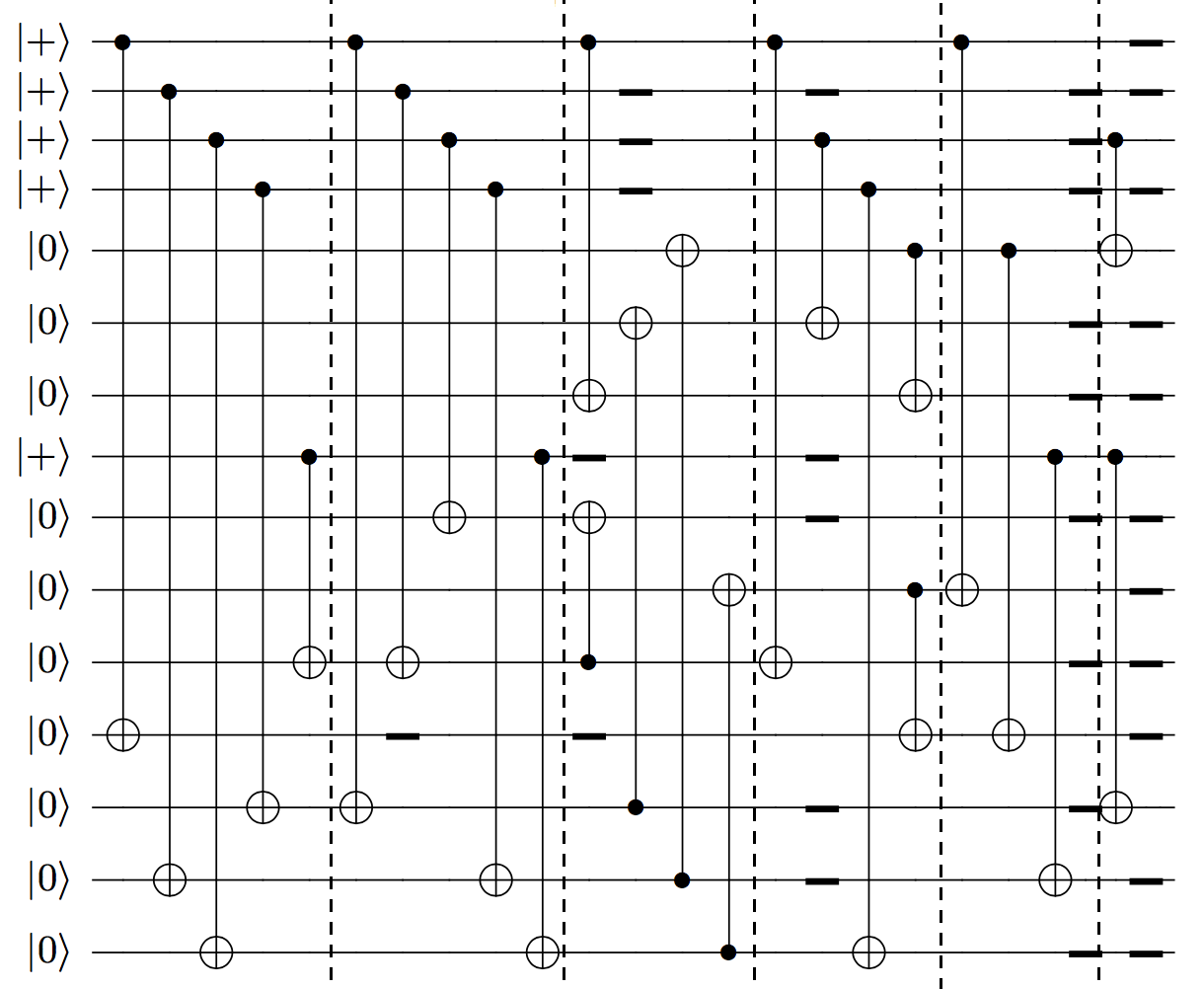}
\caption{}
\label{fig:Optimized15QubitPlusprep}
\end{subfigure}
\caption{Optimized encoding circuits for the 15-qubit Reed-Muller code using the stabilizer overlap method for $\ket{\overline0}_{15}$ and a computer search algorithm for $\ket{\overline+}_{15}$. $\ket{\overline0}_{15}$ contains 22 CNOT gates and 27 resting qubit locations. $\ket{\overline+}_{15}$ contains 25 CNOT gates and 31 resting qubit locations.} 
\label{fig:Optimized15QubitCircuits}
\end{figure}

Next we apply the stabilizer overlap method to obtain optimized encoded $\ket{\overline{0}}_{15}$ and $\ket{\overline{+}}_{15}$ states of the 15-qubit Reed-Muller code. Fig.~\ref{fig:Latin15QubitOprep} and Fig.~\ref{fig:Latin15QubitPlusprep} give the encoding circuits for the $\ket{\overline{0}}_{15}$ and $\ket{\overline{+}}_{15}$  states using Steane's Latin rectangle method. There are 28 CNOT gates in $\ket{\overline{0}}_{15}$ and 32 CNOT gates in $\ket{\overline{+}}_{15}$. For the circuit $\ket{\overline{0}}_{15}$, the stabilizer overlap method removes six CNOT gates. For the $\ket{\overline{+}}_{15}$, instead of using the stabilizer overlap method to reduce the number of CNOT gates, we consider all pairs of CNOT gates that have the same target qubit. For a given pair, we replace the pair by a single CNOT with a different control qubit. We test all 13 different possible controls until we obtain a correct $\ket{\overline{+}}_{15}$ state. Applying the latter technique, we were able to remove seven CNOT gates from the $\ket{\overline{+}}_{15}$ state (see Figs.~\ref{fig:Optimized15QubitOprep} and~\ref{fig:Optimized15QubitPlusprep}). 

Concatenating the optimized circuits of Fig.~\ref{fig:First7qubitCircuitsApp} with the circuits of Fig.~\ref{fig:Optimized15QubitCircuits} and taking the 7-qubit code as the outer code, we obtain the $\ket{\overline{0}}_{105}$ and $\ket{\overline{+}}_{105}$ ancilla states. The 49-qubit code ancilla states are obtained by concatenating the states of Fig.~\ref{fig:7QubitCircuit49} with those of Fig.~\ref{fig:Optimized15QubitCircuits} (see Table~\ref{tab:LocationNumbers} for an enumeration of all the types of locations in 49 and 105-qubit ancilla states). A 1-EC contains four $\ket{\overline{0}}$ states, four $\ket{\overline{+}}$ states, eight encoded CNOT gates, two storage locations and four encoded $Z$-measurement and $X$-measurement locations (see Fig.~\ref{fig:SteaneECcircuit}). Using the values of Table~\ref{tab:LocationNumbers}, the total number of locations in a 1-EC circuit of the 105-qubit code is thus 7110.

\begin{table}
\begin{centering}
\begin{tabular}{|c|c|c|c|c|}
\hline 
& $\ket{\overline0}_{105}$ & $\ket{\overline+}_{105}$ & $\ket{\overline0}_{49}$ & $\ket{\overline+}_{49}$\tabularnewline
\hline 
\hline 
Number of CNOT's & 298 & 301 & 112 & 115\tabularnewline
\hline 
Number of Resting qubits & 246 & 250 & 248 & 252\tabularnewline
\hline 
Number of Physical $\ket{0}$ states & 74 & 73 & 34 & 33\tabularnewline
\hline
Number of Physical $\ket{+}$ states & 31 & 32 & 15 & 16\tabularnewline
\hline 
Total & 649 & 656 & 409 & 416\tabularnewline
\hline 
\end{tabular}
\par\end{centering}
\caption{\label{tab:LocationNumbers}Enumeration for all the types of locations for the  $\ket{\overline0}_{105}$, $\ket{\overline+}_{105}$, $\ket{\overline0}_{49}$ and $\ket{\overline+}_{49}$ states.}
\end{table}

Threshold estimates for the 7-qubit CSS code and the 23-qubit Golay code were calculated in Ref.~\cite{AGP06,PR12}. The threshold calculation was limited by the exRec with the largest number of locations which, in both cases, corresponded to the CNOT-exRec. Given the non-transversal nature of the $T$-gate in the 7-qubit code and the Hadamard gate in the 15-qubit code, the threshold calculation is no longer be limited by the exRec with the largest number of locations. As was shown in \ref{Threshold results for adversarial noise}, the threshold calculation is instead limited by the Hadamard-exRec. It is thus worthwhile to analyze the Hadamard circuit in more detail.

\begin{figure}%[htbp]
\centering
\begin{align*}
\includegraphics{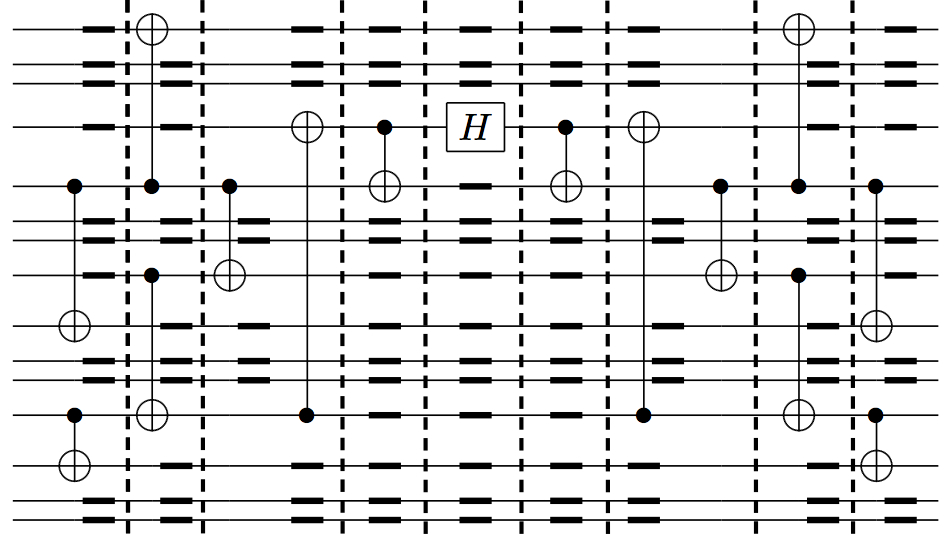}
\end{align*}
\caption{Logical Hadamard~$H$ circuit for $[[15,1,3]]$~Reed-Muller code. Logical~$H$ for the 105-qubit code is implemented fault-tolerantly by applying each non-fault-tolerant logical~$H$ gates in parallel. Note that there are a total of nine time steps. Consequently, the 49-qubit code Hadamard circuit will contain a single physical Hadamard and 8 resting qubit locations on blocks 2-5.}
\label{fig:HadCircuit}
\end{figure}

The encoding circuit for the Hadamard gate must map the stabilizer generators of the 15-qubit code to an element of the stabilizer group. Furthermore, since $HXH=Z$ and $HZH=X$, the logical $X$ operator of the 15-qubit code must be mapped to the logical $Z$ operator and vice-versa. The circuit in Figure~\ref{fig:HadCircuit} satisfies these properties. To derive such a circuit, we wrote a program in Matlab which inserted CNOT gates at random locations within the 15 physical qubits and propagated all stabilizer generators and logical operators through the circuit. If all operators transformed appropriately as described above, then the locations of the CNOT gates were recorded. The best circuit that we found contains a total of 14 CNOT gates, one physical Hadamard gate and has a depth of 9 time steps (see Table~\ref{tab:NumLocationsHadamard} for a complete enumeration of the locations in the Hadamard circuit for the 15-qubit, 49-qubit and 105-qubit code). However, it is still an open question whether a circuit using fewer CNOT gates with a smaller depth can be found.  

\begin{table}
\begin{centering}
\begin{tabular}{|c|c|c|c|}
\hline 
& $H_{15}$ & $H_{49}$ & $H_{105}$ \tabularnewline
\hline 
\hline 
Number of CNOT's & 14 & 42 & 98\tabularnewline
\hline 
Number of Resting qubits & 106 & 350 & 742\tabularnewline
\hline 
Number of Physical $H$ states & 1 & 7 & 7\tabularnewline
\hline
Total & 121 & 399 & 847\tabularnewline
\hline 
\end{tabular}
\par\end{centering}
\caption{\label{tab:NumLocationsHadamard}Enumeration of all the types of locations for the 15-qubit, 49-qubit and 105-qubit Hadamard gate. Restricting to the 105-qubit code, since a 1-EC circuit contains 7110 locations, the total number of locations in the Hadamard-exRec is 15067. These quantities will be relevant in the adversarial noise threshold calculation of the Hadamard-exRec as well as in the resource overhead calculation.}
\end{table}

One important aspect of the circuit in Fig.~\ref{fig:HadCircuit} is that input $Z$ errors from the LEC are much more likely to lead to a logical error at the output of the Hadamard circuit than input $X$ errors. To illustrate this, we consider two different cases. In the first case, we insert a single $X$ error at the input of the Hadamard circuit and propagate the error throughout the circuit to determine if a logical error occurred at the output of the circuit, which occurs in the case of qubits 4, 8 and 12. In the second case, we perform the exact same operations but with a single $Z$ error. In this case, qubits 1, 4, 5, 8, 9, 12 and 13 produce different logical faults at the output of the Hadamard circuit compared to when input $X$ errors were considered. It should then be expected that $Z$ errors will play an important role when calculating the noise threshold for the concatenated scheme. 

For the 105-qubit code, logical Hadamard is obtained by applying the non-transversal logical Hadamard of Fig~\ref{fig:HadCircuit} to each of the encoded 15-qubit codeblocks. For the 49-qubit code, the logical Hadamard gate is obtained by applying the non-transversal logical Hadamard of Fig~\ref{fig:HadCircuit} to blocks 1, 6 and 7. Blocks 2-5 will consist of a physical Hadamard gate along with 8 resting qubit locations (since there are a total of 9 time-steps in Fig~\ref{fig:HadCircuit}).

 As was explained above, a single error can lead to a logical fault when propagating through the 15-qubit Hadamard circuit. However, since the Hadamard circuit is transversal for the 7-qubit CSS code, a weight one error will lead to at most a \textit{single} logical fault on one of the codeblocks which will be corrected by the outer code. Thus the Hadamard circuit for the 49 and 105-qubit code is fault-tolerant. 

\begin{figure}%[htbp]
\centering
\begin{align*}
\includegraphics{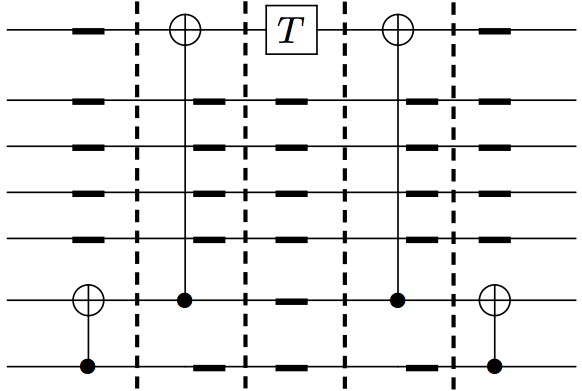}
\end{align*}
\caption{Logical $T$ circuit for $[[7,1,3]]$~CSS code. All stabilizer generators map to elements in the stabilizer group. Furthermore, $\overline{X}\to \overline{X}(I+\overline{Z})$ and $\overline{Z}\to \overline{Z}$ as required.}
\label{fig:TgateCircuit}
\end{figure}

Next we analyze the construction of the $T$-gate circuit for the 7-qubit code following the ideas of Ref.~\cite{JL14,CJL16}. Similarly to the Hadamard circuit, to obtain a circuit that correctly encodes the $T$-gate for the 7-qubit code, all stabilizer generators must be mapped to elements of the stabilizer group. Furthermore, since $TXT^{\dagger}=\frac{1}{\sqrt{2}}(X+Y)$ and $TZT^{\dagger}=Z$, we require that the logical operators for the 7-qubit code transforms as $\overline{X}\to \overline{X}(I+\overline{Z})$ and $\overline{Z}\to \overline{Z}$. The circuit in Fig.~\ref{fig:TgateCircuit} satisfies the above properties. It should be noted that regardless of the error model studied, we treated the transformation of $X$ errors in our simulations by taking an adversarial approach to their transformation to~$X$ or~$Y$ errors, depending on the other errors in the circuit. That is, we considered both the case when the $X$~error transformed to either $X$~or~$Y$, and treated the worst case logical error outcome based on the choice of error. This sufficed to prove a lower bound on the appropriate threshold.

Logical $T$ for the 49 and 105-qubit code is constructed from the circuit in Fig.~\ref{fig:TgateCircuit} with each codeblock encoded using the 15-qubit Reed-Muller code (for the 49-qubit code, blocks 2-5 only contain one qubit). The construction is not fault-tolerant on the outer code since errors can spread between codeblocks. However, since the underlying logical gates are transversal on the 15-qubit codeblocks, a single error would propagate to at most a single error on each codeblock which would be corrected by the inner code.

\end{document}